\documentclass[prb,aps,twocolumn,eqsecnum,showpacs,superscriptaddress,amsmath,amssymb]{revtex4}

\usepackage{graphicx}
\usepackage{psfrag}
\usepackage{epsfig}
\usepackage{color}
\usepackage{subfigure}
\definecolor{dred}{rgb}{0.7,0.0,0.0}
\usepackage{dcolumn}% Align table columns on decimal point
\usepackage{bm}% bold math

\begin{document}

\title { Spectral properties of orbital polarons in Mott insulators 
}

\author {     Krzysztof Wohlfeld }
\affiliation{ Max-Planck-Institut f\"ur Festk\"orperforschung,
              Heisenbergstrasse 1, D-70569 Stuttgart, Germany }
\affiliation{ Marian Smoluchowski Institute of Physics, Jagellonian
              University, Reymonta 4, PL-30059 Krak\'ow, Poland }

\author {     Maria Daghofer }
\affiliation{ Max-Planck-Institut f\"ur Festk\"orperforschung,
              Heisenbergstrasse 1, D-70569 Stuttgart, Germany }
\affiliation{ Materials Science and Technology Division, Oak
  Ridge National Laboratory, Oak Ridge, Tennessee 37831, USA and
  Department of Physics and Astronomy, The University of Tennessee,
  Knoxville, Tennessee 37996, USA}

\author {     Andrzej M. Ole\'{s} }
\affiliation{ Max-Planck-Institut f\"ur Festk\"orperforschung,
              Heisenbergstrasse 1, D-70569 Stuttgart, Germany }
\affiliation{ Marian Smoluchowski Institute of Physics,
              Jagellonian University, Reymonta 4,
              PL-30059 Krak\'ow, Poland }

\author {     Peter Horsch }
\affiliation{ Max-Planck-Institut f\"ur Festk\"orperforschung,
              Heisenbergstrasse 1, D-70569 Stuttgart, Germany }

\date{\today}

\begin{abstract}
We address the spectral properties of Mott insulators with orbital 
degrees of freedom, and 
investigate cases where the orbital symmetry leads to 
Ising-like superexchange in the orbital sector. The paradigm of a 
hole propagating by its coupling to quantum fluctuations, known 
from the spin $t$--$J$ model, then no longer applies. We find instead 
that when one of the two orbital flavors is immobile, as in the 
Falicov-Kimball model, trapped orbital polarons coexist with free 
hole propagation emerging from the effective three-site hopping in 
the regime of large on-site Coulomb interaction $U$. 
The spectral functions are found analytically in this case within the 
retraceable path approximation in one and two dimensions. On the 
contrary, when both of the orbitals are active, as in the model for
$t_{2g}$ electrons in two dimensions, we find propagating polarons with 
incoherent scattering dressing the moving hole and renormalizing 
the quasiparticle dispersion. Here, the spectral functions, 
calculated using the self-consistent Born approximation, 
are anisotropic and depend on the orbital flavor. 
Unbiased conclusions concerning the spectral properties are 
established by comparing the above results for the orbital $t$--$J$ models 
with those obtained using the variational cluster approximation or 
exact diagonalization for the corresponding Hubbard models.
The present work makes predictions concerning the essential features 
of photoemission spectra of certain fluorides and vanadates.
\end{abstract}


\maketitle

\section{Introduction}
\label{sec:intro}

A hole propagates coherently in the valence band of a band insulator 
with an unrenormalized one-particle dispersion 
(i.e., determined by electronic structure calculations), whereas hole
propagation in a Mott insulator is a nontrivial many-body problem.
The paradigm here is a hole doped in the half filled one-band Hubbard 
model that is a minimal model used to describe the parent compounds of 
high-$T_c$ superconductors. Such a hole forms a defect in the 
antiferromagnetic (AF) background, and its coherent propagation may 
appear then only on a strongly renormalized energy scale.\cite{Ima98} 
Naively, i.e., considering the N\'eel state induced by Ising-like spin 
interactions, one expects that a propagating hole would disturb the AF 
background and generate a string of broken bonds, with ever increasing 
energy cost when the hole creates defects moving away from its initial 
position. This suggests hole confinement as realized already four 
decades ago.\cite{Bul68} 
Nevertheless, the quantum nature of this problem leads to a new 
quality: a hole in the AF Mott insulator can propagate coherently 
on the energy scale $J$ which controls AF quantum fluctuations,
\cite{Sch89,Kan89,Mar91} because they heal the defects arising on the 
hole path. Crucial for this observation is the presence of transverse 
spin components  $\propto (S_i^+S_j^-+S_i^-S_j^+)$, 
responsible for quantum fluctuations 
in the SU(2)-symmetric Heisenberg exchange interactions.

In contrast, the orbital interactions which induce alternating orbital 
(AO) order are known to be more Ising-like\cite{vdB04} (classical) and 
quantum fluctuations are then either substantially reduced,\cite{vdB99} 
or even absent. Perhaps the most prominent example of robust orbital 
order occurs for degenerate $e_g$ orbitals in the ferromagnetic (FM) 
planes of LaMnO$_3$.\cite{Dag01} Orbital interactions are there induced 
both by the lattice (due to cooperative Jahn-Teller effect) and by the
superexchange\cite{Fei99} --- both of them are classical and
Ising-like, so the quantum fluctuations are to a large extent 
suppressed. Therefore, the orbital order is more robust than the spin 
order in two-dimensional (2D) models,\cite{Dag07} and a higher hole
concentration is required to destroy it.\cite{Hor99}
In the three-dimensional case off-diagonal (interorbital) hopping 
($e_g$ orbital flavor is here not conserved\cite{Zaa93}) may 
even lead to an orbital liquid phase already at rather low hole 
doping.\cite{Fei05}

Another difference to the SU(2) spin model is that a hole in a 
ferromagnet with AO order of $e_g$ orbitals can propagate coherently 
without introducing any string states. This coherent propagation arises 
here due to the $e_g$ interorbital hopping, but is strongly renormalized 
by orbital excitations.\cite{vdB00} However, in the $t_{2g}$ systems 
the interorbital hopping is forbidden by symmetry,\cite{Zaa93} and 
the superexchange is purely Ising-like, so the above established 
mechanisms of coherent hole propagation in the regime of large on-site 
Coulomb repulsion $U$, i.e., for $t\ll U$ where $t$ is the hopping 
element, are absent. One may then wonder whether a hole doped to a Mott 
insulator with the $t_{2g}$ AO order would then be confined.\cite{Dag08} 

The present paper is motivated by the above important difference 
between the spin physics and the orbital physics, 
especially prominent in the hole motion in the $t_{2g}$ orbital systems. 
\cite{Dag08} In spin models, the 
Ising-like superexchange represents only a rather poor approximation 
for the SU(2)-symmetric Heisenberg spin exchange (as obtained in the 
$t$-$J$ model by its derivation from the Hubbard model \cite{Cha77})
and could merely serve as a starting point for the full SU(2)-symmetric 
calculations. \cite{Wro08}
In contrast, in the systems with $t_{2g}$ orbital degeneracy the Ising 
superexchange follows from a similar derivation as an accurate 
description of charge excitations in the second order of the 
perturbation theory in the regime of $t\ll U$, in cases when only two 
orbitals are active (see below). Strictly speaking, in these realistic 
orbital systems the Ising-like superexchange occurs in a two-band 
model when precisely one orbital flavor permits the hopping along each 
bond, as only then the orbital flavor cannot be exchanged and the 
pseudospin $(T^+_iT^-_j+T^-_iT^+_j)$ operators are absent in the 
respective $t$--$J$ model.\cite{noteti}
Such a situation is found not only in the above mentioned $t_{2g}$ 
model but also in three other distinct orbital models (see below). 
Hence, in the following paragraphs we give a brief 
overview of these \textit{four models with realistic Ising 
superexchange\/} for which the spectral function of a single hole 
doped into the half-filled ground state will be studied in this paper.

As a first example we introduce the spinless Falicov-Kimball (FK) 
model which describes itinerant $d$ electrons coupled to localized 
$f$ electrons. On sites occupied by an $f$ electron, the $d$ electrons 
feel a strong on-site Coulomb repulsion $U$. The FK model can be solved 
exactly in infinite dimension,\cite{Fre03} where it leads to a complex 
phase diagram including periodic ground states as well as a regime of 
phase separation. When the energies of two involved orbitals are 
degenerate (which is not the case in $4f$ or $5f$ materials), 
one finds that electron densities in the two orbitals are the same 
($n_d=n_f$), and second order perturbation theory leads in the regime 
of large $U$ to a strong-coupling model with one mobile flavor. 
One finds therefore the ground state with the AO order formed by
sites occupied by $d$ and $f$ electrons on the two sublattices. The 
spectral density for the mobile $d$ electrons can be obtained as well 
in one~\cite{Bec07} and two dimensions,\cite{Mas06} but computation of 
the $f$ spectral density is quite involved even at infinite dimension.
\cite{Fre05} Below we will discuss exact one-dimensional (1D) and 
approximate two-dimensional (2D) analytic results for the relevant 
FK models.

A second and different realization of the effective low-energy model 
with Ising superexchange was proposed recently\cite{Dag08} for FM 
planes in transition metal oxides with $t_{2g}$ orbital degeneracy: 
In this case, both orbital flavors are equivalent (the third one is
inactive) and both allow for electronic hopping, however, each one 
permits the hopping along one axis only. Despite the 1D character of 
the kinetic energy in such a model, the ground state at half filling 
has 2D AO order, stabilized by the Ising superexchange. Electron 
propagation, on the other hand, is strictly 1D, so a hole replacing an 
electron with either orbital flavor may
only move in one direction by the hopping $t$. Such a situation might 
be realized in Sr$_2$VO$_4$, where the crystal field splits the $t_{2g}$ 
orbitals \cite{Mat05} and one finds indeed AO order\cite{Ima05} in the 
weakly FM planes.~\cite{Noz91} A different possible realization of such 
a model is found in cold-atom systems,\cite{Jan05} with strongly 
anisotropic hopping in the spinless $p$-orbital Hubbard model.\cite{p_Hubb}

The third realization concerns systems with peculiar $e_g$ AO and is 
somewhat subtle. As described above in the "well-known" systems with 
active $e_g$ orbitals (such as the perovskite manganites) the hole can 
propagate coherently due to the interorbital hopping.\cite{Fei05}
However, one can identify two peculiar cases of $e_g$ AO order where 
the interorbital hopping is strongly reduced between the orbitals 
occupied in the ground state, and the question of the hole confinement 
in the Ising superexchange model is of high relevance. The first one is 
realized in the FM planes of K$_2$CuF$_4$,\cite{Hid83} or in the 
recently investigated Cs$_2$AgF$_4$.\cite{Wu07} 
While the AO order is formed in these cases by $e_g$ orbitals, crystal 
field stabilizes their particular linear combinations with alternating 
$x^2-z^2/y^2-z^2$ $e_g$ orbitals,\cite{McL06} and thus suppresses the 
interorbital hopping present, for instance, in the ground state of the
manganites. In addition, the phase factors of two orbitals along each
bond allow for the hopping only along one direction in the plane,
\cite{Zaa93} so one arrives at a situation similar to that 
found for strongly correlated electrons in $t_{2g}$ orbitals
--- it will be discussed in the Appendix \ref{app:fluo}. Thus, the model
called $t_{2g}$ model throughout this paper is expected to describe 
a wider class of transition metal oxides.

Finally, another case where the AO order could in principle lead to 
the hole confinement is the situation encountered in the 1D $e_g$ 
model in which the interorbital hopping and quantum fluctuations are 
suppressed by symmetry. Since 
this model is actually equivalent to the 1D FK model discussed above,
we will refer to it later simply as to the "1D model". Its 
study was stimulated by the experimental findings in the lightly
hole-doped vanadates such as La$_{1-x}$Sr$_x$VO$_3$.\cite{Fuj08} 
There the ground state for $x=0.1$ is an insulating three-dimensional 
FM phase with AO order at $x=0.10$. The physical origin of the Mott 
insulating state in the case of such a finite hole doping is not yet
understood. Again, a natural question could be whether the peculiar 
stability of the Mott insulating state is related to the presence of 
the AO order. Since it is obvious that the hole can move in the AO 
state in the plane (due to the interorbital hopping, see above) it is 
interesting to verify whether the AO order along the third direction 
could block the hole motion.

The four above cases are, to our knowledge, the only straightforward
models in which the full superexchange is purely classical, i.e., 
where Ising superexchange follows from the orbital symmetry, and is 
not an approximation to the Heisenberg Hamiltonian.\cite{Poi04} The 
superexchange models, however, have to be extended by the second order 
three-site hopping terms in each case. Such effective hopping terms 
arise in the same order of the perturbation theory when holes are
present, and are necessary for a faithful representation of the spectral 
weight distribution in the one-particle and in the optical spectra of 
the underlying spin Hubbard model.\cite{Esk94} Therefore, also in the 
present orbital $t$-$J$ models with Ising superexchange similar terms 
are expected to play an 
important role and cannot be neglected. Moreover, while in the spin case 
such terms are in conflict with the quantum fluctuations and give thus 
only minor quantitative corrections to the coherent hole motion,
\cite{vSz90} in the orbital (pseudospin) models with Ising superexchange 
they become of crucial importance as they are the only possible source 
for coherent carrier propagation (the Trugman loops\cite{Tru88} with the 
hole repairing the defects on its path are here absent)
and dictate possible coherent processes.\cite{Dag08}
To get a better understanding of the balance between the coherent
and incoherent processes in the orbital strong-coupling models
(i.e. $t$-$J$ models with three-site terms)
we analyze their spectral properties in some detail below.

In order to arrive at a comprehensive and rather complete understanding 
of the elemental processes which accompany hole propagation in the 
orbital models, we not only combine numerical and analytical approaches
but also we calculate the spectral properties both of the strong-coupling
and of the respective Hubbard models. Hence, we first 
determine the Green's functions using the self-consistent Born 
approximation (SCBA) and the analytical treatments applied to the 
above mentioned orbital strong-coupling models. This allows us to identify 
the dominant mechanism responsible for the quasiparticle (QP) behavior. 
We note that for finite doping such methods as the slave boson approach, 
the path integral formalism, or the numerical approaches, are 
much more suitable for the $t$-$J$-like models than the 
SCBA.\cite{Gre06} However, in the one hole limit the SCBA 
gives reasonably good results which are in agreement with 
other methods.\cite{Mar91,Gre06}.
We then compare these results to those obtained for the 
respective orbital Hubbard model. Here, in some cases we determine 
Green's functions numerically by use of exact 
diagonalization on small clusters, in others we use the variational 
cluster approach (VCA),\cite{Aic03} where a cluster is solved exactly 
and then embedded into a larger system. 
This variational approach is based on cluster perturbation theory 
developed in the last decade,\cite{Gro93,Mas98,Sen00} and
corresponds to taking the self-energy from a small cluster and 
optimizing it with respect to mean-field terms arising due to 
the AO order. The embedding via the self-energy approach\cite{Pot03} 
allows us to include long-range (orbital) ordering 
phenomena by optimizing a fictitious field due to the AO order. 
This method is appropriate for orbital Hubbard models 
with on-site interactions. Since the exact solution 
on the cluster is obtained for the full Hamiltonian, it contains all
potentially relevant processes like, e.g., the three-site hopping. 

The paper is organized as follows. The 1D orbital model is introduced 
in Sec. \ref{sec:1Dmodel} whereas its extension to the 2D FK model is 
discussed in Sec. \ref{sec:FalKim}. In both sections, we introduce the 
respective Hubbard-like Hamiltonian, derive from it the appropriate 
strong-coupling Hamiltonian, and calculate analytically the hole Green's 
function for the AO state at half filling. Next, we introduce an 
exactly-solvable 1D model with three-atom units along the chain 
(Sec. \ref{sec:1Dlegs}), called the 1D 'centipede' model. The latter 
model (which was not mentioned above) serves merely as a didactic tool 
and explains the essence of string excitations present in the 2D model 
with $t_{2g}$ orbital flavors, discussed thoroughly in Sec. 
\ref{sec:t2g}. Here, again we start from the orbital Hubbard model, derive its 
strong-coupling version, and calculate the hole Green's function for the 
AO state at half filling, using two approximate methods described above: 
the SCBA and the VCA. 
In Sec. \ref{sec:NNN} we include longer-range hopping in the 2D $t_{2g}$
model (as expected in real materials) and discuss the main experimental 
implications of our study by calculating the photoemission spectra of 
certain vanadates and fluorides.
General conclusions are presented in Sec.~\ref{sec:conclusions}. The 
analysis is supplemented by the Appendix \ref{app:fluo}, where we derive 
the effective strong-coupling model for the above mentioned fluorides 
and prove that the $t_{2g}$ model discussed in Sec.~\ref{sec:t2g} 
may indeed be applied to the hole motion in the systems with 
a particular type of $e_g$ orbital order.

%%%%%%%%%%%%%%%%%%%%%%%%%%%%%%%%%%%%%%%%%%%%%%%%%%%%%%%%%
%                         1D model
%%%%%%%%%%%%%%%%%%%%%%%%%%%%%%%%%%%%%%%%%%%%%%%%%%%%%%%%%
\section{1D orbital model with Ising superexchange}
\label{sec:1Dmodel}

\subsection{Effective strong-coupling model}

As explained in Sec. \ref{sec:intro}, the Ising-like superexchange 
follows if only one orbital flavor permits hopping along each bond,
and the spins are polarized in the FM state.
The simplest case which captures the essential features of the 
effective strong-coupling model with Ising-like superexchange 
follows from the 1D orbital Hubbard model
\begin{equation}
\label{eq:hubb_1d}
H_{\rm 1D}=-t\sum_{i}(a^{\dagger}_{i}a^{}_{i+1} + \textrm{h.c.})
+U \sum_i n_{ia} n_{ib}\;,
\end{equation}
where $a^{\dagger}_{i}$ ($b^{\dagger}_{i}$) creates a spinless 
electron with 
orbital flavor $a$ ($b$) at site $i$, and $\{n_{ia},n_{ib}\}$ are 
electron density operators. On-site Coulomb repulsion $U$ is the 
energy of a doubly occupied state (it arises as a linear combination 
of the Coulomb and Hund's exchange in the respective high-spin
configuration\cite{vdB99}), and $t$ is the nearest neighbor (NN) 
hopping element. 
Only electrons with orbital flavor $a$ are mobile while the other 
ones with flavor $b$ cannot hop. To simplify, we call below the $a$ and
$b$ orbitals mobile and immobile ones, respectively. This situation 
corresponds to (spinless) interacting $e_g$ electrons in the FM chain,
\cite{Dag04} or to the 1D (spinless) FK model with degenerate orbitals. 

In the regime of large $U$, i.e. for $t\ll U$, second order 
perturbation theory leads to the effective strong-coupling Hamiltonian
with Ising-like superexchange
\begin{equation}
\label{H_1D}
{\cal H}_{\rm 1D} = H_{t}+H_{J}+H_{\rm 3s}\,,
\end{equation}
where
\begin{eqnarray}
\label{Ht_eg} 
H_{t} &=&
-t \sum_{i} \left(\tilde{a}^\dag_i \tilde{a}^{}_{i+1}+ \mbox{h.c.}\right)\,, \\
\label{HJ_eg} 
H_{J} &=& \frac12 J \sum_{i }
\left(T^z_i T^z_{i+1} - \frac{1}{4}\tilde{n}_i\tilde{n}_{i+1}\right),\\
\label{H3s_eg} 
H_{\rm 3s}  &=& -\tau \sum_{i}
\left(\tilde{a}^\dag_{i-1}\tilde{n}^{}_{ib}\tilde{a}^{}_{i+1} 
+\mbox{h.c.}\right)\;.
\end{eqnarray}
Here a tilde above a fermion operator indicates that the Hilbert space 
is restricted to unoccupied and singly occupied 
sites, e.g. $\tilde{a}^\dag_i=a^\dag_i(1-n_{bi})$. The pseudospin
operators are defined as follows
\begin{equation}
\label{Tz}
T^z_i = \frac{1}{2} \left(\tilde{n}_{ib}-\tilde{n}_{ia}\right)\,,
\end{equation}
and the superexchange constant $J$ and the effective hopping parameter
$\tau$ are given by
\begin{equation}
\label{J}
J=\frac{4t^2}{U}, \hskip 1.2cm \tau=\frac{t^2}{U}\,.
\end{equation}
We introduced above the parameter $\tau$ in order to 
distinguish below between the terms which originate from the pseudospin 
superexchange and the hopping processes arising from the superexchange via 
the three-site terms that lead to the second or third neighbor effective 
hopping and contribute to the hole dispersion in the strong-coupling
regime. Note that $\tau$ is of 
the same order $\propto t^2/U$ as $J$, so {\it a priori\/} these terms 
cannot be neglected. But similar as for the constrained hopping term
(\ref{HJ_eg}), their contribution is proportional to hole doping $x$. 
The 1D $t$-$J$ orbital model ${\cal H}_{tJ}=H_t+H_{J}$, 
i.e. without the three-site hopping $H_{\rm 3s}$, was solved 
exactly before\cite{Dag04} and all excitations occurred to be 
dispersionless. Here we generalize this exact solution to the 
full strong-coupling Hamiltonian (\ref{H_1D})
{\em including\/} the three-site terms, and show that the spectral 
functions for both orbital flavors are then distinctly different.

\subsection{Analytic Green's functions}
\label{sec:1dresults}

We calculate below the \emph{exact} Green's functions $G_a(k,\omega)$ 
and $G_b(k,\omega)$ which demonstrate whether and how a hole added 
to an $a$ (or $b$) orbital may propagate coherently along a 1D chain 
with the AO order. Interestingly, both functions can be determined
\emph{analytically} using retraceable path approximation
\cite{Bri70} (RPA), which becomes here exact because 
closed loops are absent in the 1D system.\cite{noteis} 

An important simplification as compared with the spin case is the 
knowledge of the exact ground state $|0\rangle$ at half filling. 
As the Hamiltonian given by Eq. (\ref{H_1D}) does contain then only 
the Ising superexchange, the N\'eel state 
\begin{equation} 
\label{gs}
|0\rangle=\prod_{i\in A} a_i^{\dagger}\prod_{j\in B} b_j^{\dagger}\,
|{\rm vac}\rangle\,,
\end{equation}
with $a$ orbitals occupied on the sublattice $A$ and $b$ orbitals 
occupied on the sublattice $B$ is an exact ground state. Here
$|{\rm vac}\rangle$ is the true vacuum state with no electrons, 
while $|0\rangle$ is the physical vacuum at half filling.

\begin{figure}[t!]
   \includegraphics[width=0.4\textwidth]{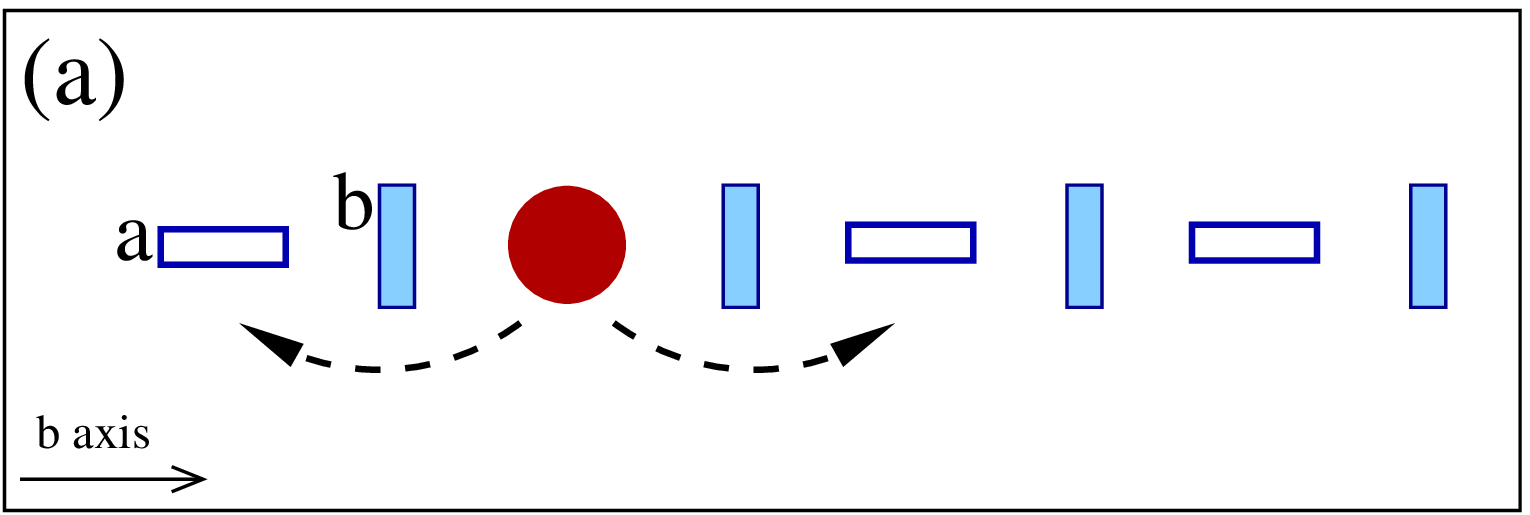}
\\[1.0em]
   \includegraphics[width=0.4\textwidth]{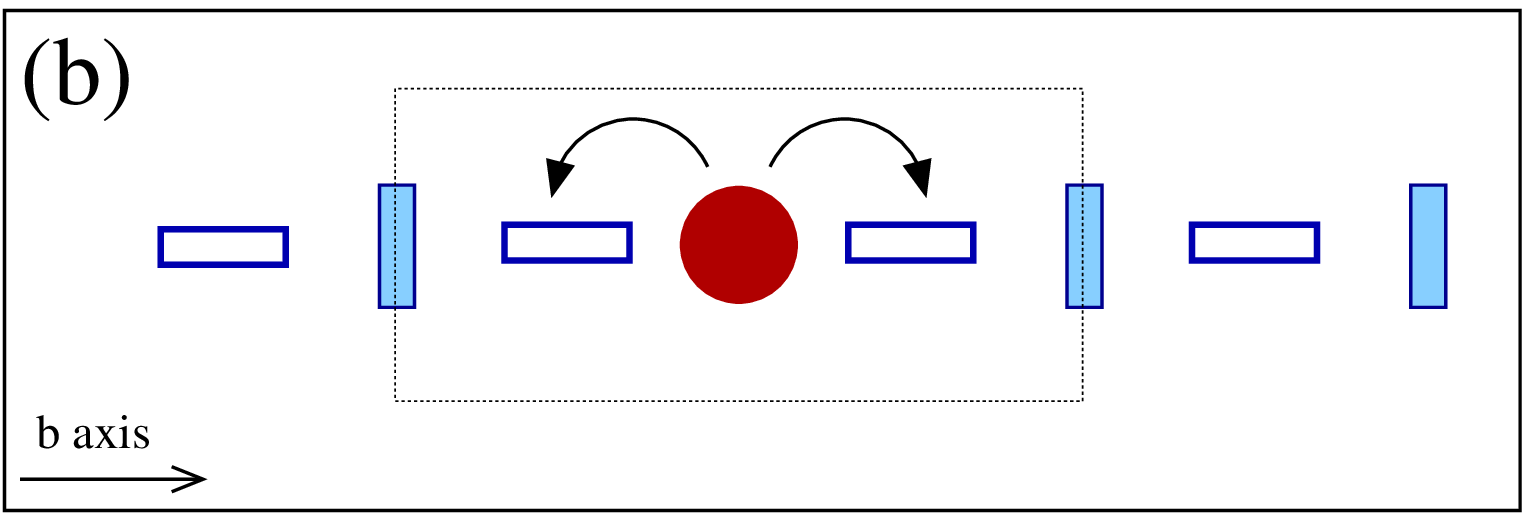}
\\[1.0em]
   \includegraphics[width=0.4\textwidth]{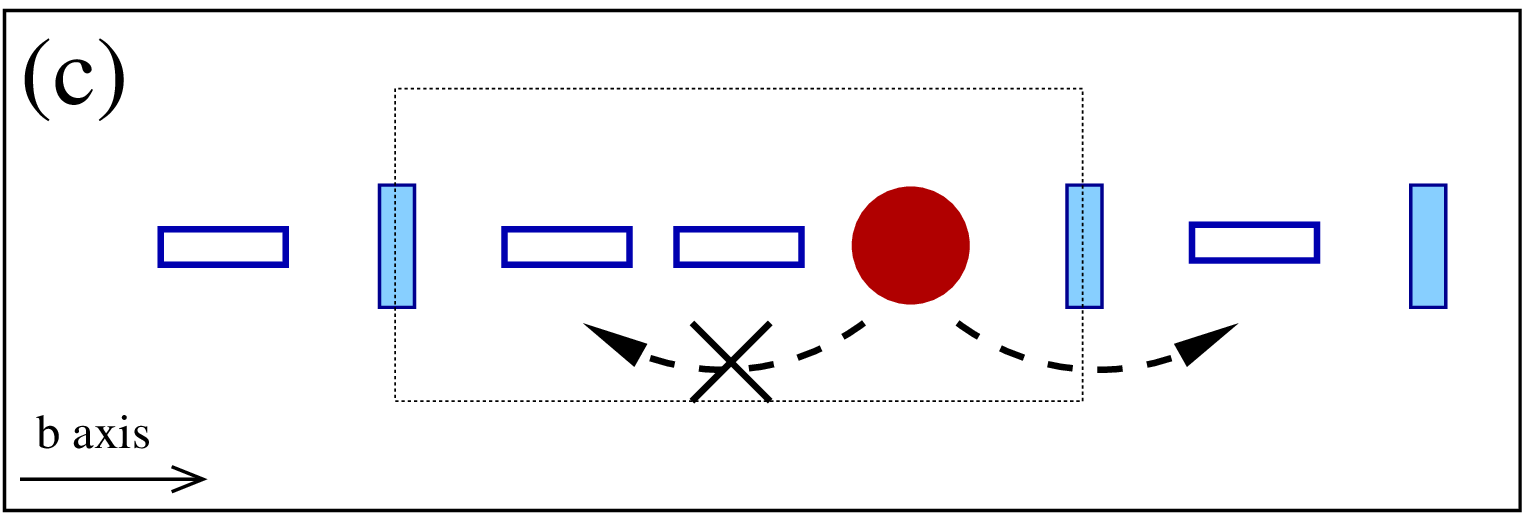}
\\[1.0em]
   \includegraphics[width=0.41\textwidth]{holot1d}
\caption{(Color online)
Hole propagation in the 1D strong-coupling model (\ref{H_1D}). 
Two top panels show a hole doped into: 
(a) mobile $a$ orbitals (empty boxes), and 
(b) immobile $b$ orbitals (filled boxes). 
Solid (dashed) arrows indicate possible hopping processes with hopping
elements $t$ and $\tau$, respectively; in case of a hole added to the 
$b$ orbital the latter process occurs only after the initial hopping
by $t$, see panel (c). Panel (d) shows the exact spectral functions 
$A_a(k,\omega)$ and $A_b(k,\omega)$ of a hole 
added into the $a$ orbital (middle dispersive feature between 
$\omega=-0.4t$ and $\omega=0$) and the $b$ orbital (two side dispersionless maxima)
as obtained from the 1D strong-coupling model (\ref{H_1D}). 
Parameters: $J=0.4t$, $\tau=0.1t$, and peak broadening $\delta=0.01t$.
The spectral functions obtained using the VCA for the 1D Hubbard model 
(\ref{eq:hubb_1d}) with $U=10t$ (not shown) are {\em identical\/} to 
the exact result. 
} 
\label{fig:1d:1}
\end{figure}

We start with the Green's function for the hole doped in the mobile $a$ orbital,
\begin{equation} 
\label{eq:defga}
G_a(k,\omega)=\lim_{\delta\to 0}\,\left\langle0\left|\,a_k^{\dagger}
\frac{1}{\omega + {\cal H}_{\rm 1D} - E_0+i\delta}\,a_k^{}\right|0\right\rangle\,,
\end{equation}
where $E_0$ is the energy of the physical vacuum at half filling $|0\rangle$, 
$a_k^{\dagger}$ is a Fourier transform of the 
$\{a_j^{\dagger}\}$ operators with $j\in A$, and the hole is 
created by the operator
\begin{equation} 
\label{eq:ak+}
a_k^{}=\sqrt{\frac{2}{N}}\sum_{j\in A}e^{-ikR_j}a_j^{}\,,
\end{equation}
with $N/2$ being the number of sites in one sublattice. 
By construction, the above operator creates a hole (annihilates 
an electron) with momentum $k$ on the $A$ sublattice. After a hole is 
created, one finds that the state $a_k|0\rangle$ in Eq. (\ref{eq:defga}) 
is an eigenstate of the Hamiltonian (\ref{H_1D}). The hopping 
$\propto t$ is blocked by the constraint in the Hilbert space, and the 
only two terms that contribute in this state are: 
(i) the superexchange term (\ref{Ht_eg}) which gives the energy 
$\frac12 J$ of two missing bonds, and 
(ii) the three-site hopping term (\ref{H3s_eg}) which contributes to
the $k$ dependence due to the processes shown in Fig. \ref{fig:1d:1}(a) 
after Fourier transformation. As a result, one finds
\begin{equation} 
\label{eq:1d:ga}
G_a (k,\omega)= \frac{1}{\omega+\frac12 J+2\tau\cos(2k)}\;.
\end{equation}
Note that $\tilde{n}_{ib}\equiv 1$ in $H_{\rm 3s}$, as in this case all 
the sites with $j\in B$ are occupied by $b$ electrons in the ground 
state (\ref{gs}). The hole spectral function,
\begin{equation} 
\label{eq:1d:a}
A_a(k,\omega)=-\frac{1}{\pi}\,\mbox{Im}\,G_a(k,\omega)\,,
\end{equation}
consists of a single dispersive state, shown as the middle peak in Fig. 
\ref{fig:1d:1}(d). As expected, the hole is mobile thanks to the 
three-site terms and it propagates coherently with the unrenormalized 
bandwidth $W=4\tau$. The result obtained here is {\it identical\/} 
with the one found using the VCA for the corresponding Hubbard model 
(\ref{eq:hubb_1d}) (see also Fig.~5 of Ref.~\onlinecite{Dag08}).
This confirms that both the orbital Hubbard model (\ref{eq:hubb_1d}) and 
its strong-coupling version {\it with\/} three-site terms (\ref{H_1D}) 
are equivalent and describe precisely the same physics in the regime of 
$t\ll U$.

%%%%%%%%%%%%%%%%%%%%%%%%%%%%%%%%%%%%%%%%%%%%%%%%%%%%%%%%
%%                       figure 1
%%%%%%%%%%%%%%%%%%%%%%%%%%%%%%%%%%%%%%%%%%%%%%%%%%%%%%%%

When one attempts to calculate the Green's function for a hole doped in 
the immobile $b$ orbital, 
\begin{equation} 
\label{eq:defgb}
G_b(k,\omega)=\lim_{\delta\to 0}\,\left\langle0\left|\,b_k^{\dagger}
\frac{1}{\omega+{\cal H}_{\rm 1D}-E_0+i\delta}\,b_k^{}\right|0\right\rangle\,,
\end{equation}
one finds immediately that the state 
\begin{equation} 
\label{bk}
|\psi_k^{(1)}\rangle\equiv b_k^{}|0\rangle=
\sqrt{\frac{2}{N}}\sum_{j\in B}e^{-ikR_j}b_j^{}|0\rangle\,, 
\end{equation}
is not an eigenstate of the Hamiltonian ${\cal H}_{\rm 1D}$.
Here a hole is doped in each Fourier component in an occupied $b$ 
orbital at site $j$ in the ground state with AO order (\ref{gs}). 
When a hole is doped it can delocalize to its neighbors in the 1D
chain, as depicted in Fig.~\ref{fig:1d:1}(b), so one has to introduce 
appropriate basis of states obtained when the single hole delocalizes
along the 1D chain. The hopping $H_t$ acting on
$|\psi_k^{(1)}\rangle$ generates the first (normalized) state 
\begin{equation} 
\label{bk1}
|\psi_k^{(2)}\rangle\equiv \frac{1}
{\sqrt{N}}\sum_{j\in B}e^{-ikR_j}(a_{j-1}^{}+a_{j+1}^{})
a_j^{\dagger}b^{}_j|0\rangle\,, 
\end{equation}
with the hole delocalized to the neighboring $j-1$ ($j+1$) sites of the 
$A$ sublattice, i.e., to the the left (right) from the initial hole 
position $j$ in each Fourier component $|b_j\rangle$ included in 
Eq. (\ref{bk}). The remaining states $\{|\psi_k^{(n)}\rangle\}$ with 
$n>2$, which occur in the continued fraction expansion needed to 
evaluate the Green's function $G_b(k,\omega)$ (see below), are generated by 
acting $(n-2)$ times on $|\psi_k^{(2)}\rangle$ with the three-site 
hopping term $H_{\rm 3s}$. In this way one finds the set of symmetric 
states, with a superposition of the hole propagating forward (either to 
the left or to the right from the initial defect), i.e., along the same 
direction as that given by the first hop which led to 
$|\psi_k^{(2)}\rangle$, cf. Fig. \ref{fig:1d:1}(c). This structure of 
the basis set explains the absence of the $k$ dependence in the 
Green's function for $b$ orbitals, so we adopt the simplified 
notation $G_b(\omega)$ below.

In the infinite basis generated by the above described procedure, 
the Hamiltonian matrix of the Hamiltonian (\ref{H_1D}) reads: 
\begin{align}\label{eq:ham_lr}
&\langle\psi_k^{(m)} |\,\omega+{\cal H}_{\rm 1D}-E_0 | \psi_k^{(n)}\rangle= 
\nonumber \\
&= \left(
\begin{array}{ccccc}
\omega + J /2 & \sqrt{2} t & 0 & 0 & ... \\
\sqrt{2} t & \omega + 3J/4 &  \tau & 0 & ... \\
0 & \tau & \omega + J & \tau & ... \\
0 & 0 &  \tau & \omega + J & ... \\
... & ... & ... & ... & ... \\
\end{array}
\right).
\end{align}
In order to calculate the relevant Green's function $G_b(\omega)$,
we need the $(1,1)$ element of the inverse of this matrix. Due to
the tridiagonal form of the Hamiltonian, this can be done even for an 
infinite Hilbert space, and we arrive at a continued fraction result:
\begin{align}
\label{green1D}
G_b(\omega)&=\left\{\left\langle\psi_k^{(m)}\left|\omega+{\cal H}_{\rm 1D}-E_0
\right|\psi_k^{(n)}\right\rangle^{-1}\right\}_{1,1}
\nonumber \\
&= \left\{\omega + \frac12 J-\frac{2t^2}{\omega+\frac{3}{4}J-
  \frac{\tau^2}{\omega+J-\frac{\tau^2}{\omega+J-\dots}}}\right\}^{-1},
\end{align}
where the whole self-similar part can be summed up to the
self-energy which does not depend on $k$:\cite{Bri70}
\begin{equation}
\Sigma(\omega)\equiv\frac{\tau^2}{\omega+J-\frac{\tau^2}
  {\omega+J-\frac{\tau^2}{\omega+J-...}}}=\frac{\tau^2}
  {\omega+J-\Sigma(\omega)}.
\end{equation}
This, together with Eq. (\ref{green1D}), leads to a quadratic equation 
for $\Sigma(\omega)$ with two solutions:
\begin{equation}
\Sigma(\omega) = \frac12\left\{(\omega+J)\pm\sqrt{(\omega+J)^2-4
\tau^2}\,\right\}. 
\label{eq:tildeG}
\end{equation}
The proper sign may be determined using the Green's function 
$G_b(\omega)$ obtained before\cite{Dag04} in the limit of $\tau=0$,
\begin{equation}
G_b^{(0)}(\omega)=
\left\{\omega +\frac12 J -\frac{2t^2}{\omega+\frac34 J}\right\}^{-1}\,.
\label{eq:green1D0}
\end{equation}
In this limit the self-energy vanishes, $\Sigma(\omega)=0$, and 
the Green's function has two poles at energies 
\begin{equation} 
\label{pole1D}
\omega=-\frac{5}{8}J\pm\sqrt{2}t\,
\sqrt{1+\frac{1}{128}\left(\frac{J}{t}\right)^2}.
\end{equation}
Finally, we arrive at the general result for $\tau>0$:
\begin{equation}
G_b(\omega)= \left\{\omega+\frac 12
  J-\frac{4t^2}{\omega+\frac12 J \mp\sqrt{(\omega+J)^2-4
      \tau^2}}\right\}^{-1},
\label{eq:green1D}
\end{equation}
where the sign convention is fixed by comparing this result with
the Green's function $G_b^{(0)}(\omega)$ (\ref{eq:green1D0}):
This implies that one has to select $-$ ($+$) sign for $\omega<-J$ 
($\omega>-J$), respectively.

Due to the obtained analytic structure of $G_b(\omega)$
the hole spectral function 
\begin{equation} 
\label{eq:1d:b}
A_b(\omega)=-\frac{1}{\pi}\,\mbox{Im}\,G_b(\omega),
\end{equation}
shown in Fig. \ref{fig:1d:1}(d), also does not depend on $k$. For the 
realistic parameters with $\tau<t$ it consists of two poles and the 
incoherent part centered around $\omega=-J$. This latter contribution 
has rather low intensity and is thus invisible on the 
scale of Fig. \ref{fig:1d:1}(d), and the two peaks absorb 
almost the entire intensity. This result resembles the case of $\tau=0$ 
(\ref{pole1D}), and might appear somewhat unexpected -- we analyze it 
in the following Section.

\subsection{Hole confinement in a three-site box}
\label{sec:box}

First, we comment on the absence of the $k$ dependence in the spectral 
function $A_b(\omega)$ (\ref{eq:1d:b}). It suffices to analyze the hole 
in a $b$ orbital at any finite value of $J$ which induces the AO ground 
state (\ref{gs}). The hole can only move incoherently, because once it 
moves away from the initial site $j$ [see Fig. \ref{fig:1d:1}(b) and 
(c)], it creates a defect in the AO state which blocks its hopping by 
the three-site processes over the site $j$, see Eq. (\ref{bk}). 
Consequently, the hole 
may hop only in the other direction, i.e. away from the defect in the
AO state, and in order to absorb eventually this orbital excitation it 
has to come back to its original position, retracing its path. In this
way a forward and backward propagation along the 1D chain interfere 
with each other, resulting in the fully incoherent spectrum of Fig. 
\ref{fig:1d:1}(d).

Looking at the spectral function $A_b(\omega)$ of a hole doped into 
the $b$ orbital at finite $\tau=0.1t$ shown in Fig. \ref{fig:1d:1}(d), 
one may be somewhat surprised that the result indicates only two final 
states of the 1D chain. These are the bonding and the antibonding 
state of a hole confined within a three-site box, and discussed 
in detail in Ref. \onlinecite{Dag04} in the limit of $\tau=0$. 
One finds that the two excitation energies obtained for the present 
parameters, $\omega=-1.67t$ and $\omega=1.17t$, are indeed almost 
unchanged from those given by Eq. (\ref{pole1D}) at $\tau=0$.
We note that the third nonbonding state has a different symmetry and 
thus gives no contribution to $A_b(\omega)$. 

%%%%%%%%%%%%%%%%%%%%%%%%%%%%%%%%%%%%%%%%%%%%%%%%%%%%%%%
%%                         figure 2
%%%%%%%%%%%%%%%%%%%%%%%%%%%%%%%%%%%%%%%%%%%%%%%%%%%%%%%
\begin{figure}[t!]
\includegraphics[width=8cm]{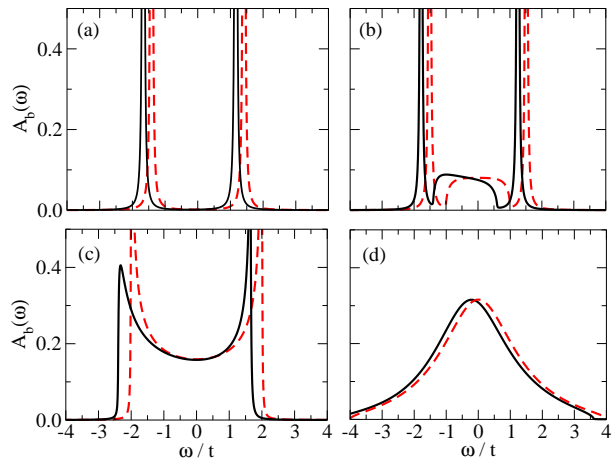}
\caption{(Color online)
Spectral function $A_b(\omega)$ of a hole doped into 
the $b$ orbital in the 1D model with:
(a) $\tau=0$, 
(b) $\tau=0.5t$,
(c) $\tau=t$, and 
(d) $\tau=2t$.
Dotted (solid) lines for $J=0$ ($J=0.4t$), respectively, with
broadening $\delta=0.01t$.
}
\label{fig:1d:2}
\end{figure}

Altogether, one finds that in the realistic regime of parameters with 
$\tau=J/4$, the incoherent part of the spectrum is extremely small and 
thus invisible in the scale of Fig. \ref{fig:1d:1}(d). This implies 
that the hole is still practically trapped within the three-site box 
depicted on Fig. \ref{fig:1d:1}(b), in spite of the potential 
possibility of its delocalization by finite $\tau$. Only when the value 
of the three-site hopping $\tau$ is considerably increased, the hole 
can escape from the three-site box, and may delocalize over the entire 
chain. 

A systematic evolution of the spectral function $A_b(\omega)$ 
with increasing $\tau$ is depicted in Fig. \ref{fig:1d:2}. One observes 
that the incoherent spectral weight grows with increasing $\tau$ and is 
already visible in between the two maxima for $\tau=0.5t$. When the 
three-site hopping term approaches $\tau=t$, the spectrum changes in a 
qualitative way --- both peaks are absorbed by the continuum, and the
spectral density resembles the density of states of the 1D chain with the
NN hopping. For the extremely large effective hopping $\tau\simeq 2t$ 
the two peaks corresponding to the energies given by 
Eq. (\ref{pole1D}) are entirely gone, and the spectrum 
corresponds to the incoherent delocalization of the hole over the 1D 
chain. Note also that finite $J$ results only in an overall shift of the 
spectra due to the energy cost of the hole excitation in the ordered 
ground state (\ref{gs}).

%%%%%%%%%%%%%%%%%%%%%%%%%%%%%%%%%%%%%%%%%%%%%%%%%%%%%%%%%%%%%%
%                  2D: Falicov-Kimball model
%%%%%%%%%%%%%%%%%%%%%%%%%%%%%%%%%%%%%%%%%%%%%%%%%%%%%%%%%%%%%%
\section{2D spinless Falicov-Kimball model}
\label{sec:FalKim}

\subsection{Effective strong-coupling model}
\label{sec:2dfk_ham}

There are two essentially different ways to generalize the 1D orbital
Hubbard model with one passive orbital flavor to two dimensions
in such a way that the superexchange remains still Ising-like. 
Either (i) one allows that the electrons with mobile flavor $a$ can hop 
along all the bonds, i.e. in both directions in the square lattice, or 
(ii) one allows that $a$ electrons can hop along the bonds parallel to 
the $b$ axis, and $b$ electrons can hop along the bonds parallel to the 
$a$ axis. The first scenario leads to a special case of the 2D FK
model (see below), while the second one describes spinless electrons in 
$t_{2g}$ orbitals of a FM plane and will be analyzed in 
Sec.~\ref{sec:t2g}.

In analogy to the 1D model of Sec. \ref{sec:1Dmodel}, the 2D FK model 
describes interacting electrons in mobile $a$ and immobile $b$ 
orbitals, 
\begin{equation}
\label{eq:hubb_2d_fk}
H_{\rm FK}= -t\sum_{\langle ij\rangle} (a^{\dagger}_ia^{}_j +
\mbox{h.c.}) +U \sum_i n_{ia} n_{ib}\;.
\end{equation}
Here we used the same notation as in Eq. (\ref{eq:hubb_1d}), and 
$\langle ij\rangle$ are the bonds (pairs of NN sites) in the 2D 
lattice. This Hamiltonian shows complex physics\cite{Bry08} and phase 
separation\cite{Mas06} away from half filling. In contrast to the 
usual situation with large energy difference between $f$ and $d$ 
orbitals,\cite{Fre03} we will consider degenerate $a$ and $b$ 
orbitals. Then the ground state at half filling (i.e., one electron 
per site) and large Coulomb interaction $U$ is relatively 
straightforward to investigate, and one finds the robust AO order 
rather then phase separation.

Again, we can perform second order perturbation theory in the regime 
of $t\ll U$ as above. For the present square lattice there are two 
types of three-site terms --- they contribute: 
(i) along $a$ and $b$ axes due to forward (linear) processes, and also 
(ii) connect next-nearest neighbor (NNN) sites along the diagonals of
each plaquette in the 2D lattice, along two $90^{\circ}$ paths. The 
resulting strong-coupling effective Hamiltonian reads
\begin{equation}
\label{eq:tj_fk}
{\cal H}_{\rm FK} = H_{t}+H_{J}+H_{\rm 3s}^{(l)}+H_{\rm 3s}^{(d)},
\end{equation}
where
\begin{eqnarray}
\label{Ht_FK} 
H_{t}&=& -t \sum_{\langle ij\rangle}
(\tilde{a}^{\dagger}_i\tilde{a}^{}_j+\mbox{h.c.} ),\\
\label{HJ_FK} 
H_{J} &=& \frac12 J \sum_{\langle ij\rangle }
\left(T^z_i T^z_j - \frac{1}{4}\tilde{n}_i\tilde{n}_j\right)\,,\\
\label{H3s_FK1}
H_{\rm 3s}^{(l)}
 &=& - \tau \sum_{i} \left(\tilde{a}^\dag_{i-\bf{\hat{a}}}
\tilde{n}^{}_{ib}\tilde{a}^{}_{i+\bf{\hat{a}}}\!+\!\mbox{h.c.}\right) \nonumber \\
&&- \tau \sum_{i} \left(\tilde{a}^\dag_{i-\bf{\hat{b}}}\tilde{n}^{}_{ib}
\tilde{a}^{}_{i+\bf{\hat{b}}}+\!\mbox{h.c.}\right)\!, \\
\label{H3s_FK2}
H_{\rm 3s}^{(d)}
 &=& - \tau \sum_{i} \left(\tilde{a}^\dag_{i\pm\bf{\hat{b}}}
\tilde{n}^{}_{ib}\tilde{a}^{}_{i\pm\bf{\hat{a}}}\! +\!
\mbox{h.c.}\right) \nonumber \\ 
&&- \tau \sum_{i} \left(\tilde{a}^\dag_{i\mp\bf{\hat{b}}}
\tilde{n}^{}_{ib}\tilde{a}^{}_{i\pm\bf{\hat{a}}}\! +\!
\mbox{h.c.}\right)\!\!.
\end{eqnarray}
Here $\bf{\hat{a}}$ and $\bf{\hat{b}}$ are the unit vectors along 
the axes $a$ and $b$, while $H_{\rm 3s}^{(l)}$ and $H_{\rm 3s}^{(d)}$
terms stand for linear and diagonal processes in the three-site
effective hopping. The parameters $J$ and $\tau$ are defined as in Eqs. 
(\ref{J}), the orbital (pseudospin) operators $T^z_i$ are defined as in 
Eq. (\ref{Tz}), and again the tilde above the fermion operators 
indicates that the Hilbert space is restricted to the unoccupied 
and singly occupied sites.

\subsection{Analytic Green's functions}
\label{sec:res_2d_fk}

\begin{figure}
\includegraphics[width=8cm]{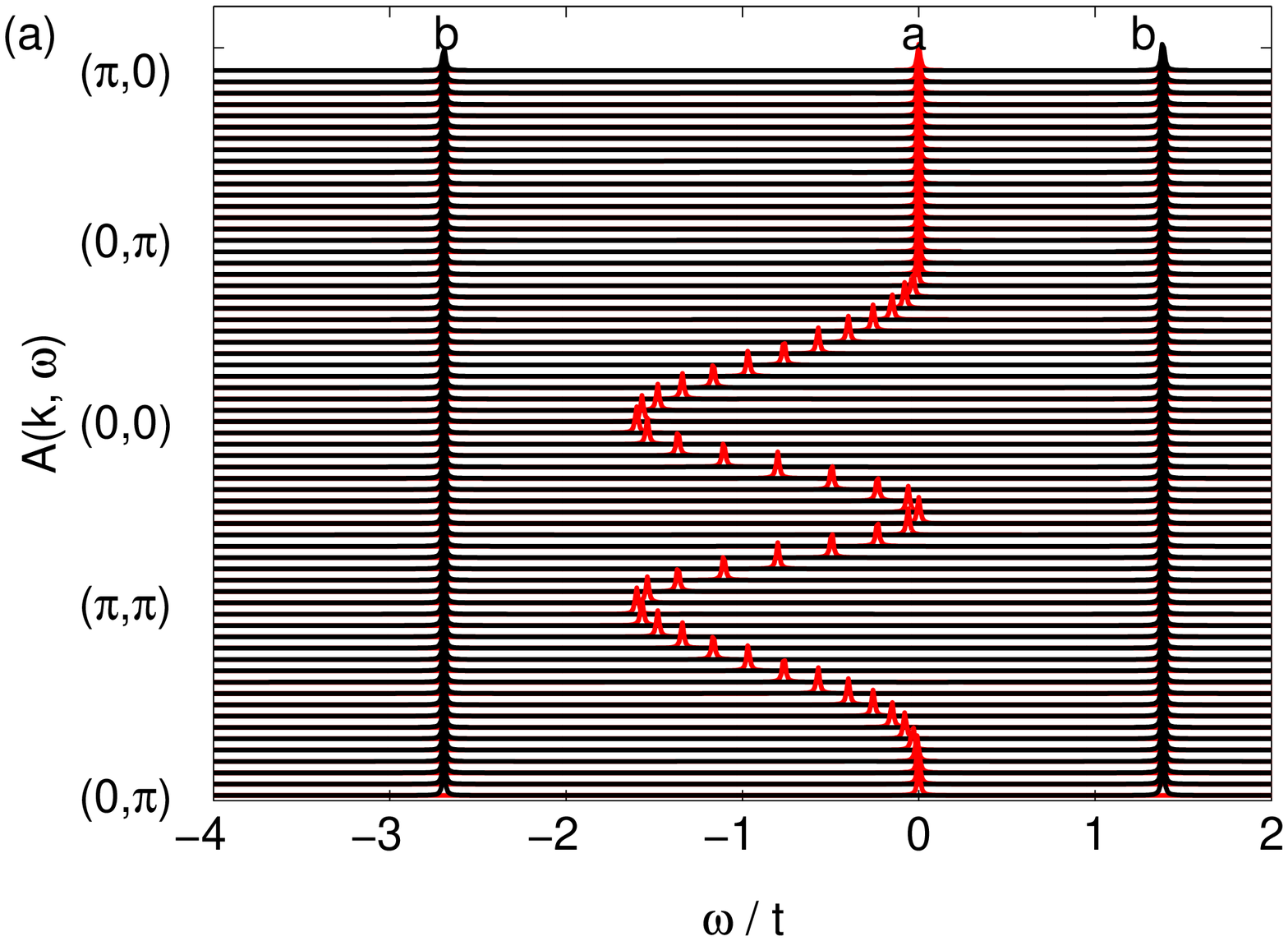}
\\[1.0em]
\includegraphics[width=8cm]{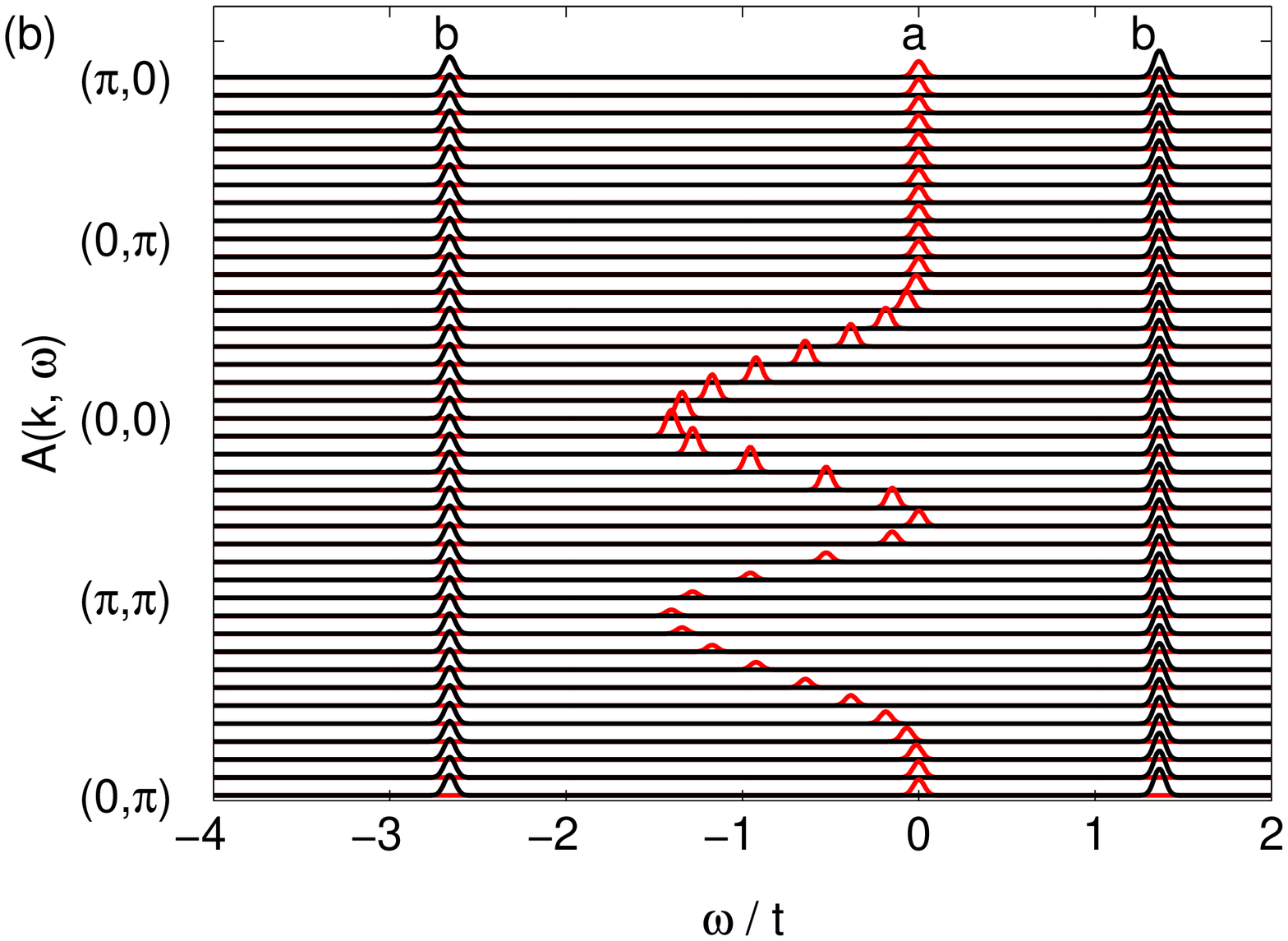}
\caption{(Color online) 
The spectral functions for the 2D FK model, as obtained 
for mobile $a$ orbitals (middle dispersive feature between 
$\omega=-1.6$ and $\omega=0$) and immobile $b$ orbitals
(two side dispersionless maxima): 
(a) with the RPA of Ref. \onlinecite{Bri70} for the 
2D FK strong-coupling 
model (\ref{eq:tj_fk}) with $J=0.4t$ and $\tau=0.1t$, and 
(b) by a numerical diagonalization of a $20\times 20$ cluster for 
the 2D FK Hubbard model (\ref{eq:hubb_2d_fk}) with $U=10t$.
Peak broadening $\delta=0.01t$.
\label{fig:fk_2d}}
\end{figure}

The Green's function $G_a({\bf k},\omega)$ for a hole in mobile
$a$ orbitals, defined by Eq. (\ref{eq:defga}), can be calculated 
as straightforwardly as in the 1D model of Sec. \ref{sec:1Dmodel}, 
and one finds:
\begin{equation}
G_a({\bf k},\omega)=\frac{1}{\omega+J+\varepsilon_{\bf k}}\;,
\end{equation}
where the hole dispersion relation is given by
\begin{equation}\begin{split}
    \label{eq:band_eff}
 \varepsilon_{\bf k}= -4\tau\left\{(\cos k_x+\cos k_y)^2-1\right\}\,.
\end{split}\end{equation}
As in the 1D model, see Eq. (\ref{eq:1d:ga}), the hole propagates
freely, resulting in a bandwidth of $W=16\tau=4J$. 
Indeed, in the strong-coupling model (\ref{eq:tj_fk}) the hopping to 
the sites occupied by $b$ electrons is blocked by the constraint. The 
spectral function $A_a({\bf k},\omega)$ (\ref{eq:1d:a}) obtained for 
the hole in $a$ orbitals consists thus of a single pole, 
as shown in Fig.~\ref{fig:fk_2d}(a). 

The dispersion relation of Eq. (\ref{eq:band_eff}) can be compared 
with the one of the lower Hubbard band obtained\cite{Mas06} for the 
FK model (\ref{eq:hubb_2d_fk}) in the regime of $U\gg t$, where one 
finds dispersion
\begin{eqnarray}
\label{eq:band_FK_Hubb}
\epsilon_{\bf k}&=&\frac{1}{2}U\left\{1-\sqrt{1+
\frac{16t^2}{U^2}(\cos k_x+\cos k_y)^2}\right\}\nonumber \\
&\simeq& -4\tau(\cos k_x+\cos k_y)^2\,.
\end{eqnarray}
This result is the same (up to a nonsignificant constant) as the one
obtained in the strong-coupling limit, see Eq. (\ref{eq:band_eff}).
The Green's function obtained for the Hubbard-like model 
(\ref{eq:hubb_2d_fk}) is shown in Fig.~\ref{fig:fk_2d}(b). While it 
qualitatively agrees with the result derived in the strong-coupling 
limit (\ref{eq:band_eff}), it is here renormalized and gives a somewhat 
reduced bandwidth of the hole band. This indicates finite probability 
of double occupancies which hinder the three-site effective hopping
processes, and reduce the order parameter from its classical value 
as given in the N\'eel state (\ref{gs}), see also Sec. \ref{sec:overqp} 
for a similar discussion concerning the 2D $t_{2g}$ orbital model. 

%%%%%%%%%%%%%%%%%%%%%%%%%%%%%%%%%%%%%%%%%%%%%%%%%%%%%%%
%%                         figure 3
%%%%%%%%%%%%%%%%%%%%%%%%%%%%%%%%%%%%%%%%%%%%%%%%%%%%%%%

\begin{figure}[t!]
\psfrag{A}{}
\psfrag{B}{}
\psfrag{t}{$t$}
\psfrag{J4}{$\tau$}
\subfigure[]{\hspace*{-2.5em}\includegraphics[width=0.23\textwidth]
  {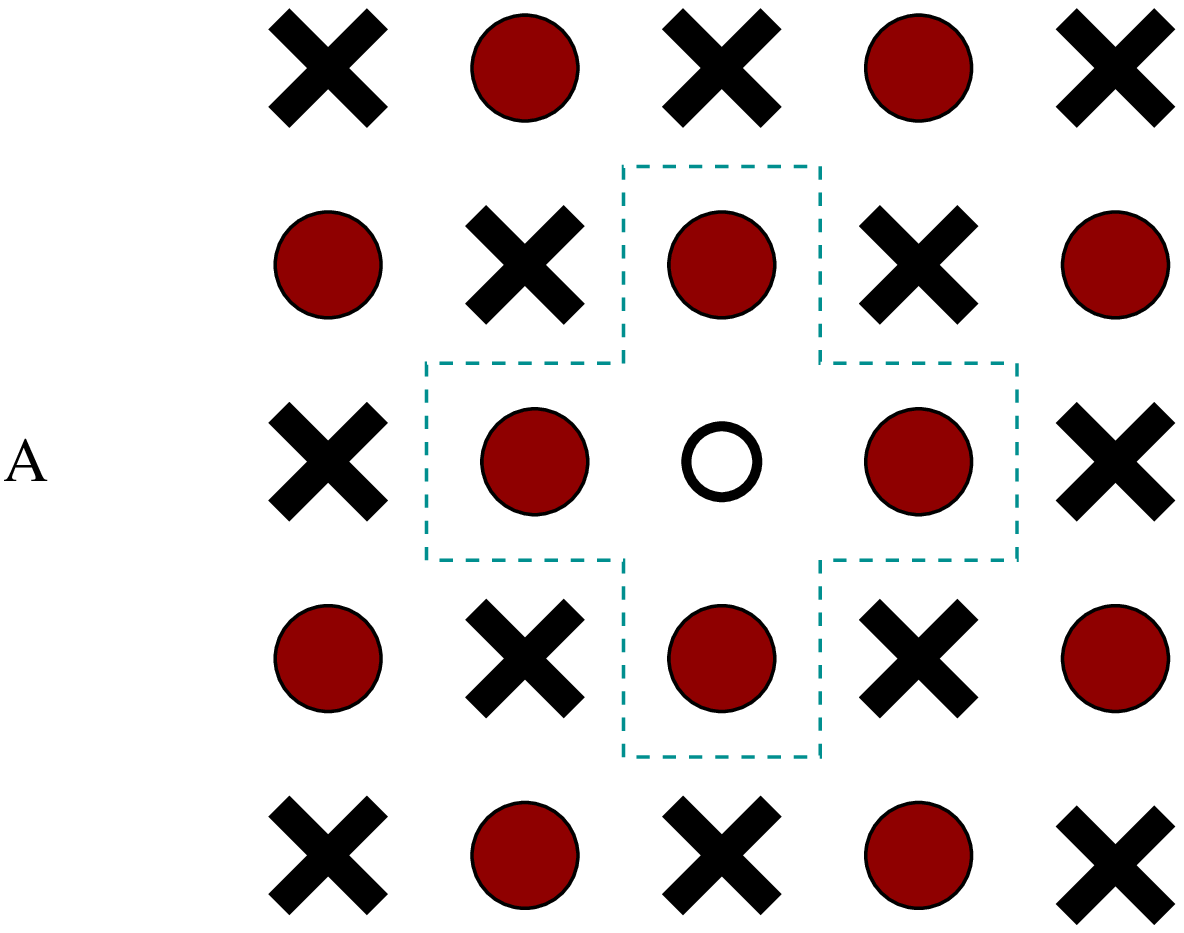}\label{fig:latt_pol}}\hfill
\subfigure[]{\hspace*{-2.5em}\includegraphics[width=0.23\textwidth]
  {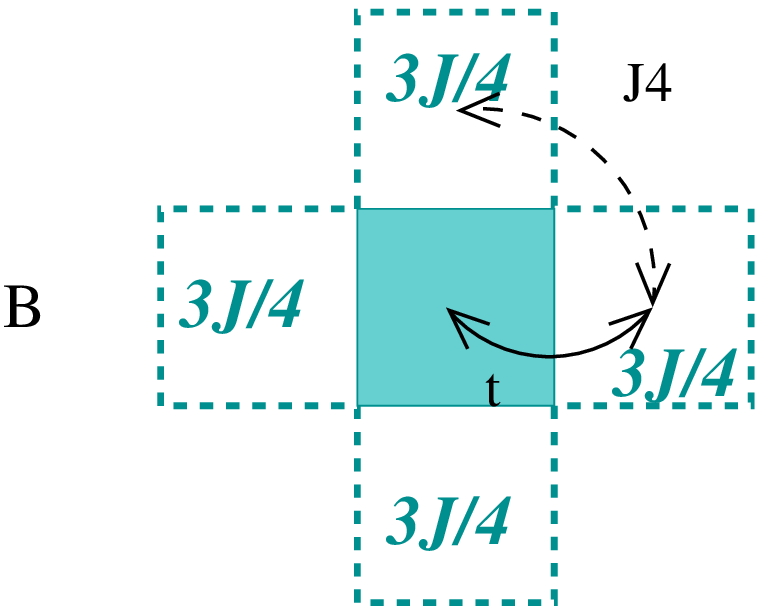}\label{fig:ham_pol}}
\caption{(Color online) 
Schematic view of a five-site polaron which occurs after the removal 
of one $b$-electron in the 2D FK model (\ref{eq:tj_fk}). 
(a) Initial state with a hole
(empty circle) on $B$ sublattice --- crosses represent immobile $b$ 
electrons and filled circles mobile $a$ electrons. Five-site polaron
is indicated by the dotted line. (b) The hole may delocalize within
it to the nearest neighbors by hopping $t$ (solid line), generating 
three broken bonds which cost energy $3J/4$ for each hopping process. 
The hole can hop directly between the external sites (dashed line)
by second order three-site hopping $\tau=J/4$ via a doubly occupied 
site (outside the polaron).
\label{fig:polaron}}
\end{figure}

As in the 1D model of Sec. \ref{sec:1dresults}, we use here the RPA to 
calculate the Green's function for a hole inserted into the immobile 
$b$ orbital. However, the RPA is no longer exact in two dimensions, 
because also paths with closed loops are possible when electrons with 
one flavor are allowed to hop in both directions.
We use a similar basis of states $\{|\psi_{\bf k}^{(n)}\rangle\}$ as 
for the 1D calculation of Sec.~\ref{sec:1dresults} to describe 
a single hole doped to the plane with the AO order. Starting from 
the N\'eel state as in Eq. (\ref{gs}), the first two 
states are defined as follows: 
\begin{eqnarray} 
\label{FKbk}
|\psi_{\bf k}^{(1)}\rangle\!\!&\equiv&\! b_{\bf k}^{}|0\rangle=
\sqrt{\frac{2}{N}}\sum_{j\in B}e^{-i{\bf k}{\bf R}_j}b^{}_j|0\rangle\,,
\\
\label{FKbk1}
|\psi_{\bf k}^{(2)}\rangle\!\!&\equiv&\! 
\frac{1}{\sqrt{2N}}\sum_{j\in B}e^{-i{\bf k}{\bf R}_j}\nonumber \\
&\times&\!\!
\left(a_{j-\bf{\hat{a}}}^{}+a_{j+\bf{\hat{a}}}^{}
+a_{j-\bf{\hat{b}}}^{}+a_{j+\bf{\hat{b}}}^{}\right)a_j^{\dagger}b^{}_j|0\rangle\,.
\end{eqnarray}
Here the first state $|\psi_{\bf k}^{(1)}\rangle$ denotes the Fourier 
transformed states with hole doped into the immobile orbital at the 
initial position in the ground state of the 2D lattice, with AO order 
between two sublattices $A$ and $B$, as in (\ref{gs}), see Fig.
\ref{fig:polaron}(a). This state may delocalize by the hopping $t$
which interchanges the hole with an occupied $a$ orbital to the left, 
right, down, or up from the initial site, resulting in the symmetric 
state with four external sites of the five-site polaron depicted in 
Fig. \ref{fig:polaron}(b) --- the resulting state is denoted above by 
$|\psi_{\bf k}^{(2)}\rangle$ (\ref{FKbk1}). 

At this point, we have to introduce an approximation and we will 
consider only these states $|\psi_{\bf k}^{(n)}\rangle$ with $n>2$ which 
are created on a Bethe lattice by 9 possible forward going steps from
$|\psi_{\bf k}^{(n-1)}\rangle$. Therefore, all such states 
(with $n>2$) are generated by the three-site effective hopping $\tau$, 
so each of them means that the hole has moved forward by $(n-2)$ steps 
from the symmetric state $|\psi_{\bf k}^{(2)}\rangle$, being a linear
combination of the configurations with the hole at one of the external 
sites in the five-site polaron (Fig. \ref{fig:polaron}). 
Hence, we make two approximations here, i.e., we assume that: 
(i) no closed loops occur (RPA), and 
(ii) the number of forward going steps is chosen to be $9$ which is 
the most probable number of possible forward-going three-site 
steps on a square lattice with the AO order.\cite{notebet} 
Let us emphasize, however, that the closed loops which are neglected
here do not lead to the delocalization of the hole as these are not the 
so-called Trugman loops\cite{Tru88} where the hole could repair the 
defects by a circular motion around a plaquette in the square lattice.
The Hamiltonian [Eq. (\ref{eq:tj_fk})] matrix written in this basis is
as follows:
\begin{align} \label{eq:tj_fk_m}
&\langle\psi_{\bf k}^{(m)}|\,\omega+{\cal H}_{\rm FK}-E_0 
|\psi_{\bf k}^{(n)}\rangle= 
\nonumber \\
&= \left(
\begin{array}{ccccc}
\omega + J  & 2 t & 0  & 0 & ... \\
2 t & \omega + \frac{7}{4}J+2\tau &  3\tau & 0 & ... \\
0 & 3\tau & \omega + 2J  & 3\tau & ... \\
0 & 0 & 3\tau & \omega + 2J  &  ... \\
... & ... & ... & ... & ...  \\
\end{array}
\right).
\end{align}
Again, as in Eq. (\ref{eq:ham_lr}), the ${\bf k}$-dependence is absent,
and the Green's function $G_b(\omega)$ can be calculated using continued 
fraction method in a similar way as in the 1D case (cf. Sec. 
\ref{sec:1dresults}) and we obtain
\begin{eqnarray}
G_b(\omega)\!\!&=&\!\! \left\{\omega+J\right.
\nonumber\\
&-&\!\!\left.\frac{8t^2}{\omega+\frac 32 J+4\tau  \mp
    \sqrt{(\omega+2J)^2-36\tau^2}}\right\}^{-1}\!\!,
\label{eq:green2dfk}
\end{eqnarray}
where we select $-$ $(+)$ sign for $\omega<-2J$ $(\omega>-2J)$,
respectively.

%%%%%%%%%%%%%%%%%%%%%%%%%%%%%%%%%%%%%%%%%%%%%%%%%%%%%%%
%%                         figure 4
%%%%%%%%%%%%%%%%%%%%%%%%%%%%%%%%%%%%%%%%%%%%%%%%%%%%%%%

The spectral function of a hole doped into the $b$ orbital is shown in 
Fig. \ref{fig:fk_2d}(a). It consists of two distinct dispersionless
peaks, and a dispersionless incoherent part with negligible spectral 
weight [invisible on the scale of Fig. \ref{fig:fk_2d}(a)] between 
them. While the incoherent spectrum disappears in the limit of vanishing 
three-site hopping $\tau$, the two dispersionless peaks survive and the 
distance between them becomes $4t$ for $J\to 0$. Hence, the doped hole 
is not only immobile but also trapped within the five-site orbital 
polaron depicted in Fig. \ref{fig:latt_pol} -- only four sites can be
reached from the central site by NN hopping $t$, so the ground state can
be found in the truncated basis 
$\{|\psi_{\bf k}^{(1)}\rangle,|\psi_{\bf k}^{(2)}\rangle\}$ (see below).
This situation resembles very much the 1D case discussed previously, 
and indeed the same discussion as the one for the hole confinement in 
a three-site box presented in Sec. \ref{sec:box} applies here.
In the next Section we discuss in detail the quantitative arguments 
which suggest the hole confinement in this five-site polaron.

We would like to emphasize that the spectral functions obtained using 
the RPA for the strong-coupling model Eq. (\ref{eq:tj_fk}) are almost 
identical to the ones obtained numerically by exact diagonalization of 
the Hubbard model (\ref{eq:hubb_2d_fk}) on a $20\times 20$ lattice, 
cf. Fig. \ref{fig:fk_2d}(b). This means that the crude assumption of 
walks without closed loops made within the RPA approximation 
(i.e., replacement of the square lattice by the Bethe lattice) is 
{\it a posteriori\/} well justified for the strong-coupling model 
defined by Eq. (\ref{eq:tj_fk}). We provide also more arguments which 
complete our understanding of this result in the next Section.
Furthermore, this means that not only the RPA method is correct
but also the two models (Hubbard and the strong-coupling model) 
are fully equivalent and describe 
the same physics in the considered regime of parameters.

Finally, we note that the results obtained by the VCA (not shown) are 
very similar to the ones of exact diagonalization, if we use periodic 
boundary conditions. While open boundaries are usually optimal for the 
VCA,\cite{Aic03} they can `cut' the five-site polaron and lead to 
signals at wrong frequencies. In a large enough cluster, these 
contributions from polarons with less than five sites would have 
vanishing weight, but for the cluster sizes considered here, 
self-energies with periodic boundary conditions have to be used to 
eliminate them.

\subsection{Localized five-site orbital polaron}
\label{sec:disc_2d_fk}

The following comparison shows that the two dominant peaks of the
Green's function for $b$ orbitals $G_b(\omega)$ can be well reproduced
by taking into account the polaron alone, i.e., by considering just
a cluster of five sites depicted in Fig.~\ref{fig:ham_pol} and
neglecting the rest of the lattice. In this case the problem can be
solved by diagonalizing the Hamiltonian in the basis consisting of two
states defined in the last section: 
$|\psi_{\bf k}^{(1)}\rangle$ (\ref{FKbk}) and 
$|\psi_{\bf k}^{(2)}\rangle$ (\ref{FKbk1}).
This means that the infinite matrix of Eq. (\ref{eq:tj_fk_m}) for the
hole doped into the central site of the polaron
reduces to the $2\times2$ matrix and one obtains the energies of
two poles of the Green's function $G_b(\omega)$, corresponding to
the bonding and antibonding state within the five-site polaron:
\begin{align}\label{eq:peaks}
\omega_{1,2} = -\frac{11}{8} J - \tau \pm
\sqrt{4t^2+\frac{9}{64} J^2 + \frac{3}{4}J \tau +\tau^2}\,.
\end{align}

Assuming $J=0.4t$ and $\tau=0.1t$ in Eq. (\ref{eq:peaks}) we obtain
$\omega_{1,2}=\{1.37t,-2.67t\}$.  This compares very favorably with 
the results obtained for the strong-coupling model (\ref{eq:tj_fk}):
(i) within the RPA (see Sec. \ref{sec:res_2d_fk}),
$\omega_{1,2}=\{1.38t,-2.69t\}$, and 
(ii) using the numerical analysis of this model on the $20\times 20$ 
lattice, which gives $\omega_{1,2}=\{1.40t,-2.74t\}$ (not shown). 
We stress that the excellent agreement between all these methods 
demonstrates that the RPA (i.e., full continued fraction) turns out to 
be only slightly better than the calculation restricted to the five-site 
polaron of Fig. \ref{fig:ham_pol}. It means that the probability
of the configurations with the hole outside the five-site polaron is 
indeed very low, and it explains why the RPA assumption of having no 
walks with closed loops works here so well. Lastly, we note that all 
these results agree quite well with the numerical ones for the itinerant 
FK model (\ref{eq:hubb_2d_fk}), cf. Fig. \ref{fig:fk_2d}(b) with the 
peaks situated at $\omega_1=1.42t$ and $\omega_2=-2.56t$.

The present five-site orbital polaron resembles the five-site spin 
polaron identified in Monte Carlo studies for the 2D Kondo model.
\cite{Dag04a} For example, as for the spin polaron in the Kondo model, 
the spectral density of the orbital polaron is comprised of two 
dispersionless peaks with a distance of $4t$ for $J\to 0$. There is, 
however, one difference: Here not only the hole can move by direct NN 
hopping $t$ between the central site and the four external sites, but 
there is also a second order three-site diagonal hopping 
(\ref{H3s_FK2}) which couples directly the neighboring external sites
of the polaron (see Fig.~\ref{fig:ham_pol}), and contributes to the 
energy of the $|\psi_{\bf k}^{(2)}\rangle$ state. Actually, due to the 
inclusion of these processes (which enable the smallest loops on the
lattice, with two $t$ and one $\tau$ hopping processes) we could obtain 
the above mentioned perfect agreement between the
numerical, the RPA, and the five-site polaron results for the
strong-coupling version of the FK model Eq. (\ref{eq:tj_fk}). 
Otherwise, e.g. for $J=0.4t$ the energies of the two peaks in the RPA 
(five-site polaron) calculation would be equal to $\{1.47t,-2.58t\}$ 
[$\{1.46t,-2.56t\}$], respectively, and would only rather poorly agree 
with the numerical results of Eq. (\ref{eq:tj_fk}). 

Summarizing, the holes doped into the immobile orbitals of the FK
model are almost entirely localized within the five-site orbital
polaron of Fig.~\ref{fig:ham_pol}. In order to calculate the energy of 
this polaron correctly one has to take into account the energies of 
the processes which involve four external polaron sites. In addition, 
we note that the widely used SCBA\cite{Mar91} 
(used for the orbital strong-coupling model in the next Section) does not work 
so well for the case of hole doped into the immobile orbital of the FK 
model, as it does not respect the constraint on the hole motion. 
It incorrectly uses an on-site energy of $J$ instead of $3J/4$ for the 
excitations at external sites of the polaron. As shown above, the hole 
spends almost all its time inside the polaron and hence this 
underestimation of the energy heavily influences the energies of the 
poles of the Green's function in this case [e.g. the lowest peak for 
$J=0.4t$ is situated almost at $-3t$ in the SCBA calculations (not shown)].

\section{string excitations in the 1D model}
\label{sec:1Dlegs}

\begin{figure}[t!]
\includegraphics[width=0.4\textwidth]{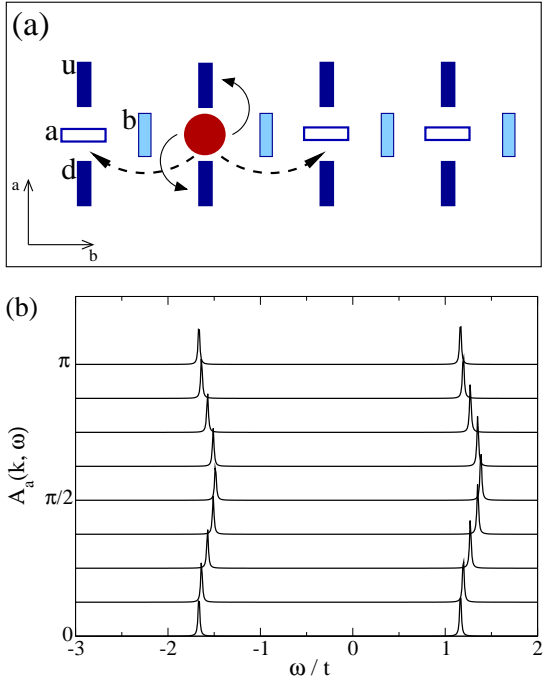}
\\[1.0em]
\includegraphics[width=0.4\textwidth]{holot5b}
\caption{(Color online) 
Propagation of a hole added into the $a$ orbital in the centipede 
strong-coupling model (\ref{eq:ext}):
(a) schematic picture of a hole doped at site $a$ and its possible 
delocalization via hopping
$t$ (solid lines) and three-site effective $\tau$ term (dashed lines);
(b) spectral function $A_a(k,\omega)$.
Parameters: $J=0.4t$, $\tau=0.1t$, peak broadening $\delta=0.01t$.
The chain is oriented along the $b$ axis, and 
nonequivalent positions of the orbitals which do not permit hopping
along this direction are labelled $b$, $u$ and $d$ in panel (a).
}
\label{fig:ext:1}
\end{figure}

The 1D model (Sec. \ref{sec:1Dmodel}) and the 2D FK model (Sec. 
\ref{sec:FalKim}) bear the same generic features: 
(i) a hole generated in the so-called mobile orbital always 
leads to the dispersive spectrum with the full unrenormalized bandwidth, 
and 
(ii) a hole doped into the so-called immobile orbital is 
localized, leading to a non-dispersive spectral function. In this 
Section we investigate the consequences of string excitations which may 
arise when both orbital states allow only 1D hopping, as in the case of 
two $t_{2g}$ orbitals lying in two vertical planes with respect to the 
considered plane. Thus we will study the 1D model with electrons 
hopping between $yz$ and $zx$ orbitals in $(a,b)$ plane --- the model 
has only $2N$ sites for the chain of length $N$, see Fig. 
\ref{fig:ext:1}. We will show that even the shortest possible strings 
with the length of one bond which can be excited here when the hole 
moves in this geometry are sufficient to generate some characteristic 
features recognized later in the spectral properties of the 2D $t_{2g}$ 
model (see Sec. \ref{sec:t2g}).

The 1D model of Fig. \ref{fig:ext:1}(a) consists of a chain along $b$ 
axis, with the Hamiltonian as described by Eq. (\ref{eq:hubb_1d}), and two sites 
being the NNs of every second site of the chain along the $a$ axis, 
which could represent radicals added to a linear molecule. 
We use here the convention introduced before for the $t_{2g}$ orbital 
systems,\cite{Kha01,Ole05} that $a$ and $b$ orbitals stand for $yz$ and 
$zx$ $t_{2g}$ orbitals, respectively, that permit the electron hopping 
along the $b$ and $a$ axis in the $(a,b)$ plane. The Hamiltonian 
of the present (called here {\it centipede\/}) model is,
\begin{eqnarray}
\label{eq:1dlegs}
H_{\rm c}&=&-t\sum_{i}\left\{b^{\dagger}_{2i}(b^{}_{2i,u}+b^{}_{2i,d}) 
+ \textrm{h.c.}\right\}\nonumber \\
&-&t\sum_{i}(a^{\dagger}_{i}a^{}_{i+1} + \textrm{h.c.})
+U \sum_i n_{ia} n_{ib}\;.
\end{eqnarray}
The hopping along the bonds parallel to the $a$ axis is allowed only to 
the orbitals $b$, with the corresponding creation operators 
$\{b^{\dag}_{2i,u},b^{\dag}_{2i,d}\}$, see Fig. \ref{fig:ext:1}(a).
To simplify notation, we call these orbitals $u$ and $d$, and introduce
the following operators:
\begin{equation}
\label{eq:1dud}
u^{\dag}_{2i}\equiv b^{\dag}_{2i,u}, \hskip 1cm
d^{\dag}_{2i}\equiv b^{\dag}_{2i,d}\;.
\end{equation}

%%%%%%%%%%%%%%%%%%%%%%%%%%%%%%%%%%%%%%%%%%%%%%%%%%%%
%%                     figure 5
%%%%%%%%%%%%%%%%%%%%%%%%%%%%%%%%%%%%%%%%%%%%%%%%%%%%

In the limit of large $U$ ($U\gg t$) the occupied orbitals form AO order along the 
chain and we select the N\'eel state with $b$ ($u$ and $d$) orbitals 
occupied on the external sites, as shown in Fig. \ref{fig:ext:1}, as we
are interested in their effect on the hole propagation when it was 
doped to an $a$ orbital. This leads to the following strong-coupling
version of the 1D centipede model (\ref{eq:1dlegs}):
\begin{eqnarray}
{\cal H}_{\rm c}\!&=&
-t\,\sum_{i} \{(\tilde{u}^\dag_{2i}+\tilde{d}^\dag_{2i})\tilde{b}^{}_{2i}
+\textrm{h.c.}\} \nonumber \\
&-&\tau\sum_{i} (\tilde{a}^\dag_{2i}\tilde{n}_{2i+1,b}\tilde{a}^{}_{2i+2}
                +\textrm{h.c.}) \nonumber \\
&-&\frac{3}{4}J\sum_{i} (\tilde{u}^\dag_{2i}\tilde{u}^{}_{2i}+\tilde{d}^\dag_{2i}
\tilde{d}^{}_{2i})
\,.
\label{eq:ext}
\end{eqnarray}
On the one hand, the superexchange interaction for all the bonds within the 
centipede was not included in Eq. (\ref{eq:ext}) as it results only in a
rather trivial energy shift of the spectra obtained from the Green's 
function $G_a(k,\omega)$ which is of interest here,\cite{notenob} cf. 
Sec. \ref{sec:1dresults}. On the other hand,
the last term in Eq. (\ref{eq:ext}) was added to simulate the
creation of string excitations which occur in the full 2D model of
Sec. \ref{sec:t2g} (see also discussion below).

Whereas the second term in Eq. (\ref{eq:ext}) is once again the 
three-site hopping derived before in the 1D model (\ref{H3s_eg}) [cf. 
Fig. \ref{fig:ext:1}(a)], the other two terms describe the possibility 
of creating defects in the AO order when the hole leaves the spine of 
the centipede (i.e., moves away from the $a$ sites) by creating strings of 
length one, just as it may happen in the $t_{2g}$ 2D model, see Sec. 
\ref{sec:t2g}. Here the hole can leave the chain to its NN orbital 
$u_{2i}$ or $d_{2i}$ [cf. sites attached to the chain along the $a$ 
axis shown in Fig. \ref{fig:ext:1}(a)]. Such defects are created by 
hopping $t$ and costs energy $3J/4$ in each case. Hence, the present 1D 
model represents an extreme reduction of the full $t_{2g}$ 2D model,
allowing only the strings of length one, and each defect has to be 
deexcited before the hole can hop to another three-site unit along the 
chain. Note however, that the energies of these string 
excitations are properly chosen and are just the same as in the 
full 2D model of Sec. \ref{sec:t2g}.

%%%%%%%%%%%%%%%%%%%%%%%%%%%%%%%%%%%%%%%%%%%%%%%%%%%%
%%                    figure 6
%%%%%%%%%%%%%%%%%%%%%%%%%%%%%%%%%%%%%%%%%%%%%%%%%%%%
\begin{figure}[t!]
\includegraphics[width=7.5cm]{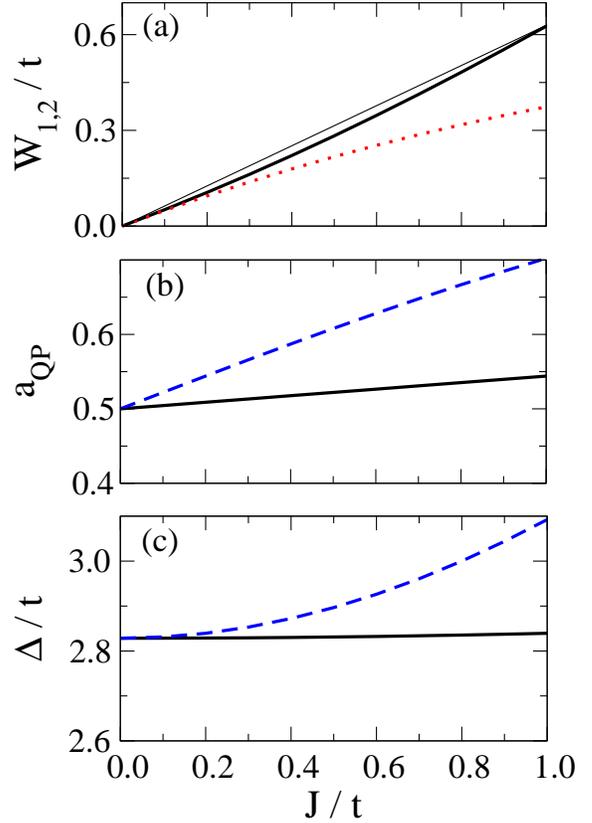}
\caption{(Color online)
Characteristic features in the spectra obtained for the 1D 
centipede model (Fig. \ref{fig:ext:1}) for increasing $J$:
(a) the bandwidth $W_{1,2}$, 
(b) the spectral weight $a_{\rm QP}$, and 
(c) the distance $\Delta$ between the two peaks.
The solid (dotted) line in (a) corresponds to the first (second) 
dispersive peak in $A_a(k,\omega)$ whereas the solid (dashed) lines 
in the lower panels show results for $k=0$ ($k=\pi / 2$), respectively. 
The light solid line in (a) is merely a guideline for the eye to show 
that the bandwidth of the first peak is a function with a positive 
second derivative.
Parameter: $\tau=J/4$. }
\label{fig:ext:2}
\end{figure}

The model given by Eq. (\ref{eq:ext}) constitutes a one-particle problem 
(after inserting $\tilde{n}_{2i+1, b}\equiv 1$ which is consistent with 
the Ising nature of the superexchange) and hence can be solved exactly. 
We will consider the Green's function $G_a(k,\omega)$ for $a$ orbitals,
defined similarly as in Eq. (\ref{eq:defga}), and a hole excitation is 
created again by the operator $a_k$ of Eq. (\ref{eq:ak+}). The continued 
fraction terminates after the second step and one finds the exact Green's 
function
\begin{equation}
G_a(k,\omega) = \frac{1}{\omega+2\tau\cos(2k)
-\frac{2t^2}{\omega+\frac{3}{4}J}}\,,
\label{eq:greenext}
\end{equation}
leading to the corresponding spectral function $A_a(k,\omega)$, defined 
as in Eq. (\ref{eq:1d:a}). The numerical results obtained with $J=0.4t$ 
are shown in Fig. \ref{fig:ext:1}(b). Instead of a single dispersive 
state of Fig. \ref{fig:1d:1}(d), the spectral function consists here of 
two dispersive peaks, separated by a gap of roughly $2\sqrt{2}t$. This 
demonstrates that the larger hopping $t$ suppresses at first instance 
the hopping along the chain by the element $\tau$, and a hole doped 
into the $a$ orbital delocalizes in first place over the three-site 
unit, discussed in Sec. \ref{sec:box}, consisting of a hole and two $b$ 
($u$ and $d$) orbitals. Therefore, the hole behaves effectively as a 
defect created at a $b$ site in the 1D chain of Sec. \ref{sec:1Dmodel}. 
This explains that the maxima of $A_a(k,\omega)$ are found again for a 
bonding and antibonding state, similar to the structure of $A_b(\omega)$ 
in Sec. \ref{sec:1dresults}. 
However, at present the corresponding states gain weak dispersion
because the hole may as well delocalize along the chain by the 
three-site hopping $\tau$. Note also that the low-energy (right) peak 
has slightly higher dispersion (leading to a broader band) than the left 
one. This case illustrates that the 1D dispersion is broader for the QP 
state but is also shared by the feature at higher energy. This 
observation will help us to interpret the spectra for the 2D 
$t_{2g}$ model in Sec. \ref{sec:t2g}.

In addition we also calculated some characteristic features of
the spectra of the centipede model, cf. Fig. \ref{fig:ext:2}.
They will mostly serve for a comparison with the respective 
results of the 2D $t_{2g}$ model, presented in Sec. \ref{sec:overqp}.
However, let us only remark that the renormalization of the bandwidth, 
shown in Fig. \ref{fig:ext:2}(a) follows from an intricate
interplay between coherent hole propagation and the string 
excitations. With increasing $\tau=J/4$ the free bandwidth increases
but at the same time the energies of the defects (generated by the 
hole when it moves to 'lower' or 'upper' sites) are $\propto J$; 
hence, the bandwidth does not depend in a linear way on $J$,
cf. Fig. \ref{fig:ext:2}(a). Physically this means that the hole 
motion is gradually more and more confined to just the 1D path along 
the chain with increasing $J$ (and keeping $\tau=J/4$).

%%%%%%%%%%%%%%%%%%%%%%%%%%%%%%%%%%%%%%%%%%%%%
%           2D: $t_{2g}$ orbital model
%%%%%%%%%%%%%%%%%%%%%%%%%%%%%%%%%%%%%%%%%%%%%
\section{2D model for $t_{2g}$ electrons}
\label{sec:t2g}

\subsection{Effective strong-coupling model}
\label{sec:sc}

After analyzing the spectral properties of the simpler 1D model and 2D 
FK model, we consider below the model relevant for transition metal 
oxides with active $t_{2g}$ orbitals, when the crystal field splits them 
into $e_g$ and $a_1$ states, and the doublet $e_g$ is filled by one
electron per site. This occurs for the $d^1$ configuration (e.g. in 
the titanates) when the $e_g$ doublet has lower energy than the $a_1$ state, 
or for $d^2$ configuration when the $e_g$ states have higher energy and
are considered here, while the $a_1$ state is occupied by one electron 
at each site and thus inactive (as in the high-spin ground state of 
the $R$VO$_3$ perovskites,\cite{Hor03} where $R$ stands for a rare earth
element). To be specific, we consider
electrons with two $t_{2g}$ orbital flavors, $yz\equiv a$ and 
$zx\equiv b$, moving within the $(a,b)$ plane. In contrast to the FK 
model with two nonequivalent orbitals and only one orbital flavor 
contributing to the kinetic energy (Sec. \ref{sec:FalKim}), both 
$t_{2g}$ orbitals are here equivalent and electrons can propagate 
conserving the orbital flavor by the NN hopping $t$, but only along one 
direction in the $(a,b)$ plane.\cite{Kha01} While this results is a 
complicated many-body problem at arbitrary electron filling, the motion 
of a single hole added at half filling remains still strictly 1D.
\cite{Dag08}

The orbital Hubbard model for spinless electrons in the FM $(a,b)$ 
plane reads:
\begin{eqnarray}
\label{Ht2g} 
H_{t_{2g}}&=& 
-t\sum_{\langle ij\rangle\parallel a}(b^{\dagger}_ib^{}_j+\mbox{h.c.})
-t\sum_{\langle ij\rangle\parallel b}(a^{\dagger}_ia^{}_j+\mbox{h.c.})
\nonumber \\
&+&U\sum_i n_{ia}n_{ib},
\end{eqnarray}
where $a$ and $b$ are the orbital flavors with the same hopping $t$ 
along $b$ and $a$ axis, respectively, and $U$ stands again for the 
on-site interaction energy for a doubly occupied configuration. 
At the filling of one electron in $\{a,b\}$ orbitals per site this
interaction
corresponds to the high-spin $d^2$ (or $d^3$) state. Second order 
perturbation theory applied to this Hamiltonian in the regime of 
$t\ll U$ leads then to the strong-coupling model,
\begin{equation}
\label{t2gmodel}
{\cal H}_{t_{2g}} = H_{t}+H_{J}+H_{\rm 3s}^{(l)}+H_{\rm 3s}^{(d)},
\end{equation}
where
\begin{eqnarray}
\label{Ht} H_{t}&=&-t \sum_{ i}
(\tilde{b}^{\dagger}_{i}\tilde{b}^{}_{i+\bf{\hat{a}}} +
\tilde{a}^{\dagger}_{i}\tilde{a}^{}_{i+\bf{\hat{b}}}+\mbox{h.c.})\,,\\
\label{HJ} H_{J}&=& \frac12 J \sum_{\langle ij\rangle }
\left(T^z_i T^z_j - \frac{1}{4}\tilde{n}_i\tilde{n}_j\right)\,,\\
\label{H3s0} H_{\rm 3s}^{(l)} &=& -\tau \sum_{i}
(\tilde{b}^\dag_{i-\bf{\hat{a}}}\tilde{n}^{}_{ia}
\tilde{b}^{}_{i+\bf{\hat{a}}}+\mbox{h.c.}) \nonumber\\
&& -\tau\sum_{i}(\tilde{a}^\dag_{i-\bf{\hat{b}}}
\tilde{n}^{}_{ib}\tilde{a}^{}_{i+\bf{\hat{b}}}+\mbox{h.c.})\,,\\
\label{H3s1} H_{\rm 3s}^{(d)} &=& -\tau \sum_{i}
(\tilde{a}^\dag_{i\pm\bf{\hat{b}}}\tilde{a}^{}_i
\tilde{b}^\dag_i\tilde{b}^{}_{i\pm\bf{\hat{a}}}+\mbox{h.c.}) \nonumber\\
&&-\tau \sum_{i} (\tilde{a}^\dag_{i\mp\bf{\hat{b}}}\tilde{a}^{}_i
\tilde{b}^\dag_i\tilde{b}^{}_{i\pm\bf{\hat{a}}}+\mbox{h.c.}) \,.
\end{eqnarray}
The parameters $J$ and $\tau$ are defined as in Eqs. (\ref{J}), whereas 
the pseudospin operators $T^z_i$ are defined as in Eq. (\ref{Tz}). 
Again the tilde above the fermion operators indicates that the Hilbert
space is restricted to the unoccupied and singly occupied sites. One 
interesting observation here is that the strictly 1D kinetic energy of 
the two orbitals leads to the 2D superexchange (\ref{HJ}). As in the
spin case,\cite{Cha77} the superexchange is active only when electrons 
with two different flavors occupy the neighboring sites (one bond), but 
here only one of them can hop, which explains the prefactor $\frac12$ 
in Eq. (\ref{HJ}). 

\begin{figure}
\includegraphics[width=8cm]{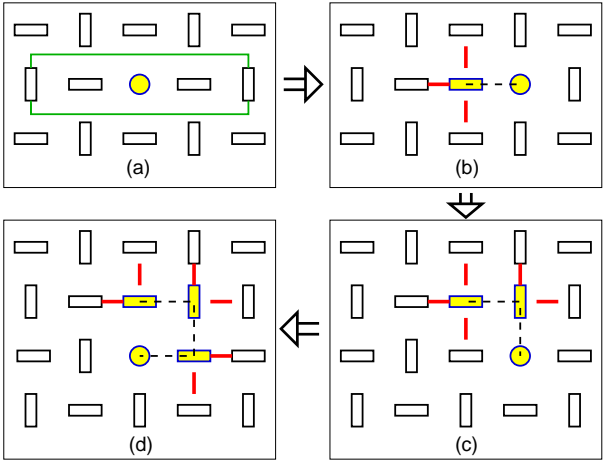}
\caption{(Color online) Schematic view of the hole motion in the
strong coupling $t_{2g}$ orbital model (\ref{Ht2g}) with AO order. 
Circles depict holes while horizontal (vertical) 
rectangles depict occupied orbitals with electrons that 
can move only horizontally (vertically), respectively. The hole 
inserted in the AO state (a) can move via NN hopping $t$, but it has 
to turn by 90$^{\circ}$ in each step along its path and leaves behind 
broken bonds leading to string excitations with ever increasing energy
(b) and (c). After moving by 270$^{\circ}$ around a plaquette (d),
the hole cannot return to its initial position as would be
necessary to complete the Trugman path.~\cite{Tru88}}
\label{fig:t2g_schem}
\end{figure}

Instead of the quantum behavior and frustration present in the 
compass model,\cite{Mos03} here one finds that the perfect AO ordered 
state $|0\rangle$ (\ref{gs}) is the ground state of the model at half 
filling. Figure \ref{fig:t2g_schem} presents in a schematic way a few 
first steps in the motion of a hole inserted at a selected site into such a 
ground state. When the hole moves via NN hopping $t$, it creates string 
excitations in each step that cannot be healed by orbital flips, 
because the orbital superexchange (\ref{HJ}) is purely Ising-like, 
see Fig. \ref{fig:t2g_schem}(c). Moreover, it cannot heal the defects 
by itself, because it cannot complete a Trugman loop\cite{Tru88} when 
the orbital defects are created and three occupied orbitals are moved 
anticlockwise on a plaquette after the hole moved clockwise by three 
steps, see Fig. \ref{fig:t2g_schem}(d).

%%%%%%%%%%%%%%%%%%%%%%%%%%%%%%%%%%%%%%%%%%%%%%%%%%%%%%%
%%                      figure 7
%%%%%%%%%%%%%%%%%%%%%%%%%%%%%%%%%%%%%%%%%%%%%%%%%%%%%%%

The structure of ${\cal H}_{t_{2g}}$ (\ref{t2gmodel}) is similar to 
that of the 2D FK model --- one obtains again three-site terms in the 
strong-coupling model along the axes (to third neighbors) 
(\ref{H3s0}) as well as along the plaquette diagonals (to second 
neighbors) (\ref{H3s1}). However, there is now an 
important difference to the 2D FK model: While three-site terms along 
the axes, given in Eq. (\ref{H3s0}), involve a single electron and
conserve orbital flavor, terms along the diagonal (\ref{H3s1}) require 
the subsequent hopping of two electrons with different orbital flavor 
in each step, so the orbital flavor (at the site where the double 
occupancy is created in the excited state) is flipped. We will see later 
that such terms flipping orbital flavor are in fact suppressed, because,
in contrast to the forward hopping along one of the cubic axes, they 
disturb the AO order in the background and thus cost energy. As such, 
they do not affect the low-energy QP state, but contribute only to the 
incoherent processes at higher energy. 

To achieve a complete understanding of the excitation spectra at half
filling, we used two complementary methods and investigated the 
orbital Hubbard model (\ref{Ht2g}) using the VCA, and the 
strong-coupling model (\ref{t2gmodel}) within the SCBA.  Both cases 
were supplemented by exact diagonalization for small ($4\times4$ and 
$4\times 6$ sites) clusters. In the following Section we formulate 
the SCBA treatment of the $t_{2g}$ model, next give results of 
the SCBA calculations in Sec.~\ref{sec_t_2g_vc_scba}, compare them
to a numerical VCA treatment in Sec.~\ref{sec:num_t2g}, and discuss
the QP properties in Sec.~\ref{sec:overqp}. Finally, the impact of 
longer-range hopping pertinent to realistic materials is treated in 
Sec.~\ref{sec:NNN}.

\subsection{Hole-orbiton coupling in the $t_{2g}$ model}
\label{sec_homa}

The calculation of spectral properties of the strong-coupling model
given by Eq. (\ref{t2gmodel}) is more involved than in the previous 
cases. On the one hand, in each such step by hopping $t$ the position of 
a hole is interchanged with an electron and a defect in the AO state 
is created (see Fig. \ref{fig:t2g_schem}). Therefore, one arrives at 
a situation analogous to a hole which tries to propagate in an 
antiferromagnet with the Ising interactions.\cite{Kan89} On the other 
hand, the important new feature which makes the present $t_{2g}$ 
problem more complex is that the electron hopping $t$ is now allowed 
for both orbital flavors. Thus, this problem cannot be solved by the 
RPA \cite{noteis}, which was used to 
determine the Green's functions $G_a(\bf{k},\omega)$ and
$G_b(\bf{k},\omega)$ of a hole doped into the $a$ and $b$ orbital in
either the 1D models or in the 2D FK model. Moreover, 
this problem cannot be reduced to any effective one-particle 
Hamiltonian that one could solve at least numerically for large 
clusters. Hence, we use below the SCBA which gives quite reliable 
results in the spin case.\cite{Mar91} Here one finds that it is well 
designed to treat this problem because several processes not included 
in the SCBA drop out of the Hamiltonian (\ref{t2gmodel}) for physical 
reasons (see below), and therefore the approximation performs remarkably 
well.

In order to implement the SCBA we have to reduce the model of Eq.
(\ref{t2gmodel}) into the polaron problem, following Ref.
\onlinecite{Mar91}. Firstly, we divide the square lattice into two
sublattices $A$ and $B$, such that all the $a$ ($b$) orbitals are
occupied in the perfect AO state in sublattice $A$ ($B$), respectively, 
see Eq. (\ref{gs}). Secondly, we rotate the orbital pseudospins on the 
$A$ sublattice [corresponding the down orbital flavor, see Eq. 
(\ref{Tz})] so that all the pseudospin operators take a positive value, 
$\langle T^z_i\rangle=1/2$, in the transformed ground state. Finally, 
we introduce boson operators $\alpha_i$ (responsible for orbital 
excitations -- orbitons\cite{vdB99}) and fermion operators $h_i$
(holons), which are related to the ones in the original Hilbert
space by the following transformation:
\begin{equation}
\label{bosons}
\tilde{b}_i\equiv h_i^\dag(1-\alpha_i^\dag\alpha_i^{}), \hskip .7cm
\tilde{a}_i\equiv h_i^\dag\alpha_i.
\end{equation}
Note, that we added the projection operators 
$(1-\alpha_i^\dag\alpha_i^{})$ to the transformation relation for the 
$\tilde{b}_i$ fermions in order to keep track of the violation of the 
local constraint that 'no hole and orbiton can be present at the same 
site', cf. constraint $C1$ in Ref. \onlinecite{Mar91}.

Before writing down the {\it polaronic\/} Hamiltonian, we
make the following approximations: 
(i) keep only linear terms in boson operators (as we use linear 
orbital-wave approximation\cite{vdB99}), 
(ii) skip $(1-h_i^\dag h_i)$ and $(1-\alpha_i^\dag\alpha_i)$ 
projection operators when deriving the effective Hamiltonian (both 
simplifications are allowed for the present case of one hole and 
Ising superexchange), and 
(iii) neglect the orbital-flipping terms Eq. (\ref{H3s1}) as 
generating the coupling between the hole and two orbitons 
and leading to higher-order processes in the perturbation theory. 
Then, after Fourier transformation, 
the Hamiltonian Eq. (\ref{t2gmodel}) reads,
\begin{equation}
\label{t2geff}
{\cal H}_{\rm eff} = H_{t}+H_{J}+H_{\rm 3s}^{(l)},
\end{equation}
with
\begin{eqnarray}
\label{Ht_pol} 
H_{t}\!&=&\! 
\frac{z}{\sqrt{N}}\sum_{{\bf k},{\bf q}}
\left\{M({\bf k}, {\bf q}) h^\dag_{{\bf k} A}h_{{\bf k}-{\bf q} B}
\alpha_{{\bf q} A}\right.  \nonumber\\
&&\hskip .5cm \left. + N({\bf k},{\bf  q}) h^\dag_{{\bf k} B}h_{{\bf
k}-{\bf q} A}\alpha_{{\bf q} B}+{\rm h.c.}\right\},
\\
\label{HJ_pol} 
H_{J} \!&=&\! \omega_0 \sum_{\bf k} \left(\alpha_{{\bf
k} A}^\dag \alpha_{{\bf k} A}^{} + \alpha_{{\bf k} B}^\dag
\alpha_{{\bf k} B}^{}\right) ,\\
\label{H3s0_pol} 
H_{\rm 3s}^{(l)} \!&=&\! \sum_{\bf k}\left\{
\varepsilon_A({\bf k}) h_{{\bf k} A}^\dag h_{{\bf k} A}^{} \!+\!
\varepsilon_B({\bf k}) h_{{\bf k} B}^\dag h_{{\bf k} B}^{}\right\},
\end{eqnarray}
where $z=4$ is the coordination number of the square lattice, the sums 
are over momenta ${\bf k}$ in the full Brillouin zone for the whole
lattice,\cite{notebz} the total number of sites in the plane is $N$, 
and indices $A$ and $B$ denote the orbiton operators in both 
sublattices. The orbiton energy $\omega_0=J$ does not depend on
momentum ${\bf k}$, and the vertices in Eq. (\ref{Ht_pol}) have 1D 
dependence on momenta:
\begin{eqnarray}
M({\bf k}, {\bf q}) = \frac{1}{2}\,t \cos (k_x-q_x)\,,  \\
N({\bf k}, {\bf q}) = \frac{1}{2}\,t \cos (k_y-q_y)\,,
\end{eqnarray}
whereas the 1D hole dispersion arising from the propagation within the 
sublattices in Eq. (\ref{H3s0_pol}) are:
\begin{eqnarray}
\varepsilon_A ({\bf k}) =  2\tau\cos (2k_y)\,, \\
\varepsilon_B ({\bf k}) =  2\tau\cos (2k_x)\,.
\end{eqnarray}

\subsection{Self-consistent Born approximation}
\label{sec_t_2g_vc_scba}

Instead of calculating hole Green's functions $G_a({\bf k},\omega)$ and 
$G_b({\bf k},\omega)$ using their definitions (see Sec. \ref{sec:1Dmodel}),
it is convenient now to express them in terms of the operators  used in 
Eq. (\ref{t2geff}). Hence, we introduce hole creation
operators on sublattice $B$
\begin{equation}
h_{{\bf k}B}^\dag = \sqrt{\frac{2}{N}}\sum_{j \in B} e^{-ikR_j} h_j^\dag\,.
\end{equation}
Next, using Eqs. (\ref{bosons}) we obtain the relation:
\begin{align}
b_{\bf k}|0\rangle&=\sqrt{\frac{2}{N}}\sum_{j\in B}e^{-i{\bf k}{\bf R}_j} 
b_j^{} |0 \rangle \nonumber \\ 
&= \sqrt{\frac{2}{N}}\sum_{j \in B} e^{-i{\bf k}{\bf R}_j} h_j^\dag  
(1-\alpha_j^\dag \alpha_j^{})|0 \rangle = h_{{\bf k}B}^\dag |0 \rangle,
\end{align}
since one does not have any pseudospin defects in the AO ordered state 
$|0\rangle$ (\ref{gs}). The latter feature is also responsible for
the fact that one cannot annihilate an electron with the 'wrong' flavor,
e.g. in the $b$ orbital on the $A$ sublattice in the ground state 
$|0\rangle$, which justifies the above definition of the Fourier 
transformation. While one still needs to perform rotation of the 
pseudospin flavor on sublattice $A$, a similar relation can be obtained 
for $h_{{\bf k}A}^\dag$ operators. Finally, we obtain that
\begin{eqnarray}
\label{GAA}
G_{AA}({\bf k},\omega)\!&\equiv&\!\lim_{\delta\to 0}
\left\langle 0\left|\,h_{{\bf k}A}\frac{1}
{\omega+\mathcal{H}_{\rm eff}-E_0+i\delta}
\,h_{{\bf k}A}^\dag \right|0  \right\rangle\,,\nonumber\\  
\\
\label{GBB}
G_{BB}({\bf k},\omega)\!&\equiv&\!\lim_{\delta\to 0}
\left\langle 0\left|\,h_{{\bf k}B}\frac{1}
{\omega+\mathcal{H}_{\rm eff}-E_0+i\delta}
\,h_{{\bf k}B}^\dag \right| 0 \right\rangle\,.\nonumber\\
\end{eqnarray}

%%%%%%%%%%%%%%%%%%%%%%%%%%%%%%%%%%%%%%%%%%%%%%%%%%%%%%%%
%%                        figure 8
%%%%%%%%%%%%%%%%%%%%%%%%%%%%%%%%%%%%%%%%%%%%%%%%%%%%%%%%
\begin{figure}[t!]
\includegraphics[width=7.5cm]{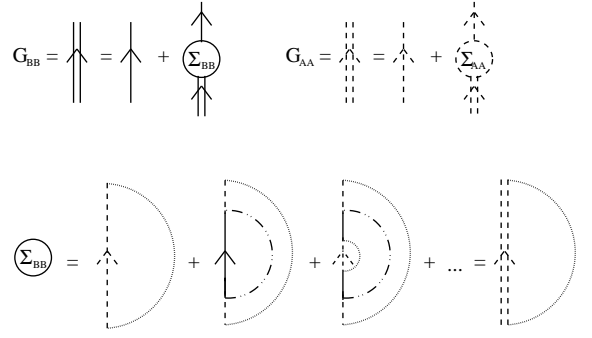}
\caption{Diagrammatic representation of the perturbative procedure 
used within the SCBA: top --- the Dyson's equation for the 
$G_{BB}({\bf k},\omega)$ and $G_{AA}({\bf k},\omega)$ Green's 
functions; bottom --- the summation of diagrams for the self-energy 
$\Sigma_{BB}({\bf k},\omega)$. The densely-dotted and the dashed-dotted 
rainbow lines in the self-energy (lower part) connect the two
vertices $N({\bf k},{\bf q})$ and $M({\bf k},{\bf q})$, respectively. }
\label{fig:Dyson}
\end{figure}

We calculate the above Green's functions (\ref{GAA}) and (\ref{GBB}) 
by summing over all possible noncrossing diagrams (i.e., neglecting 
closed loops), cf. lower part of Fig. \ref{fig:Dyson}. However, the 
crossing diagrams do not contribute here since the closed loops 
(Trugman processes) do not occur, see Fig. \ref{fig:t2g_schem}. Since 
the structure of the present problem makes it necessary that two Green's 
functions and two self-energies are considered, we write the Dyson's 
equation for each of them, as represented in Fig. \ref{fig:Dyson}:
\begin{eqnarray} 
\label{eq:Dysona}
G_{AA}^{-1}({\bf k},\omega)\!&=&\!
\left\{G_{AA}^{(0)}({\bf k},\omega)\right\}^{-1}
-\Sigma_{AA}({\bf k},\omega),
\\
\label{eq:Dysonb}
G_{BB}^{-1}({\bf k},\omega)\!&=&\!
\left\{G_{BB}^{(0)}({\bf k},\omega)\right\}^{-1}
-\Sigma_{BB}({\bf k},\omega),
\end{eqnarray}
where the free Green's functions are given by,
\begin{eqnarray} 
G_{AA}^{(0)}({\bf k},\omega)&=&
\frac{1}{\omega+J+\varepsilon_A({\bf k})}\,, \\
G_{BB}^{(0)}({\bf k},\omega)&=&
\frac{1}{\omega+J+\varepsilon_B({\bf k})}\,,
\end{eqnarray}
and the self-energies 
\begin{eqnarray} 
\label{eq:selfa}
\Sigma_{AA}({\bf k},\omega)&=&\frac{z^2}{N}\sum_{\bf q}
M^2({\bf k},{\bf q})\;G_{BB}({\bf k}-{\bf q},\omega-\omega_0)\,,\nonumber \\
\\
\label{eq:selfb}
\Sigma_{BB}({\bf k},\omega)&=&\frac{z^2}{N}\sum_{\bf q}
N^2({\bf k},{\bf q})\;G_{AA}({\bf k} -{\bf q},\omega-\omega_0)\,.\nonumber \\
\end{eqnarray}
are obtained by summing up the rainbow diagrams of Fig. \ref{fig:Dyson}.
Note that the intersublattice Green's function
$G_{AB}({\bf k},\omega)$ vanishes since it would imply that at least 
one defect was left in the sublattice $B$ after the hole was annihilated 
in the sublattice $A$, resulting in orthogonal states as there are no 
processes in the Hamiltonian which cure such defects [cf. the form of 
the Hamiltonian Eq. (\ref{t2geff}) and Fig. \ref{fig:Dyson}].

\begin{figure}
\subfigure{\includegraphics[width=7.5cm]{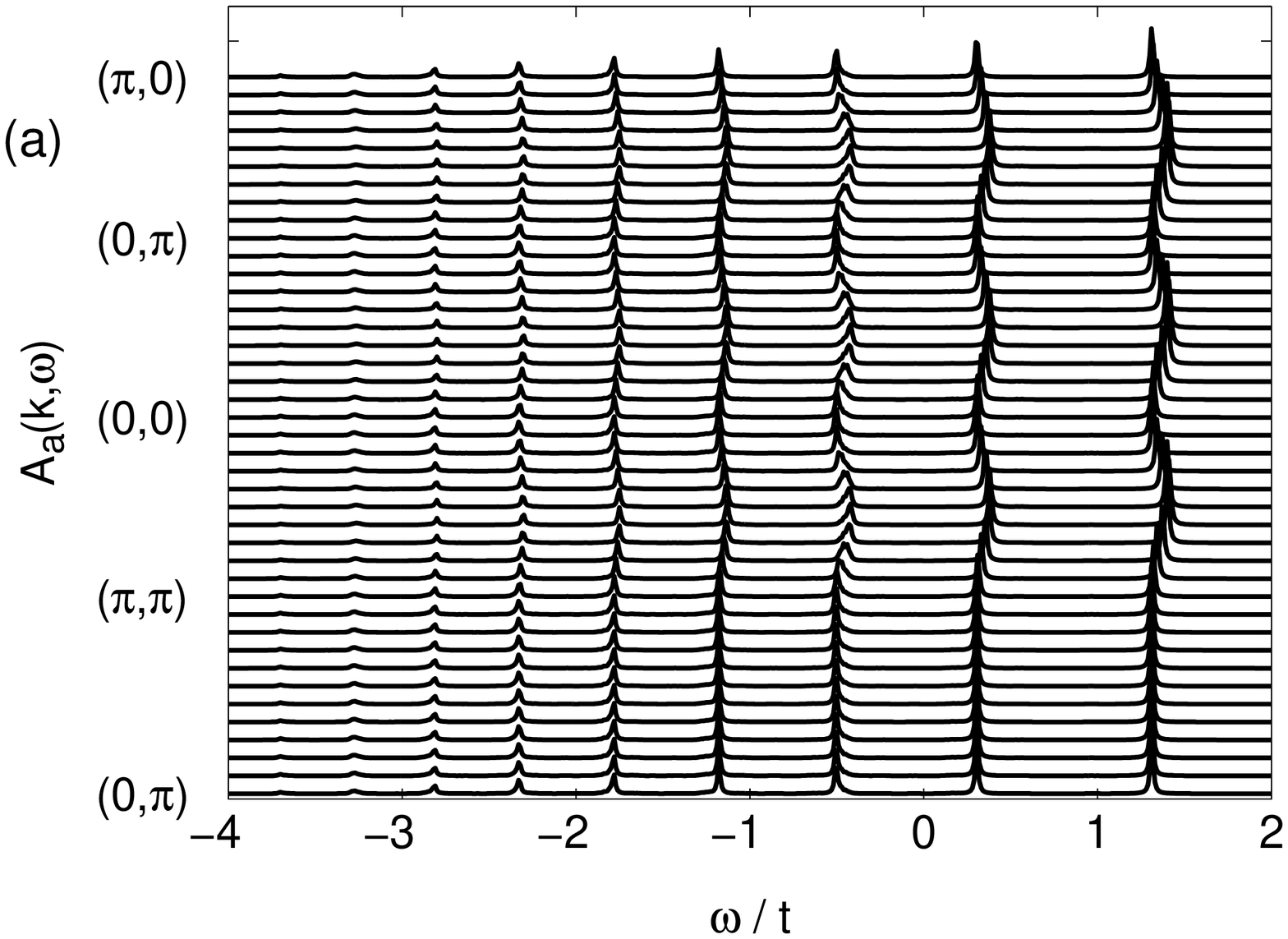}
\label{fig:spec_scba_a}}
\subfigure{\includegraphics[width=7.5cm]{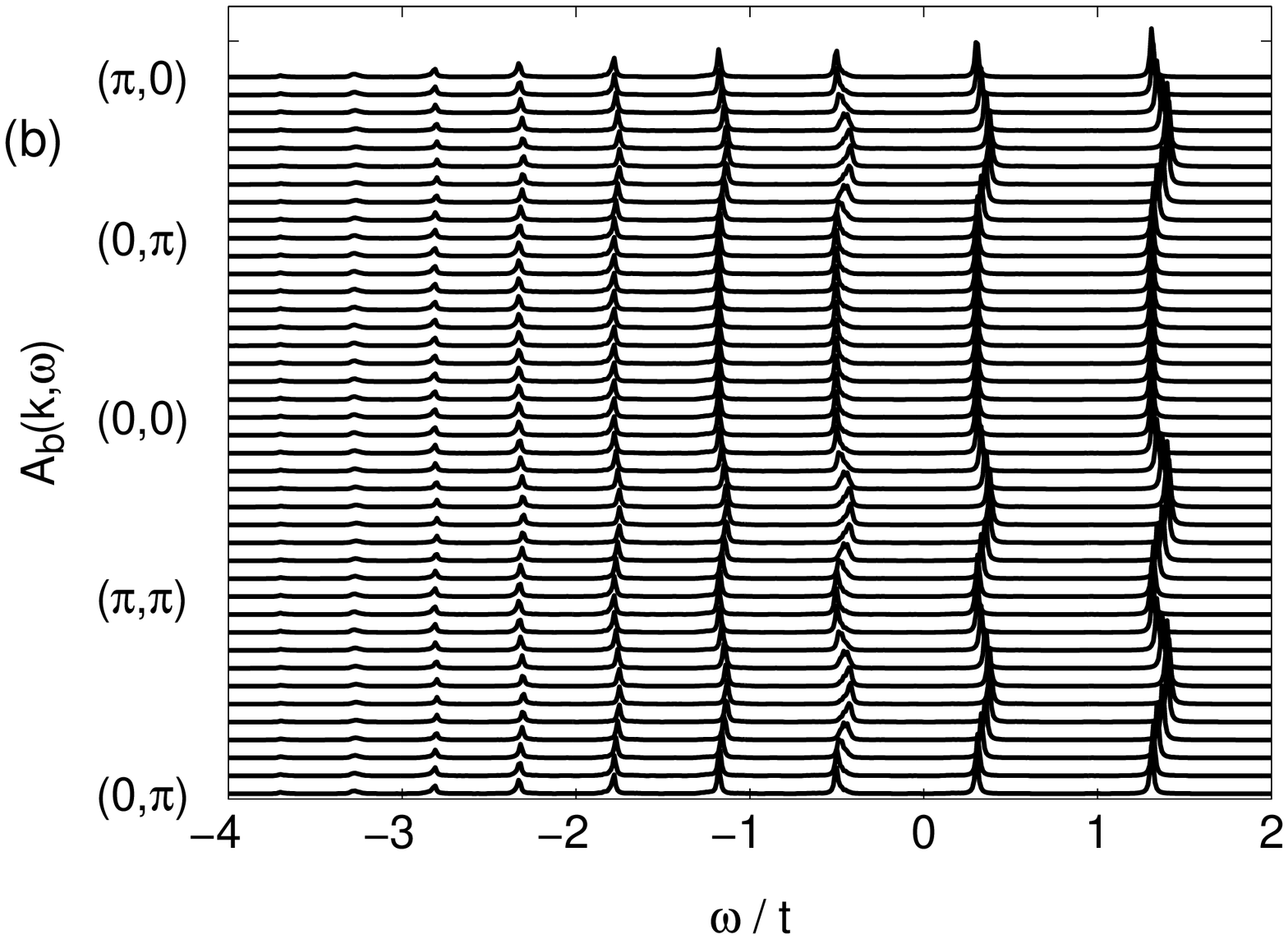}
\label{fig:spec_scba_b}}
\caption{ Spectral function as obtained in the SCBA for the effective
$t_{2g}$ model (\ref{t2geff}) for a hole doped into:
(a) $a$ orbital, and
(b) $b$ orbital. 
Parameters: $J=0.4t$, $\tau=0.1t$, and peak broadening $\delta=0.01t$. }
\label{fig:spec_scba}
\end{figure}

We solved Eqs. (\ref{eq:Dysona})--(\ref{eq:Dysonb}) together with 
Eqs. (\ref{eq:selfa})--(\ref{eq:selfb}) self-consistently on a mesh of 
$20\times 20$ ${\bf k}$-points (and checked the convergence comparing
the results with those obtained for the cluster with $32\times 32$ 
${\bf k}$-points). The spectral functions defined for the sublattices
\begin{eqnarray} 
\label{eq:2d:ab}
A_a({\bf k},\omega)&=&-\frac{1}{\pi}\,\mbox{Im}\,G_{AA}({\bf k},\omega)\,,\\
A_b({\bf k},\omega)&=&-\frac{1}{\pi}\,\mbox{Im}\,G_{BB}({\bf k},\omega)\,,
\end{eqnarray}
are displayed in Fig. \ref{fig:spec_scba}. As discussed in detail in 
Ref. \onlinecite{Dag08}, the spectral density consists of dispersive 
ladder-like spectrum suggesting that the hole doped into any of the 
two orbitals is mobile. The dispersion is particularly pronounced for
the first (low-energy) excitation which we identify as a QP state. 
One finds that its dispersion is strictly 1D and is dictated by the 
orbital flavor at the site where the hole was added, i.e. no dispersion 
occurs in the complementary direction. 
For example, a hole added to the $a$ orbital moves (thanks to the 
three-site terms) only along the $b$ direction. However, such a hole 
moving along the $b$ direction due to the three-site terms could also 
undergo incoherent scattering on orbital excitations, and in addition 
performs "excursions" to the $B$ sublattice due to the $t$ processes, 
which create string-like states. The peculiar interrelation of these 
two types of (coherent and incoherent) propagation (which we
discuss in detail in Sec. \ref{sec:overqp}) leads to the spectra 
depicted in Fig. \ref{fig:spec_scba}.

%%%%%%%%%%%%%%%%%%%%%%%%%%%%%%%%%%%%%%%%%%%%%%%%%%%%%
%%                     figure 9
%%%%%%%%%%%%%%%%%%%%%%%%%%%%%%%%%%%%%%%%%%%%%%%%%%%%%

The lack of the QP dispersion in one direction, e.g. along the $a$ 
direction for a hole doped into the $a$ orbital, is at first instance
counterintuitive:
One could imagine that it should be allowed that the hole doped into 
the $a$ orbital switches to a neighboring site of the $B$ sublattice 
by the $t$ process, and then propagates freely along the $a$ axis by 
the three-site effective hopping $\tau$ without generating any further
defects. This might lead to some dispersion in the spectra along the 
$k_x$ direction. However, the hole 
always has to return to the original site where it has been doped 
as it has to erase the defect it created 
in the first $t$ step when it moved to the other sublattice (otherwise, 
the hole annihilation operator would not permit to return to the 
ground state). Note that this behaviour is similar to the 
hole confinement in a three--site cluster as calculated 
for the hole doped into the $b$ orbital in the 1D model, 
cf. Fig. \ref{fig:1d:1}.\cite{notesub} As a result of such processes, 
one finds very small incoherent (and ${\bf k}$-independent) spectral 
weight in the spectra of Fig. \ref{fig:spec_scba}, which remains 
invisible in the present scale. This discussion demonstrates also that 
the spectra found for the 2D $t_{2g}$ orbital model are dominated 
by the 1D physics explained in Secs. \ref{sec:1Dmodel} and \ref{sec:1Dlegs}.

Next, three remarks which concern the validity of our results are in
order here. Firstly, note that if we skip the flavor-conserving 
three-site terms (\ref{H3s0_pol}), the calculated spectral functions 
(not shown) reproduce the well-known ladder spectra and are equivalent 
to those calculated for the Ising limit of the spin $t$-$J$ model.
\cite{Mar91} 
This means that the zig-zag-like hole trapping in the orbital case 
is physically similar to the standard hole trapping in the spin case 
(apart from the modified energy scale due to a different value of the
superexchange, the ladder spectra are similar in both cases), whereas 
for the free hole movement obviously it matters whether the dispersion 
relation is 1D or 2D. Moreover, this also means that in this special 
case ($\tau=0$) the spectra are the same for holes doped into either 
of the orbitals as the Green's functions are the same for both 
sublattices. However, even in this case it is not allowed to assume 
{\it a priori\/} that $A=B$ and $G_{AA}({\bf k},\omega)=
G_{BB}({\bf k},\omega)$. In fact, these are two sublattices with two 
distinct orbital states occupied in the ground state at half filling, 
and each orbital has entirely different hopping geometry. This does not 
happen in the standard spin case with isotropic hopping, and for this 
reason one can eliminate there the sublattice indices.

%%%%%%%%%%%%%%%%%%%%%%%%%%%%%%%%%%%%%%%%%%%%%%%%%%%%%
%%                     figure 10
%%%%%%%%%%%%%%%%%%%%%%%%%%%%%%%%%%%%%%%%%%%%%%%%%%%%%
\begin{figure}
  \centering
  \psfrag{a}{(a)}
  \psfrag{b}{(b)}
  \psfrag{c}{(c)}
  \psfrag{d}{(d)}
  \psfrag{e}{(e)}
  \psfrag{f}{(f)}
  \includegraphics[width=0.2\textwidth]
    {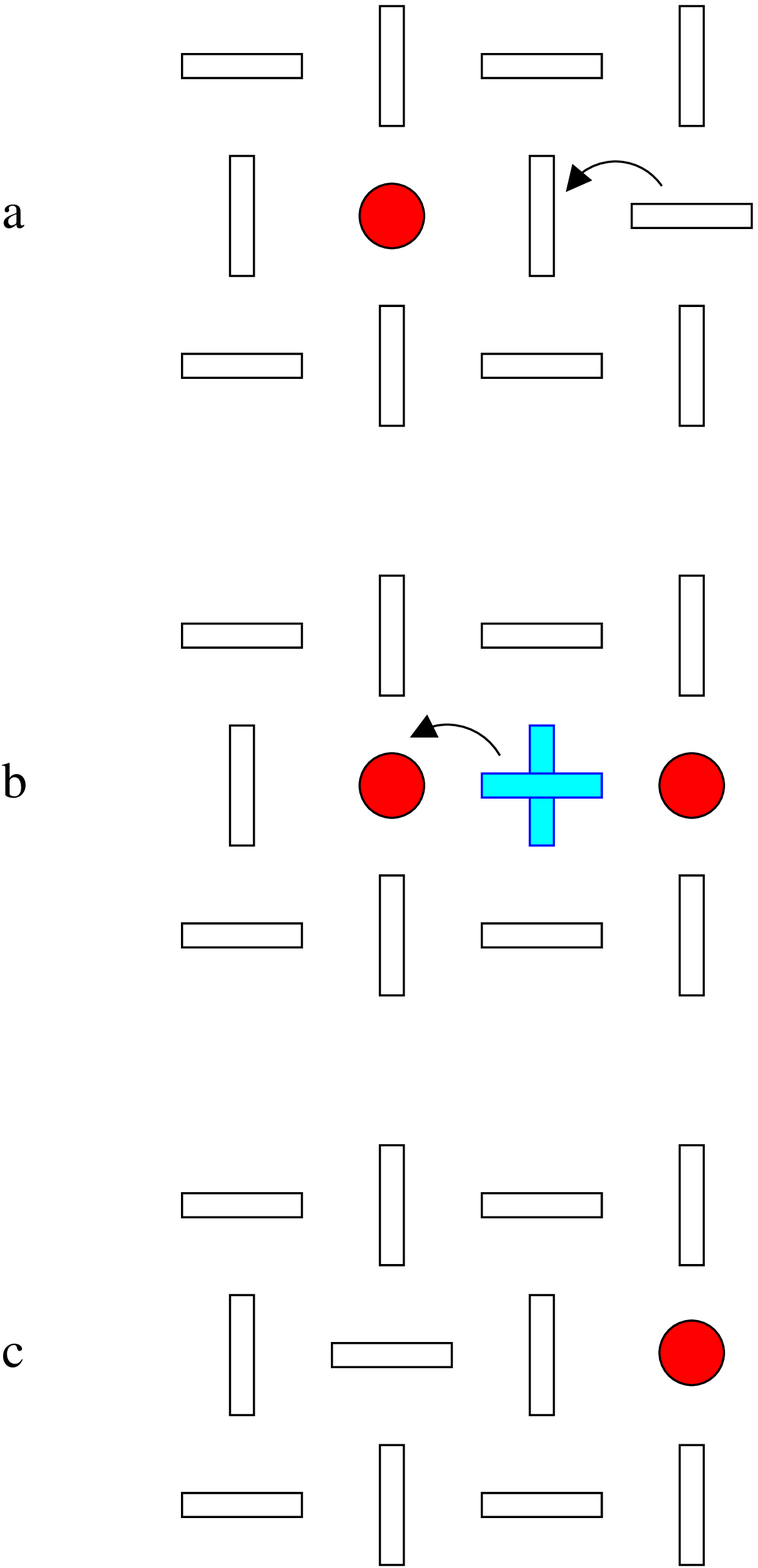}\hfill
  \includegraphics[width=0.2\textwidth]
    {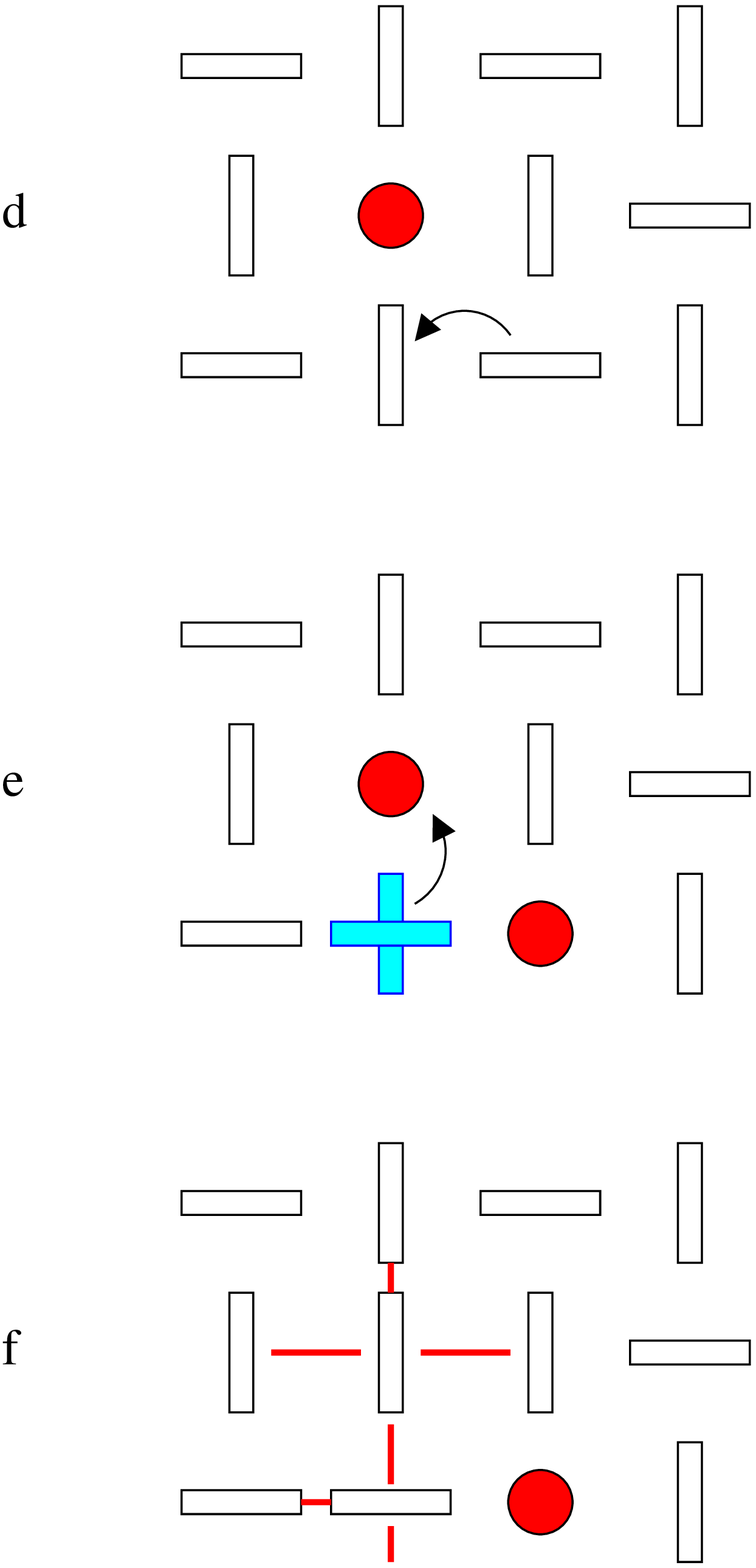}\\
\caption{(Color online) 
Schematic representation of two three-site terms in the $t_{2g}$ 
orbital model (\ref{t2gmodel}). 
Circles depict holes while horizontal (vertical) 
rectangles depict occupied orbitals with electrons that 
can move only horizontally (vertically), respectively. Processes 
shown in panels (a)--(c) result from forward propagation (\ref{H3s0}), 
while the ones shown in panels (d)--(f) and given by Eq.. (\ref{H3s1})
create a defect in the AO order with the energy cost indicated by the 
lines (broken bonds) in (f).}
\label{fig:three_site_t2g}
\end{figure}

Secondly, to obtain the result shown in Fig.~\ref{fig:spec_scba} we 
neglected the three-site terms with $90^{\circ}$ hopping, see Eq. 
(\ref{H3s1}). One may wonder whether this approximation is justified 
whereas the formally quite similar forward hopping term (\ref{H3s0})
is crucial and is responsible for the absence of hole confinement in the
ground state with the AO order.\cite{Dag08} Hence, let us look in more 
detail at these two different kinds of three-site terms, shown in Fig. 
\ref{fig:three_site_t2g}. The first (linear) hopping term (\ref{H3s0})
transports an $a$ electron along the $b$ axis over a site occupied by
a $b$ electron. Such processes are responsible for the 1D coherent hole 
propagation. As we can see in Figs. \ref{fig:three_site_t2g}(a)--(c), the 
AO order remains then undisturbed, so these processes determine the
low-energy features in the spectra. Hopping by the other three-site 
term (\ref{H3s1}), shown in Fig. \ref{fig:three_site_t2g}(d)--(f), 
involves an orbital flip at the intermediate site, destroys the AO order 
on six neighboring bonds, and thus costs additional energy. As two 
orbitals are flipped and two excited states are generated, these 
processes go beyond the lowest order perturbation theory, and it is
consistent to neglect them in the SCBA. In any case, they could 
contribute only to the incoherent processes at high energy and not to
the low-energy QP. Indeed, this interpretation was confirmed by exact 
diagonalization performed for the strong-coupling Hamiltonian 
(\ref{t2gmodel}) on $4\times4$ and $4\times 6$ clusters, which gave 
the same results for the QP dispersion, no matter whether the 
orbital-flipping terms (\ref{H3s1}) were included or not. In addition, 
the QP dispersion found in the SCBA agrees with the numerical results 
obtained by the VCA (see below), which gives further support to the 
present SCBA.

Lastly, despite several other approximations made in writing down the 
Hamiltonian Eq. (\ref{t2geff}), the vertex part $H_t$ is {\it exact\/}, 
in contrast to the Ising interaction for spins.\cite{Mar91} The reason 
is that the constraint $C1$ mentioned above and in Ref. 
\onlinecite{Mar91}, which states that a hole {\it and\/} a boson
excitation are prohibited to occur simultaneously at the same site, 
cannot be violated here, because hopping $t$ is strictly 1D. This can be 
checked either by looking at $H_t$ and verifying that the projection 
operators $(1-\alpha^\dag_i\alpha_i^{})$ can be skipped without changing 
the physics, or by looking at the photoemission spectra in the limit of 
$J\to 0$. Whereas we did both of these checks, let us note here that for 
$J=0$ one obtains the incoherent spectrum with a bandwidth of 
$W_{\rm inc}=4\sqrt{2}t$ (not shown), which (unlike in the spin case) 
perfectly agrees with the RPA result $W_{\rm inc}=4\sqrt{z-2}t$ from 
Ref. \onlinecite{Bri70}, where $z-2$ is the number of possible forward 
going steps in the model. However, still the three-site terms 
$H_{3s}^{(0)}$ and the orbiton terms $H_J$ are not exact in Eq. 
(\ref{t2geff}) and thus we checked the present results by comparing 
them with the numerical spectra obtained for the orbital Hubbard model 
(\ref{Ht2g}) --- the results are presented in the next subsection.

\subsection{Comparison with numerical VCA results}
\label{sec:num_t2g}

Since the problem of a hole added to the background with the AO order of 
$t_{2g}$ orbitals cannot be solved exactly using analytic methods and 
the SCBA had to be employed in the last section, we used also a 
numerical approach. Actually, we compare the analytic results for the 
strong-coupling model (\ref{t2gmodel}) presented in Sec. 
\ref{sec_t_2g_vc_scba} with those obtained for the $t_{2g}$ Hubbard 
model (\ref{Ht2g}) using VCA. This enables us to compare not only the 
methods employed but also the two models which stand for the same 
physics in the strongly correlated regime. 

\begin{figure}[t!]
\includegraphics[width=7.5cm]{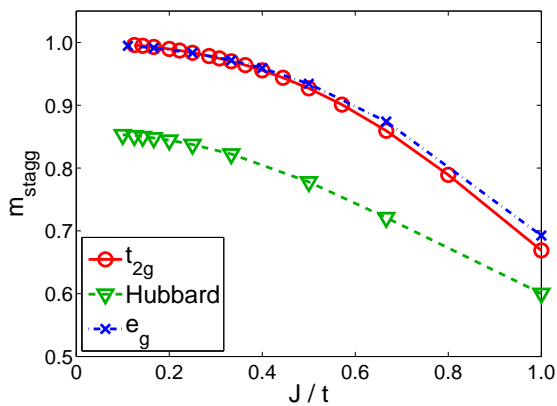}
\caption{(Color online) 
Staggered magnetization $m_{\rm stagg}$ for increasing $J/t$
as obtained for the $t_{2g}$ orbital Hubbard model, spin Hubbard model
(called Hubbard on the figure), and for the $e_g$ orbital 
Hubbard model, respectively. }
\label{fig:m}
\end{figure}

We first use the VCA to determine the staggered orbital moment in
the ground state of the $t_{2g}$ orbital Hubbard model (\ref{Ht2g}),
\begin{equation}
\label{mstag}
m_{\rm stagg} \equiv \frac{1}{N}\sum_i e^{i{\bf Q}\cdot{\bf R}_i}
\left|\left\langle\tilde{n}_{ib}-\tilde{n}_{ia})\right\rangle\right|,
\end{equation}
with ${\bf Q}=(\pi,\pi)$ corresponding to the 2D AO order. We compare this
result to the similar ones obtained for the spin Hubbard model
and for the $e_g$ orbital Hubbard model for a 2D plane (the strong-coupling 
model corrresponding to the latter situation was studied in Refs.
\onlinecite{vdB00,Dag07}.) As expected, $m_{\rm stagg}$ increases with 
decreasing $J$ (increasing $U$), see Fig. \ref{fig:m}. For $J\to 0$ 
($U\to\infty$), $m_{\rm stagg}\to 1$ for both orbital models, 
corresponding to the perfect (classical) 2D Ising-like order, while 
quantum fluctuations reduce the moment $m_{\rm stagg}$ in the spin 
model. We remark that the treatment of the quantum fluctuations within the VCA
is far from perfect (and limited by actual cluster size), so the 
staggered magnetization reaches $m_{\rm stagg}\simeq 0.85$ in the
limit $J\to 0$ (Fig.~\ref{fig:m}), and does not reproduce the value 
of $0.606$, well known from the spin-wave theory.~\cite{Mat06}

%%%%%%%%%%%%%%%%%%%%%%%%%%%%%%%%%%%%%%%%%%%%%%%%%%%%
%%                    figure 11
%%%%%%%%%%%%%%%%%%%%%%%%%%%%%%%%%%%%%%%%%%%%%%%%%%%%

In all three models, the staggered moment (\ref{mstag}) obtained using
the VCA decreases with decreasing $U$ (increasing $J$), see Fig.
\ref{fig:m}, because the kinetic energy can then generate more doubly 
occupied sites in the ground state. One finds that both the $e_g$ and 
$t_{2g}$ models give very similar results for the  staggered moment, but 
differ strongly from the SU(2) symmetric spin model, as has been shown 
before in three dimensions.\cite{Fei05} However, we note that orbital 
order is slightly weaker for $e_g$ orbitals than for $t_{2g}$. This
may be easily explained by the fact that the $e_g$ hopping is slightly 
smaller than $t$ for the relevant orbital states 
$1/\sqrt{2}(|z\rangle \pm |x\rangle)$,
while all other hopping processes are frustrated by the AO order
(\ref{mstag}). Consequently, correlations have a stronger impact on
$e_g$ electrons and induce a slightly enhanced $m_{\rm stagg}$. 
Finally, we would like to emphasize that the AO is 2D in all three 
models, in spite of the fact that the kinetic energy is strongly 
anisotropic in the orbital models and actually has a 1D nature in 
the $t_{2g}$ model, see below. 

\begin{figure}[t!]
\subfigure{\includegraphics[width=0.4\textwidth]{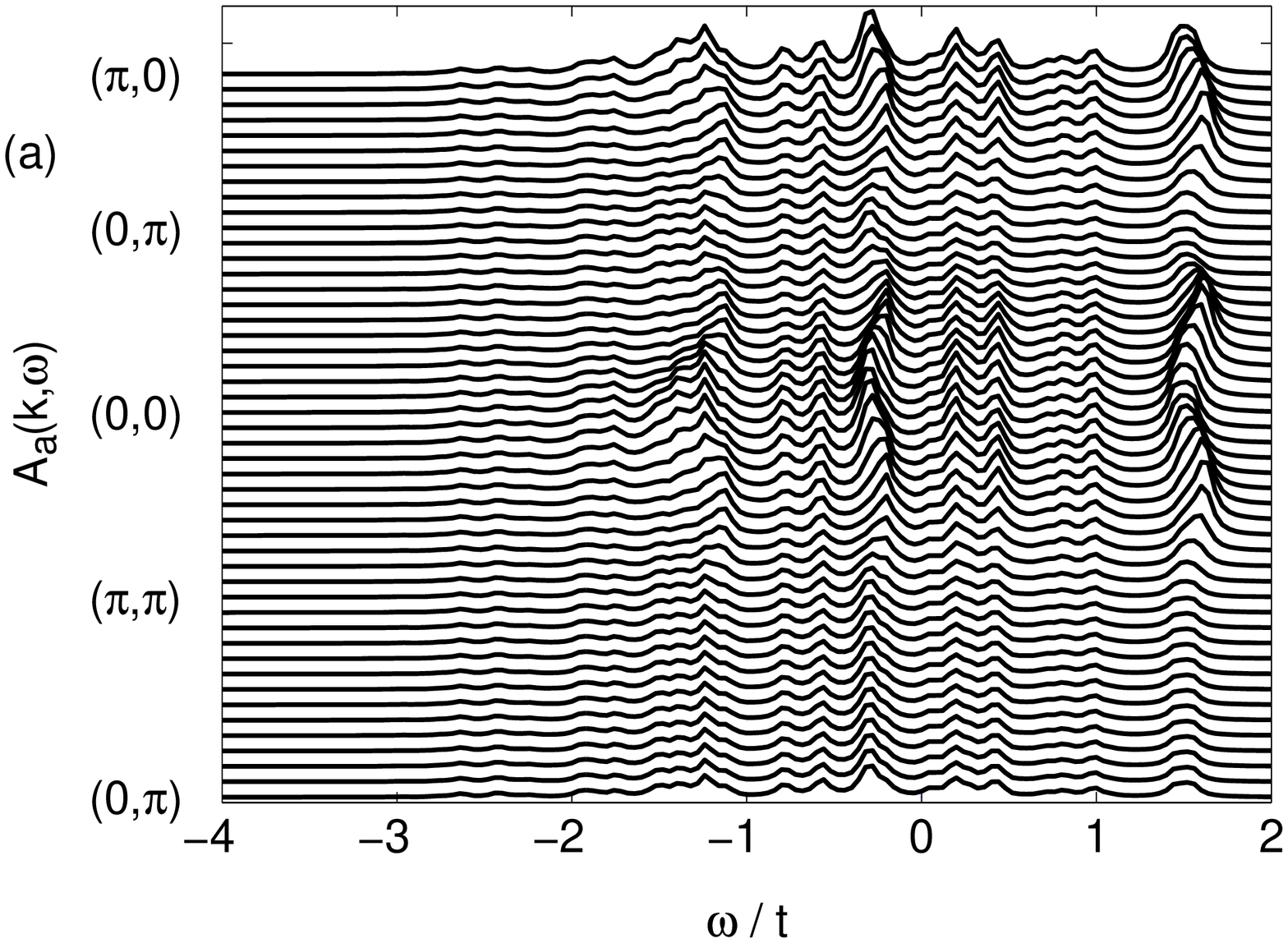}}\\[-2em]
\subfigure{\includegraphics[width=0.4\textwidth]{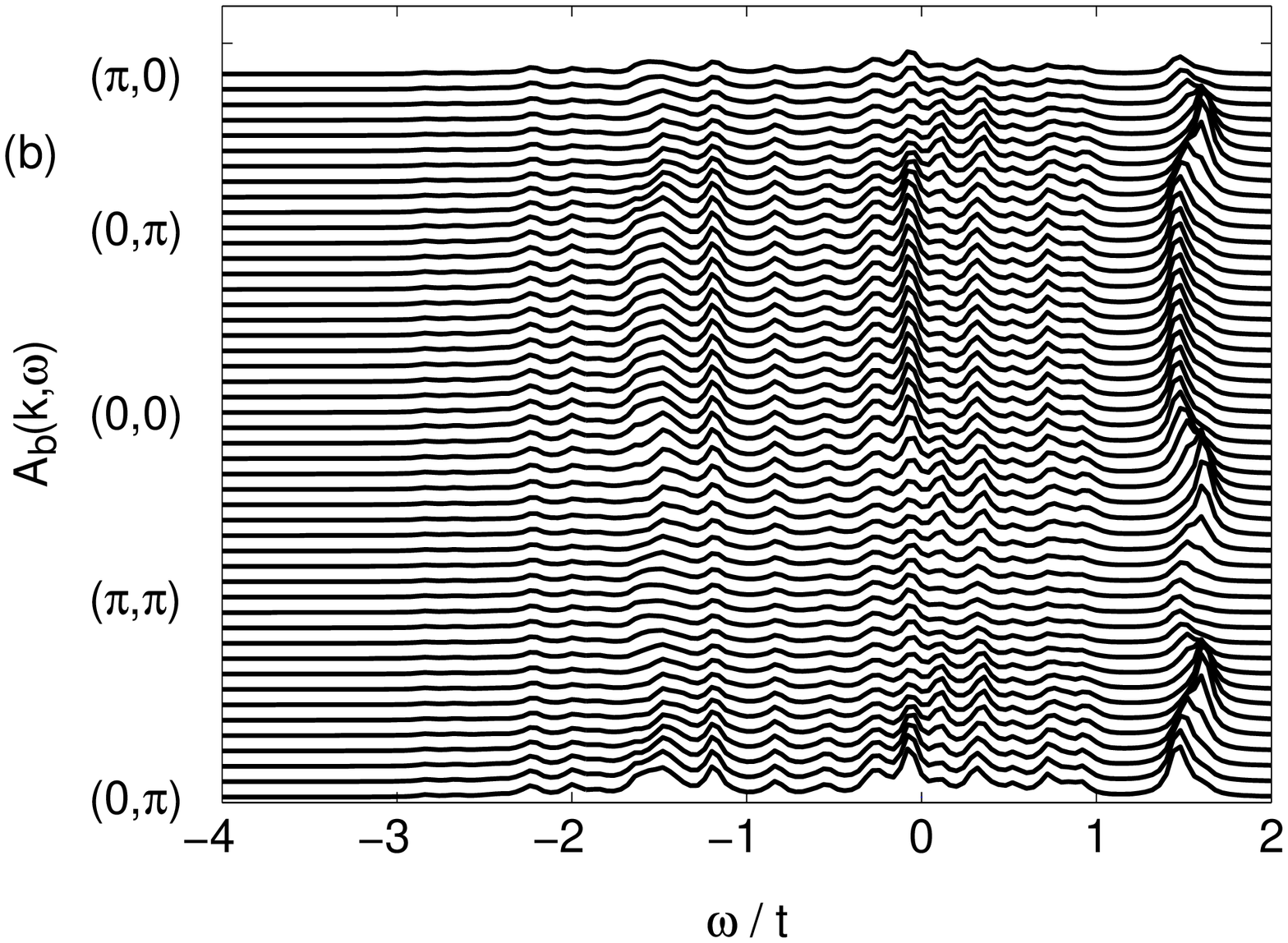}}
\caption{
Spectral function $A({\bf k},\omega)$ obtained with VCA for the 2D 
$t_{2g}$ Hubbard model (\ref{Ht2g}) for: 
(a) $a$ orbitals, and 
(b) $b$ orbitals. 
Parameter: $U=10t$.}
\label{fig:spec_cpt}
\end{figure}

Before we analyze the spectral functions, let us recall that the VCA
\cite{Aic03} is appropriate for models with on-site interactions, as for 
instance the present Hubbard model (\ref{Ht2g}) for $t_{2g}$ orbitals,
but cannot be easily implemented for models where the interacting part
connects different sites, like in the $t$-$J$ (or strong-coupling) 
model. For the present $t_{2g}$ model (\ref{Ht2g}) we use VCA with 
commonly used\cite{Aic03} open boundary conditions, which leads to
the spectral densities depicted in Fig.~\ref{fig:spec_cpt}. The
results resemble very much the SCBA results of Fig.~\ref{fig:spec_scba}
for the strong-coupling model (\ref{t2gmodel}), suggesting that not only 
both models are indeed equivalent in the strongly correlated regime, but 
also that the implemented SCBA method of Sec. \ref{sec_t_2g_vc_scba} is 
of a very good quality. The differences between them, almost exclusively 
affecting high-energy features, are discussed below. 

%%%%%%%%%%%%%%%%%%%%%%%%%%%%%%%%%%%%%%%%%%%%%%%%%%%%%
%%                     figure 12
%%%%%%%%%%%%%%%%%%%%%%%%%%%%%%%%%%%%%%%%%%%%%%%%%%%%%

On the one hand, we see that the high-energy part of the spectral density 
in Fig.~\ref{fig:spec_scba} is composed of a number of peaks with a 
dispersion almost parallel to that of the QP state. In fact, the 
spectrum corresponds almost exactly to the ladder spectrum of the 
spin $t$-$J$ model with Ising superexchange,\cite{Kan89,Mar91} 
but with a weak dispersion added to the peaks. The peaks at 
higher-energy are dispersive for the same reason as the QP state: 
After hopping a few times by NN hopping $t$ --- and creating string
excitations, see Fig.~\ref{fig:t2g_schem} --- the hole can exhibit
coherent propagation via three-site terms, leading to the observed
dispersion. On the other hand, the VCA spectrum 
(Fig. \ref{fig:spec_cpt}) does not show these distinct peaks and the 
structure of $A({\bf k},\omega)$ is richer. However, the first 
moments calculated in separate intervals of $\omega$ follow similar
dispersions obtained for the first three peaks obtained in 
$A({\bf  k},\omega)$ within the SCBA.~\cite{Dag08}  

The above difference can be
understood as following from the full Hilbert space used in the VCA
calculations which results in excitations of doubly occupied sites,
weakening of the AO order even for relatively large $U=10t$, see
Fig. \ref{fig:m}. Therefore, the spectra of Fig. \ref{fig:spec_cpt}
have more incoherent features. In addition, the three-site 
terms which create two orbiton excitations (\ref{H3s1}), that were 
neglected in the SCBA, might also influence the high-energy part
of the spectrum. The 
difference to the SCBA results might also be due to the fact that states 
with longer strings including several orbital excitations, which occur
when the hole moves by a few steps via $t$, cannot be directly 
accommodated within the 10-site cluster solved here, and cannot be
reproduced with sufficient accuracy.

Apart from the differences in the high-energy part of the spectrum, 
we also observe differences in the spectral weight distribution
(see also the detailed discussion below in Sec. \ref{sec:overqp}): 
In the VCA results (Fig.~\ref{fig:spec_cpt}) the total weight found 
in photoemission part (hole excitation) strongly depends on momentum 
${\bf k}$, while no such variation can be seen in the SCBA results in
Fig. \ref{fig:spec_scba}. This difference does not originate from
different approximate methods used, but stems from the different 
\emph{models}: In Hubbard-like models, the number of electron states 
occupied depends on the momentum ${\bf k}$.\cite{Ste91} In contrast, 
undoped $t$-$J$-like models have exactly one electron per site, which 
enforces a different sum rule and eliminates the ${\bf k}$-dependence 
from the photoemission part. 

\subsection{Discussion of quasiparticle properties}
\label{sec:overqp}

\begin{figure}
  \psfrag{0p}[Bl]{\hspace{-0.5em}$(0,\pi)$}
  \psfrag{p0}[cl]{\hspace{-1.2em}{$(\pi,0)$}}
  \psfrag{pp}[Bl]{\hspace{-1.2em}$(\pi,\pi)$}
  \psfrag{O}[cl]{\hspace*{-0.5em}$(0,0)$}
  \centering
  \subfigure{\includegraphics[width=0.47\textwidth]
    {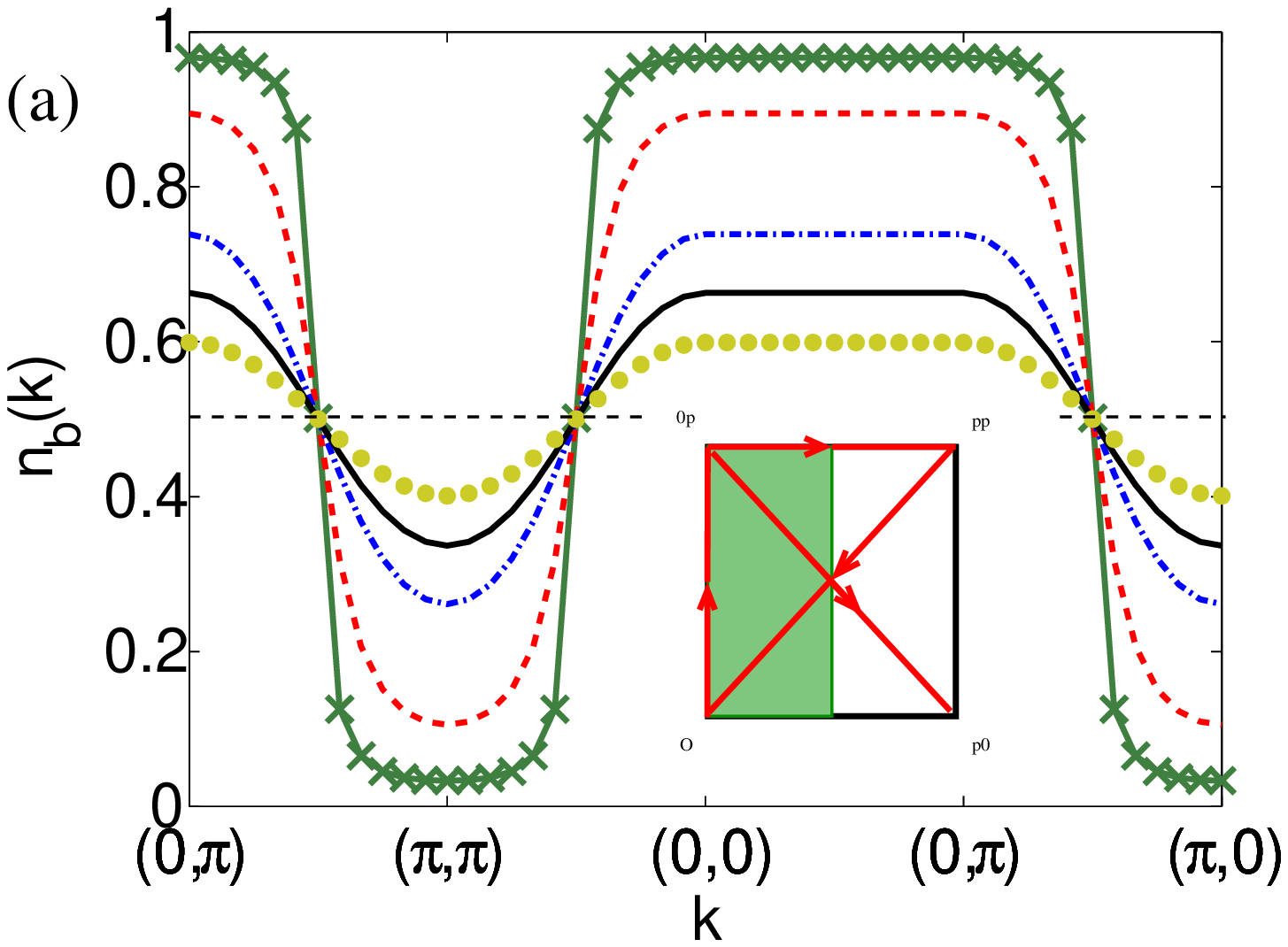}\label{fig:weight_pes_isi}}
  \subfigure{\includegraphics[width=0.47\textwidth]
    {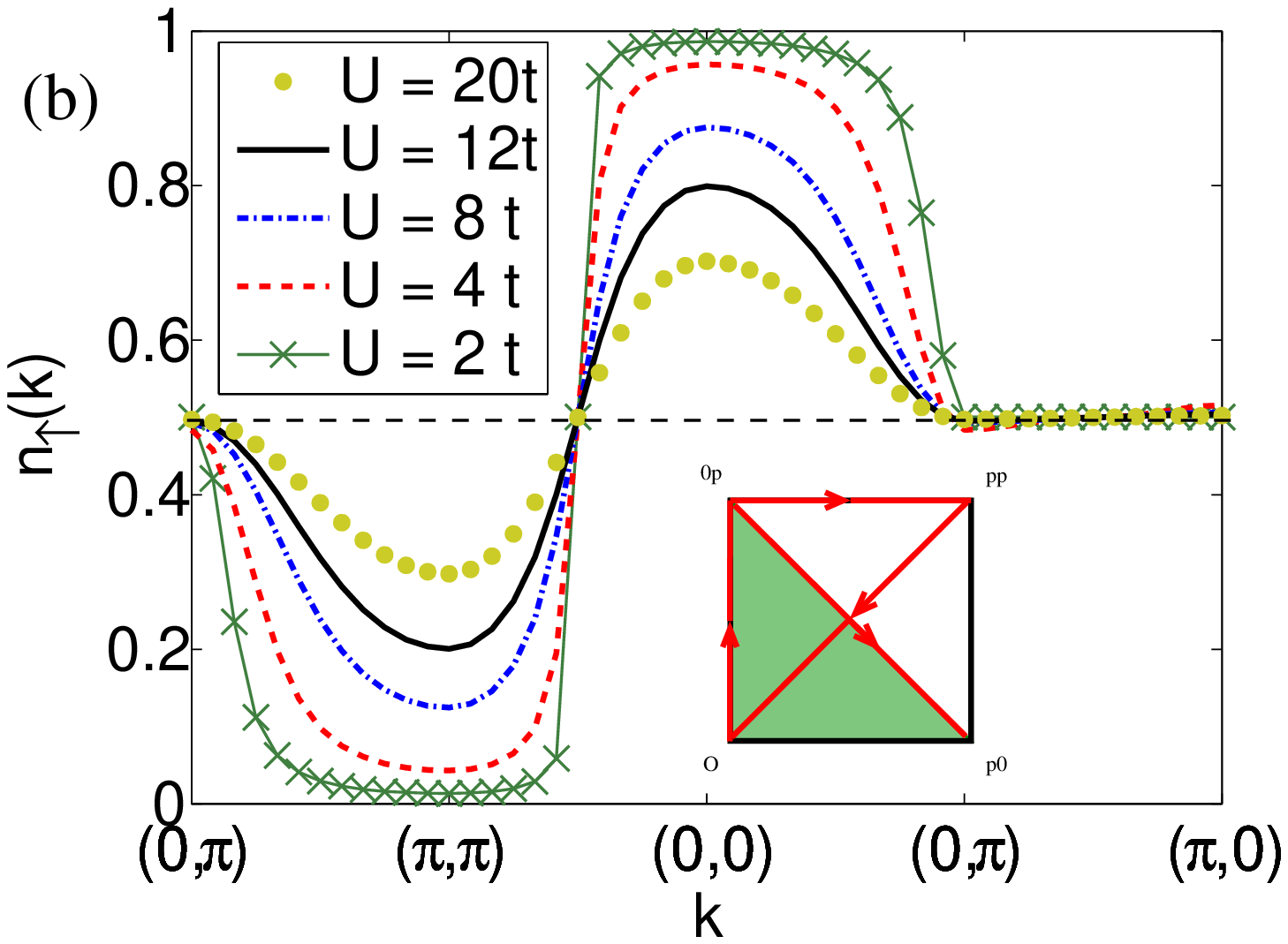}\label{fig:weight_pes_hubb}}
\caption{(Color online) Dependence of total weight found in the
photoemission spectrum on momentum $k$ for (a) a hole inserted into 
the $b$ orbital of the $t_{2g}$ model (\ref{Ht2g}) and (b) for a hole 
with spin up in the SU(2) symmetric spin Hubbard model. All data were
obtained by VCA. For illustration, we added a line at $n(k) = 0.5$, 
corresponding to the constant $n(k)$ for $t$-$J$-like models. The 
insets show the first quadrant of the first BZ: The arrows indicate 
the path taken for the main panel, the shaded area gives momenta 
with $n_b>0.5$ and $n_\uparrow > 0.5$, respectively. }
\label{fig:weight_pes}
\end{figure}

In order to get a deeper understanding of the problem mentioned in the 
last paragraph of Sec. \ref{sec:num_t2g}, let us consider first the 
overall spectral weight distribution obtained in the VCA calculations. 
It is measured by the momentum-dependent electron occupation 
\begin{equation}
\label{nk}
n_{\alpha}({\bf k}) \equiv \left\langle
c_{{\bf k}\alpha}^{\dagger}c_{{\bf k}\alpha}^{}\right\rangle,
\end{equation}
obtained for the $\alpha$-flavor in the Hubbard model, as 
for instance the $t_{2g}$ model (\ref{Ht2g}). We recall that Eq. 
(\ref{Ht2g}) which leads in the limit $U\gg t$ to the 2D $t_{2g}$ model 
(\ref{t2gmodel}) is rather different from the one obtained for the spin 
Hubbard model with the SU(2) symmetry, see Fig. \ref{fig:weight_pes}. 
One may easily identify the 
quasi-1D dependence only on $k_x$ in the $b$ orbital momentum 
dependence $n_b({\bf k})$ shown in Fig. \ref{fig:weight_pes}(a), in 
contrast to the 2D variation of $n_{\sigma}({\bf k})$ in the spin case 
with isotropic hopping of Fig. \ref{fig:weight_pes}(b). 

%%%%%%%%%%%%%%%%%%%%%%%%%%%%%%%%%%%%%%%%%%%%%%%%%%%%%
%%                     figure 13
%%%%%%%%%%%%%%%%%%%%%%%%%%%%%%%%%%%%%%%%%%%%%%%%%%%%%

The insets in Fig. \ref{fig:weight_pes} show which states are occupied
at $U=0$ in the two models, and indicate the difference between the
isotropic 2D hopping of the spin model and the 1D kinetic energy of
the orbital model. For instance, $n_{\sigma}({\bf k})=0.5$ along the
$(0,\pi)--(\pi,0)$ line for spins, while it shows full variation along
this line in the orbital case. In both cases, 
we observe strong modifications of the electron distribution with 
increasing $U$. For $U=0$, the states below the Fermi surface 
(${\bf k}\in{\cal S}_\textrm{F}$) are occupied and states above it 
(${\bf k}\not\in{\cal S}_\textrm{F}$) are empty. 
Consequently, $n_{\alpha}({\bf k})$ is given by a step function with: 
$n_{\alpha}({\bf k})=1$ for $k\in{\cal S}_\textrm{F}$ and 
$n_{\alpha}({\bf k})=0$ for $k\not\in{\cal S}_\textrm{F}$. The changes 
are particularly fast in the range of $U\sim 8t$; for $U>8t$ the 
momentum distribution function $n_{\alpha}({\bf k})$ (\ref{nk}) smears 
out and one recognizes the strong-coupling regime. However, 
the difference between
$n_{\alpha}({\bf k}=(0,0))$ and $n_{\alpha}({\bf k}=(\pi,\pi))$ is 
larger in the spin model, suggesting that the correlation effects are 
stronger in the orbital case. Indeed, this follows from the 1D character 
of the kinetic energy in the orbital model. In contrast, both 
strong-coupling models (for spin or orbital flavors) would give at half
filling a constant $n_{\alpha}({\bf k})=0.5$ even for finite $U<\infty$, 
although this result is strictly speaking correct only at $U=\infty$, 
as shown in Fig. \ref{fig:weight_pes}.

\begin{figure}[t!]
\includegraphics[width=7.5cm]{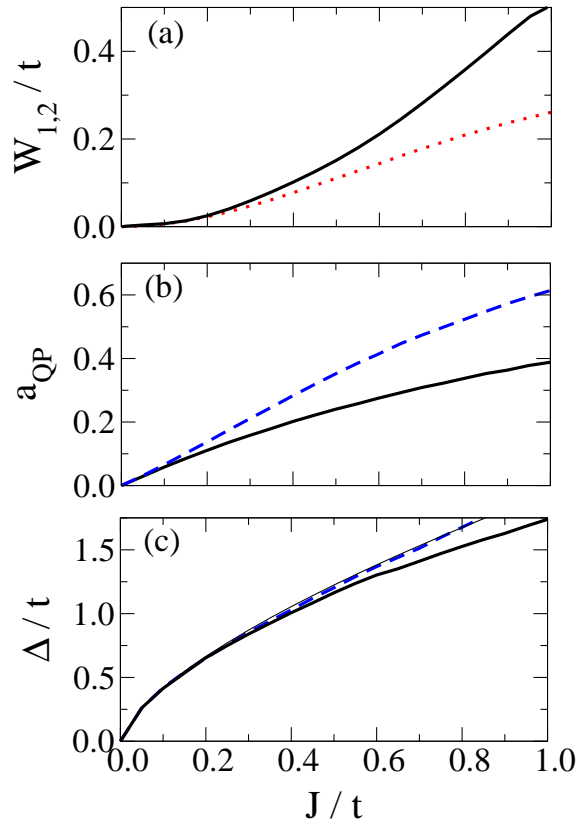}
\caption{(Color online) 
Quasiparticle properties obtained for the 2D $t_{2g}$ model within
the SCBA for increasing superexchange $J$ (with $\tau=J/4$):
(a) the bandwidth of the QP $W_1$ (solid line) and the second 
dispersive feature $W_2$ (dotted line), 
(b) the spectral weight $a_{\rm QP}$, and 
(c) the distance between the first two peaks in the spectra 
(pseudogap) $\Delta$. 
The solid (dashed) lines in (b) and (c) give the results for 
${\bf k}=(0,0)$ [${\bf k}=(\pi/2,\pi/2)$], respectively. 
The light solid line in (c) indicates the $t(J/t)^{2/3}$ law (see text). }
\label{fig:qp}
\end{figure}

After understanding the differences between the QP properties found in
the VCA and the SCBA, we concentrate solely on the 
QP properties calculated using the latter method. Hence,
following Ref. \onlinecite{Mar91}, we analyze
the characteristic features of the QP states in the 2D $t_{2g}$ model, 
such as the bandwidth $W$ and the QP spectral weight $a_{\rm QP}$. 
The energy of incoherent excitations (string states) is to some extent 
characterized by the separation between the QP state and the next 
(second) spectral feature at higher energy -- it is called here
a pseudogap $\Delta$. All these quantities increase with increasing
superexchange energy $J$ ($\tau=J/4$), see Fig. \ref{fig:qp}. 
One finds that:
(i) the bandwidth $W_1$ of the first QP peak, see Fig. \ref{fig:qp}(a), 
is proportional to $J^2$ for small $J$ ($J<0.7$) and to $J$ in the 
regime of large $J$ ($J>0.7$) --- the bandwidth renormalization is here 
distinct from the one found either in the spin SU(2) (see Ref. 
\onlinecite{Mar91}) or in the orbital $e_g$ models,\cite{vdB00}
(ii) the bandwidth $W_2$ of the second largest dispersive peak [Fig. 
\ref{fig:qp}(a)] is smaller than that for the first peak and tends to 
saturate at $W_2\sim 0.25t$ value for larger $J>t$ (not shown), 
(iii) the spectral weight $a_{\rm QP}$ of the QP peak, shown in Fig.
\ref{fig:qp}(b), grows with $J$, and 
(iv) the pseudogap $\Delta$ shown in Fig. \ref{fig:qp}(c) grows 
generally like $J^{2/3}$, while for higher $J$ values some deviation 
from this law is observed for the ${\bf k}=(0,0)$ point. Most (but not 
all) of these results are qualitatively different from the ones 
obtained for the QP states, and their momentum dependence, 
in the SU(2) Heisenberg antiferromagnet. Let us now discuss
the above mentioned QP properties in more detail.

%%%%%%%%%%%%%%%%%%%%%%%%%%%%%%%%%%%%%%%%%%%%%%%%%%%%
%%                    figure 14
%%%%%%%%%%%%%%%%%%%%%%%%%%%%%%%%%%%%%%%%%%%%%%%%%%%%

Firstly, the QP bandwidth arising from the superexchange three-site 
terms is renormalized as it is much smaller than the respective
free value, $W\ll 2J$. Even at $J=t$, the QP bandwidth is only
$W\simeq J/2$, i.e., is here reduced by a factor of 4. This is not 
surprising in the view of
incoherent processes which "dress" the propagating hole and 
increase its effective mass. Indeed, the collapse of the QP bandwidth 
in the regime of $J\to 0$ may be understood as following from numerous
incoherent string excitations which are easy in this regime as they do 
not cost much energy. A similar but considerably weaker reduction of 
the 1D dispersion by string excitations was seen before in the centipede 
model, see Fig. \ref{fig:ext:2}(a). However, in that case the 
renormalization was almost linear as the length of the string excitations 
was limited to a single step (within one of the three-atom units along 
the chain), and could not further increase with decreasing $J$. 
In addition, the dispersion of the second peak is weaker than that 
of the QP. Interestingly, the bandwidth corresponding to the 
dispersion of the second peak in the centipede model is not only weaker 
than that of the QP itself, but is also renormalized in a similar way 
to that found for the full 2D $t_{2g}$ model. Altogether, 
this suggests that the bandwidth renormalization of the coherent
hole propagation in the 2D $t_{2g}$ strong-coupling model follows from 
the creation of string states during the 1D hole propagation via 
three-site terms. Such processes are absent in the 1D and 2D FK model,
and therefore the hole moves there freely by three-site hopping terms 
and the bandwidth is unrenormalized.

Secondly, in contrast to the spin $t$-$J$ model with Ising superexchange 
interactions,\cite{Mar91} where the QP spectral weight is independent 
of ${\bf k}$, it varies here with the component ($k_x$ or $k_y$) of the 
momentum ${\bf k}$, cf. Fig. \ref{fig:qp} as well as Fig. \ref{fig:spec_scba} 
and \ref{fig:spec_cpt}. Similar to the spin 
$t$-$J$ model, the QP spectral weight is larger for the ${\bf k}$ 
values with the lowest QP energies than for the ones close to the 
maximum in QP dispersion, for instance 
$a_{\rm QP}(\pi/2,\pi/2)>a_{\rm QP}(0,0)$, see Fig. \ref{fig:qp}(b). 
Altogether, the ${\bf k}$-dependence here is however much weaker 
than in the spin case.\cite{Mar91} 
The increase of $a_{\rm QP}(\pi/2,\pi/2)$ with $J/t$ 
resembles the increase of the spectral weight for the low-energy 
peak at $k=\pi/2$ in the centipede model, see Fig. \ref{fig:ext:2}(b). 

Finally, we address the issue of the pseudogap which separates the
QP state from the first incoherent excitation. It scales almost as 
$t(J/t)^{2/3}$, see Fig. \ref{fig:qp}(c), in agreement with the result for 
the Ising spin model.
\cite{Mar91} This demonstrates that in spite of the observed
${\bf k}$-dependence of the QP properties and the pseudogap itself, 
the pseudogap originates from string excitations similar to those
generated by the hole moving in the spin background with AF order.
Note also that the spectrum of the 2D model with dense distribution 
of incoherent maxima in the range of $J\to 0$ is qualitatively
different from the 1D centipede model, shown in Fig. \ref{fig:ext:1}(b).

\section{Photoemission spectra of vanadates and fluorides }
\label{sec:NNN}

In this Section we discuss the possible implications of the results 
obtained for the $t_{2g}$ orbital model of Sec. \ref{sec:t2g} on future
experiments, and make predictions concerning the photoemission spectra 
of strongly correlated fluorides and vanadates. As before, we discuss
the strongly correlated regime with $U\gg t$. The first important 
feature to consider is the interplay of the three-site hopping with the 
longer-range $\{t_2,t_3\}$ hopping to second and third neighbors which
contributes to the electronic structure and may always be expected in 
any realistic system (for instance, due to hybridization with oxygen
orbitals). These hopping elements were neglected in both the Hubbard 
model (\ref{Ht2g}) and in the strong-coupling model (\ref{t2gmodel}), 
but they could significantly influence the spectral weight distribution. 
We will see, however, that although features induced by longer-range 
hopping are small as long as $|t_{2(3)}|<t$, they 
can be clearly distinguished from the effects of three-site hopping.

\begin{figure}[t!]
  \centering
  \includegraphics[width=0.4\textwidth]
  {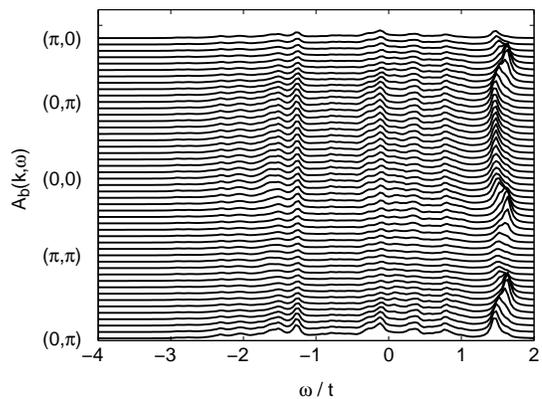}
  \caption{
Spectral density $A_b({\bf k},\omega)$ obtained within the VCA method 
for a hole inserted into $b$ orbitals of the $t_{2g}$ model
(\ref{Ht2g}), supplemented by finite NNN hopping (\ref{eq:tab}).
Parameters: $U=10t$, and $t_2=0.15t$. 
\label{fig:spec_t_ab}}
\end{figure}

The same requirements for orbital symmetry that are necessary to
obtain NN hopping, as discussed in this work, also strongly 
restrict the range of allowed longer-range hopping terms. It is 
important to recall that the $d$--$d$ hopping elements involve 
intermediate oxygen orbitals. For next 
nearest neighbor (NNN) hopping, the orbital phases of the involved 
oxygen $2p_\pi$ orbitals make all terms vanish that conserve orbital
flavor,\cite{Zaa93} and only orbital-flipping terms,
\begin{equation}
\label{eq:tab}
H_\textrm{NNN} = - t_2\sum_i \left(
a^\dag_{i\pm\bf{\hat{b}}}b^{}_{i\pm\bf{\hat{a}}}+
a^\dag_{i\mp\bf{\hat{b}}}b^{}_{i\pm\bf{\hat{a}}}+\textrm{h.c.}\right)\,,
\end{equation}
given by hopping element $t_2$, are finite. With realistic parameters, 
we arrived at the estimation of $|t_2|\sim 20$ meV, i.e.,
$|t_2|\sim J/3$. Similar to the orbital flipping three-site term 
(\ref{H3s1}), such a hopping process disturbs the AO order stabilized by 
the superexchange and induces string excitations. For this reason, its 
impact is largely confined to the high-energy part of the spectrum and 
is rather small for the low-energy QP state. This can be seen in Fig.
\ref{fig:spec_t_ab}, where we show the spectral density for 
$t_2=0.15t$ and $J=0.4t$: While the higher energy part is somewhat 
affected by finite $t_2$, the intensity and dispersion of the low-energy 
QP is almost the same as obtained for $t_2=0$, 
see Fig. \ref{fig:spec_cpt}(b).

%%%%%%%%%%%%%%%%%%%%%%%%%%%%%%%%%%%%%%%%%%%%%%%%%%%%%
%%                     figure 15
%%%%%%%%%%%%%%%%%%%%%%%%%%%%%%%%%%%%%%%%%%%%%%%%%%%%%

\begin{figure}[t!]
\centering
\subfigure{\includegraphics[width=0.4\textwidth]
    {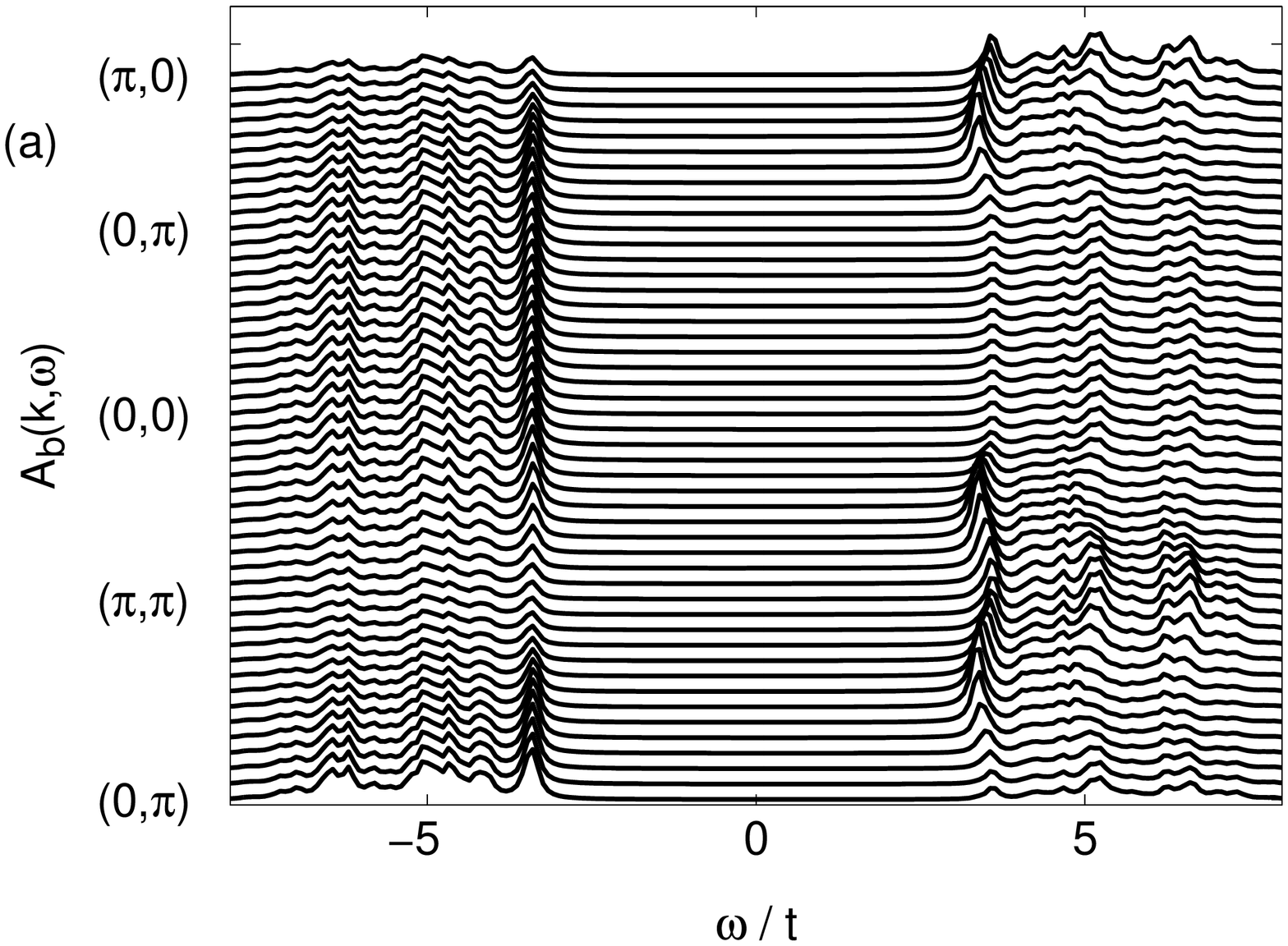}\label{fig:spec_t_prime_plus}}
\subfigure{\includegraphics[width=0.4\textwidth]
    {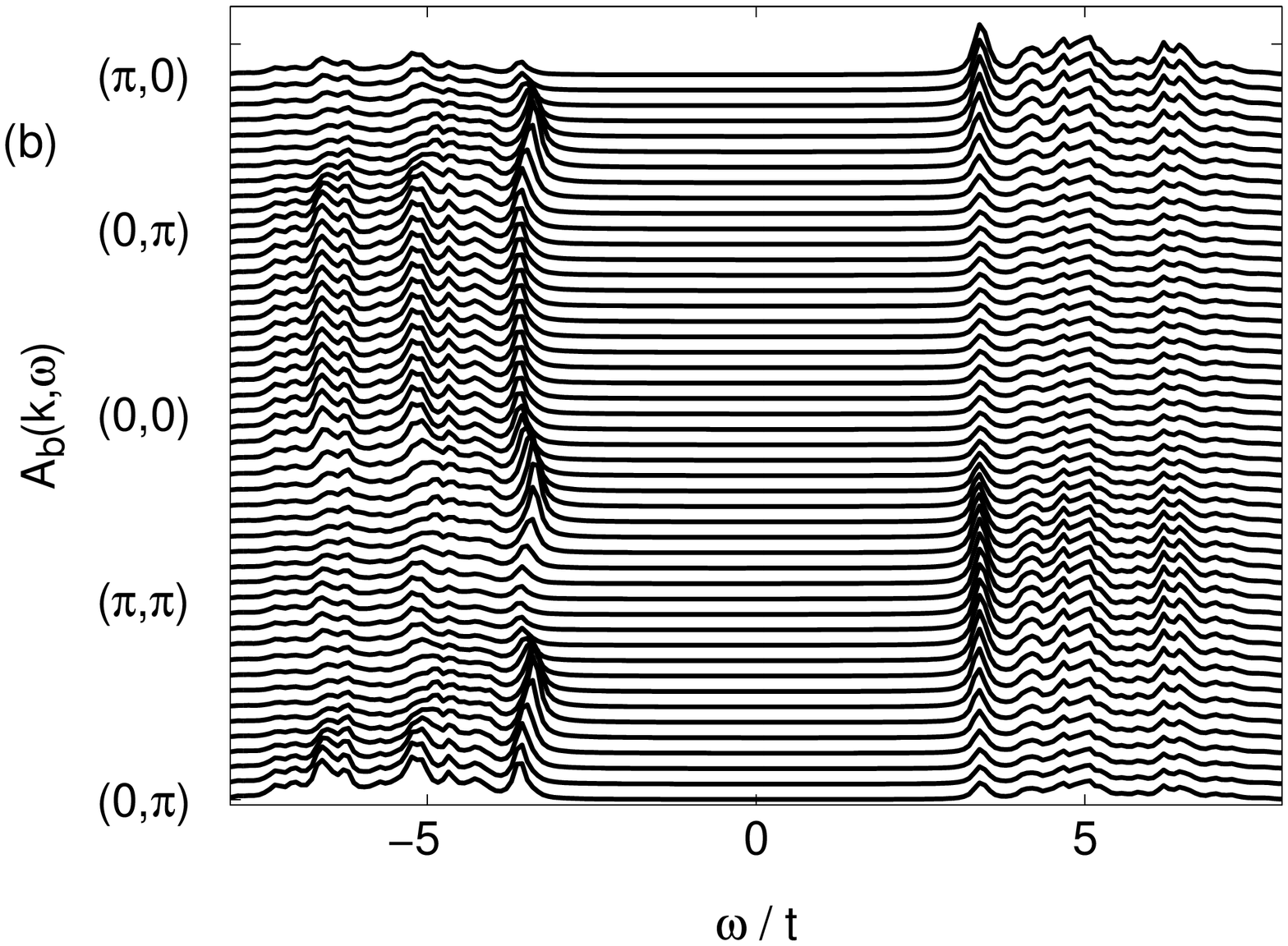}\label{fig:spec_t_prime_minus}}
\caption{
Photoemission [$(\omega-\mu)<0$] and inverse photoemission 
[$(\omega-\mu)>0$] part of the spectral density 
$A_b({\bf k},\omega)$ for a hole inserted into $b$ orbitals,
obtained within VCA for the $t_{2g}$ model (\ref{Ht2g}) with an 
additional longer-range third-neighbor hopping $t_3$ (\ref{Htp}). 
The value $t_3$ was selected to suppress dispersion arising from 
the three-site effective hopping (\ref{H3s0}) in: 
(a) the hole (photoemission) sector with $t_3=0.1t=J/4$, and 
(b) in the inverse photoemission sector with $t_3=-0.1t = -J/4$. 
Parameter: $U=10t$.
\label{fig:spec_t_prime}}
\end{figure}

The QP dispersion could also be influenced by the third-neighbor
hopping terms $t_3$, where the orbital symmetry leads to the same
anisotropy as for NN hopping: $a$ orbitals allow only hopping along
the $a$ axis, and $b$ orbitals only along the $b$ one:
\begin{equation}
\label{Htp}
H_{\rm t_3}=
- t_3 \sum_{\{imj\}\parallel a} b^\dag_ib_j
- t_3 \sum_{\{imj\}\parallel b} a^\dag_ia_j\;.
\end{equation}
Here the unit consisting of three sites $\{imj\}$, shown in Fig. 
\ref{fig:t2g_schem}(a), is parallel to one of the cubic axes in the
$(a,b)$ plane. In contrast to $t_2$ terms, these terms do not induce 
any string excitations but contribute only to the QP state itself, so
they mix with the three-site effective hopping $\tau$. To illustrate 
this effect, we have chosen $t_3=\pm J/4$ for the spectra shown in 
Fig. \ref{fig:spec_t_prime}. Note that the value of $|t_3|$ is here
larger than expected in transition metal oxides, where it is
in general smaller than the three-site hopping term $\tau=J/4$. The 
spectral density $A({\bf k},\omega)$ contains now the combined effects 
of the three-site terms $\propto\tau$ and third-neighbor hopping 
$\propto t_3$, and one finds that $t_3$, depending on its sign, 
can either amplify or weaken the QP dispersion which stems from the 
effective three-site hopping, see Fig. \ref{fig:spec_t_prime} 

%%%%%%%%%%%%%%%%%%%%%%%%%%%%%%%%%%%%%%%%%%%%%%%%%%%%%
%%                     figure 16
%%%%%%%%%%%%%%%%%%%%%%%%%%%%%%%%%%%%%%%%%%%%%%%%%%%%%

From the above example we have seen that the longer-range hopping
violates the particle-hole symmetry of the spectral functions. The 
spectra obtained for the original orbital Hubbard model (\ref{Ht2g})
with NN hopping $t$ obey the particle-hole symmetry. The three-site 
superexchange terms arise from this model, and therefore these terms 
also have to follow the particle-hole symmetry. This is in marked 
contrast to the $t_2$ terms that do not respect it,\cite{Fle97} or
to $t_3$ terms, see Fig. \ref{fig:spec_t_prime}. As a result, the 
spectra exhibit a striking \emph{particle-hole asymmetry\/} ---
reduced dispersion in the particle (inverse photoemission) sector 
corresponds to enhanced dispersion in the hole (photoemission) sector 
and {\it vice versa\/}. 

We will show now that the above asymmetry follows indeed from the 
difference between the NN and NNN hopping under particle-hole 
transformation. While this is transparent for the Hubbard model acting 
in the full Hilbert space, it is somewhat subtle for the $t$--$J$-like 
models. Thereby we focus on the $t_3$ hopping which influences
directly the QP dispersion. The operator for NN hopping can be
transformed from $\{c^{}_{\bf r},c^\dagger_{\bf r}\}$ electron 
operators to $\{h^\dagger_{\bf r},h_{\bf r}\}$ hole operators, and 
one arrives at an identical form for the kinetic energy, 
as long as a phase shift between the two sublattices is introduced:
\begin{equation}
\label{eq:trans}
h^\dagger_{\bf r} = (-1)^{(r_x+r_y)} c^{}_{\bf r}\,, \hskip .7cm 
h^{}_{\bf r} = (-1)^{(r_x+r_y)} c^\dagger_{\bf r}\,,
\end{equation}
where ${\bf r} = (r_x,r_y)$ is the lattice site. Hopping along the $a$
axis then becomes 
\begin{eqnarray}
\label{eq:trans_kin}
K_x&=&\sum_{\bf r} (c^\dagger_{\bf r}c^{}_{{\bf r} + \hat{{\bf a}}} +
c^\dagger_{{\bf r} + \hat{{\bf a}}}c^{}_{\bf r})  \nonumber \\
&=&\sum_{\bf r} \left\{\,
(-1)^{r_x+r_y}h^{}_{\bf r}(-1)^{r_x+1+r_y}h^\dagger_{{\bf r} +\hat{{\bf a}}} 
 \right.\nonumber \\
& &\hskip .5cm\left. + (-1)^{r_x+1+r_y}h^{}_{{\bf r} 
+ \hat{{\bf a}}}(-1)^{r_x+r_y}h^\dagger_{\bf r}
\;\right\} \nonumber \\
&=&-\sum_{\bf r} \left(h^{}_{\bf r}h^\dagger_{{\bf r} + \hat{{\bf a}}} +
h_{{\bf r} + \hat{{\bf a}}}h^\dagger_{\bf r}
\;\right) \nonumber \\
&=&\sum_{\bf r} \left(\;h^\dagger_{\bf r}h^{}_{{\bf r} + \hat{{\bf a}}} +
h^\dagger_{{\bf r} + \hat{{\bf a}}}h^{}_{\bf r}
\;\right)\;,
\end{eqnarray}
and analogously along the $b$ axis. The minus sign for one of the 
sublattices corresponds to a momentum shift by $\bf{q}=(\pi,\pi)$, as 
can be easily verified in the Fourier transform.
\begin{eqnarray}
\label{eq:trans_four}
h^{\dagger}_{\bf k} &=& \frac{1}{N}\sum_{\bf r}
\textrm{e}^{\textrm{i}\bf{kr}}
(-1)^{(r_x+r_y)}c^{}_{\bf r} \nonumber \\
&=&\frac{1}{N}\sum_{\bf r}
\textrm{e}^{\textrm{i}(\bf{k}+\bf{q})\bf{r}}
c^{}_{\bf r}=c^{}_{\bf{k}+\bf{q}}\,.
\end{eqnarray}
The on-site density-density interaction is not affected by the 
particle-hole transformation, apart from a shift in the chemical 
potential. 

Since the three-site hopping emerges from the Hubbard-like model with 
NN hopping, it respects particle-hole symmetry. Hence it obeys the same 
rules concerning particle-hole transformation, i.e., momentum $(0,0)$ 
for electrons is mapped to $(\pi,\pi)$ for holes. For the third-neighbor 
hopping $t_3$ (\ref{Htp}), however, the above transformation 
does not longer work, because both the creation and the annihilation 
operator act on the same sublattice. Instead the transformation vector 
would have to be ${\bf q}'=(\pi/2,\pi/2)$. Consequently, the combined
effect of explicit NNN hopping and three-site terms stemming from NN 
processes turns out to be strongly particle-hole asymmetric. 
For example, negative $t_3$ gives a band in the electron sector with 
the largest distance from the Fermi energy at momenta $(0,0)$ and 
$(\pi,\pi)$, and the values nearest to it at $(\pi/2,\pi/2)$, and the 
same is true for the three-site hopping. Consequently, the two 
dispersions add together and lead to increased total dispersion, see 
the photoemission part in Fig.~\ref{fig:spec_t_prime_minus}. On the 
contrary, in inverse photoemission the direct NNN hopping $t_3$ gives a 
maximal distance at $(\pi/2,\pi/2)$, while maximal energy is still
found at $(0,0)$ and $(\pi,\pi)$ for the three-site terms. Therefore, 
now $t_3$ and three-site hopping $\tau$ compete with each other, 
and the dispersion is weaker. For a particular choice of the model
parameters they can even cancel each other, as shown in the inverse 
photoemission part in Fig.~\ref{fig:spec_t_prime_minus}. Positive $t_3$ 
leads to the opposite result, see Fig.~\ref{fig:spec_t_prime_plus}. 
Thus, even large and unphysical values of $t_3$ not only do not destroy
the qualitative spectra predicted in the previous sections but result 
in asymmetry between the photoemission and inverse photoemission part 
of the spectra, so their contribution can easily be resolved.

The symmetry arguments leading to Eq. (\ref{eq:tab}) and Eq. 
(\ref{Htp}) remain valid also for systems with specific $e_g$ orbital 
degeneracy, as observed in certain fluorides with 2D AO order which 
involves alternating $z^2-y^2$ and $x^2-z^2$ orbitals.\cite{Wu07} 
In fact, the orbital model given by Eq. (\ref{t2gmodel}) describes also 
this case, as we show by a detailed derivation in the Appendix.
Hence, we conclude that the photoemission and inverse photoemission 
spectra for the planar vanadium oxide Sr$_2$VO$_4$ and for the planar 
K$_2$CuF$_4$ or Cs$_2$AgF$_4$ fluorides should be qualitatively similar 
to the spectral functions shown 
in Figs. \ref{fig:spec_scba} or \ref{fig:spec_cpt}.

%%%%%%%%%%%%%%%%%%%%%%%%%%%%%%%%%%%%%%%%%%%%%%%%%%%%%%%%
%                       Conclusions
%%%%%%%%%%%%%%%%%%%%%%%%%%%%%%%%%%%%%%%%%%%%%%%%%%%%%%%%
\section{Summary and Conclusions}
\label{sec:conclusions}

In this paper we analyzed only the orbital sector of the 
superexchange, which decides about the hole dynamics when spins 
are polarized in the FM ground state. We discussed all possible 
situations (see below) where the orbital symmetry leads to the
purely Ising superexchange in one and two dimensions. Exceptions 
from this rule are numerous systems
with $e_g$ orbital degrees of freedom,\cite{vdB99,Dag01,Fei99} or 
FM chains with two active orbitals,\cite{Kha01,Sir08} but we also 
provided examples of $e_g$ systems with Ising superexchange.

The 1D Hubbard-like model with two orbital flavors, but only one of 
them participating in NN hopping, served to explain the general 
principles and consequences of the Ising-like superexchange. 
Besides, this model stands for several physically relevant situations, 
including electrons moving within either $e_g$ or $t_{2g}$ orbitals 
in one dimension, and the 1D FK model.
We have shown that, particularly in all these cases, the relevant 
strong-coupling model has to include the three-site effective hopping. 
When both interorbital hopping and orbital-flip processes in the 
superexchange are absent, the three-site hopping term which arises 
from superexchange is crucial and is the only source of coherent hole 
propagation. 

We have shown that the 2D FK model with one immobile ($f$) and one 
mobile ($d$) orbital has many common features with the 1D model.
In both cases one finds only one dispersive mode for a hole inserted 
into the mobile orbital, and two non-dispersive modes for a hole doped 
in the immobile orbital. This latter hole excitation creates a trapped 
polaron, with the hole confined within a cluster consisting of 
a central site and its nearest neighbors 
(i.e., three sites in the 1D model, and five sites in the 2D model). 
While the hole can in principle escape from the polaron via three-site 
hopping process, we have shown that such processes have only very low 
spectral weight in the realistic regime of parameters, and thus the 
hole remains {\it de facto\/} trapped inside the polaron. In contrast 
to this almost perfectly localized hole, a hole in the mobile orbital
propagates freely, and its dispersion which originates from the 
three-site hopping is unrenormalized. Therefore, the two inequivalent 
orbital flavors behave in the FK models in a radically different way, 
and decouple from each other (interacting only by the on-site Coulomb 
interaction $U$, which stabilizes the AO ground state).

The model relevant for the 2D orbital physics in transition metal 
oxides leads, however, to qualitatively different results.
In the 2D $t_{2g}$ orbital model, which is also applicable to the AO 
state formed by $e_g$ orbitals in fluorides (see the Appendix), 
electrons do not separate into those confined to either sublattice 
(occupied by orbitals of particular symmetry in the ground state with
AO order), but may delocalize over the lattice and thereby undergo 
incoherent scattering on the orbital excitations, which strongly 
renormalizes and reduces the dispersion of the QP states. These QP
states arise at half filling in the regime of large Coulomb interaction 
$U$, when a hole (electon) is added to the ground state with AO order.
While electron hopping is of the purely 1D character, it selects by
symmetry possible three-site processes, which are responsible for the 
QP dispersion. Therefore, the dispersion of the QP state depends on 
the considered orbital and is again 1D, with a hole propagating 
coherently along the two crystal axes for the two orbitals. 

We emphasize that the mechanism of coherent hole propagation which 
occurs in the 2D $t_{2g}$ orbital model is completely different from 
the one known in the spin case. In orbital systems (with conserved
orbital flavors) it originates entirely from the three-site hopping 
processes, similar to hole propagation in the 1D or 2D FK model. But 
unlike in the latter models, in the 2D $t_{2g}$ case the QP bandwidth 
is strongly reduced from the value given by the amplitude of bare 
three-site hopping. We have explained this renormalization as 
following from incoherent string excitations which dress the coherent 
propagation and do not contribute additional momentum dependence.
As a special case, we have discussed the subtle interplay between the 
coherent hole propagation and string excitations in the 1D centipede 
model, where polaronic hole confinement competes with coherent 
propagation along the chain,
and which to some extent resembles the realistic 2D $t_{2g}$ case.

We discussed the impact of realistic longer-range hopping terms
(as expected in real materials such as vanadates or fluorides),
and found that the second neighbor terms are frustrated in the ground
state with AO order -- these processes would flip the orbital flavor, 
and are therefore suppressed at low energy, not affecting the QP 
dispersion. In contrast, third-neighbor hopping processes conserve 
orbital flavor and lead to a pronounced particle-hole asymmetry in the 
spectral weight distribution. In both cases, the 1D character of hole 
propagation which follows from the symmetry of involved orbitals, 
survives and determines the character of the spectral density at low 
energy.

In summary, we have demonstrated that orbital models with Ising
superexchange describe a broad class of interesting phenomena.
Spectral features resulting from such models exhibit weak momentum
dependence and are fundamentally different from those known from the 
the spin case with the SU(2)-symmetric superexchange. The predictions
of the theory presented in this paper provide an experimental challenge
for the transition metal oxides with orbital degrees of freedom, where
similar features could possibly be observed in FM planes with AO order.

\acknowledgments
We acknowledge financial support by: the Foundation for Polish Science 
(FNP), the Polish Ministry of Science and Higher Education under Project
No.~N202 068 32/1481, and the NSF under grant DMR-0706020. K.~W.
acknowledges as well support by the F. Kogutowska Foundation of the
Jagellonian University.

%%%%%%%%%%%%%%%%%%%%%%%%%%%%%%%%%%%%%%%%%%%%%%%%%%%%%%%%%%%%%%%
%                           Appendix
%%%%%%%%%%%%%%%%%%%%%%%%%%%%%%%%%%%%%%%%%%%%%%%%%%%%%%%%%%%%%%%
\appendix*
\section{Strong-coupling model for fluorides}
\label{app:fluo}

Here we show that the model developed in Sec. \ref{sec:t2g} may also
be applied to certain fluorides with FM planes and AO order.
In contrast to the $t_{2g}$ orbitally degenerate systems, in the
systems with $e_g$ orbital degeneracy the lattice distortions in the 
cubic phases are usually quite large. In particular, the \emph{static} 
distortions may counteract to some extent the AO order favored by the 
superexchange interactions, as e.g. in undoped manganites $R$MnO$_3$,
\cite{Fei99} or fluorides Cs$_2$AgF$_4$.\cite{Wu07} However, the 
crystal field does not suppress the orbital order present in these 
systems but instead it only modifies the occupied orbitals which form
the AO state. They have to be optimized in a microscopic model by 
choosing particular linear combinations of the $e_g$ orbitals, which 
form the AO order, in 
order to fit best to the superposition of the superexchange and the 
Jahn-Teller terms generated by ligand fields.\cite{Ole05} In certain 
situations this "modification" could be quite substantial and could 
even lead to such a selection of such $e_g$ orbitals that the resulting
state is modified to a ferro-type orbital order.~\cite{vdB99}

At finite crystal field splitting $\propto E_z$,
it is convenient to describe the changes in the occupied orbital 
states by making two complementary transformations at both sublattices,
\cite{vdB99} rotating the orbitals by an angle 
$\theta=\frac{\pi}{4}-\phi$ on sublattice $A$, and by an angle 
$\theta=\frac{\pi}{4}+\phi$ on sublattice $B$, so that the relative 
angle between the {\em occupied\/} orbitals is $\frac{\pi}{2}-2\phi$ 
and decreases with increasing $\phi$, i.e., with increasing $E_z$,
\begin{equation}
\left( \begin{array}{c}
 |\mu\rangle_i   \\
 |\nu\rangle_i
\end{array} \right) \! = \!
\left(\begin{array}{cc}
 \ \ \cos(\frac{\pi}{4}-\phi) & \sin(\frac{\pi}{4}-\phi) \\
    -\sin(\frac{\pi}{4}-\phi) & \cos(\frac{\pi}{4}-\phi)
\end{array} \right)\!
\left( \begin{array}{c}
 |\mbox{z}\rangle_i   \\
 |\mbox{x}\rangle_i
\end{array} \right) ,
\label{flopi}
\end{equation}
\begin{equation}
\left( \begin{array}{c}
 |\mu\rangle_j   \\
 |\nu\rangle_j
\end{array} \right) \! = \!
\left(\begin{array}{cc}
 \ \ \cos(\frac{\pi}{4}+\phi) & \sin(\frac{\pi}{4}+\phi) \\
    -\sin(\frac{\pi}{4}+\phi) & \cos(\frac{\pi}{4}+\phi)
\end{array} \right)\!
\left( \begin{array}{c}
 |\mbox{z}\rangle_j   \\
 |\mbox{x}\rangle_j
\end{array} \right) ,
\label{flopj}
\end{equation}
where the "old" orthogonal (basis) orbitals are defined as 
$|\mbox{x}\rangle_i=\frac{1}{\sqrt{2}}|x^2-y^2\rangle_i $ and 
$|\mbox{z}\rangle_i=\frac{1}{\sqrt{6}}|3z^2-r^2\rangle_i $ for
\textit{every} sublattice site $i$. Due to the above transformation
the AO order is formed now by $|\mu\rangle_i$ and $|\nu\rangle_j$ 
occupied orbitals at sublattices, $i \in A$ and $j \in B$, respectively. 
Let us stress that although the transformation defined by
Eqs. (\ref{flopi}--\ref{flopj}) is orthogonal this does not mean 
that orbitals on different sublattices, such as e.g. the occupied 
orbitals $|\mu\rangle_i$ and $|\nu\rangle_j$, 
are orthogonal for any arbitrary angle $\phi$.

For the 2D FM systems with active $e_g$ orbitals which are considered 
here, the relation between the crystal field $E_z$ and the optimal 
orbital configuration defined by the angle $\phi$ 
[see Eqs. (12) and (13) of Ref. \onlinecite{vdB99}] is given by:
\begin{equation}
\label{eq:phi}
E_z=4J \sin 2 \phi,
\end{equation}
where $J$ is the superexchange constant. In the case of fluorides such 
as Cs$_2$AgF$_4$ (Ref. \onlinecite{Wu07}) or K$_2$CuF$_4$ (Ref. 
\onlinecite{Hid83}) discussed here, the filling is one $e_g$ electron
per site and the crystal field would select the angle $\phi=\pi/12$
(for the reason of looking at this angle see below) since the convenient 
basis adapted to the actual AO order looks as follows: 
\begin{align}
\label{eq:basis}
\forall\, i\in A: \  & \left|\mu\left(\phi=\frac{\pi}{12}\right)\right\rangle_i =
\frac{1}{\sqrt{2}} |y^2-z^2 \rangle_i\equiv |x\rangle_i \,, \nonumber \\
& \ \left|\nu\left(\phi=\frac{\pi}{12}\right)\right\rangle_i 
= \frac{1}{\sqrt{6}} |3x^2-r^2 \rangle_i\equiv |z\rangle_i \,,   \nonumber \\
\forall\, j\in B: \  & \  
\left|\mu\left(\phi=\frac{\pi}{12}\right)\right\rangle_j =
\frac{1}{\sqrt{6}} |3y^2-r^2 \rangle_j\equiv |z\rangle_j \,, \nonumber \\
& \ \left|\nu\left(\phi=\frac{\pi}{12}\right)\right\rangle_j =
\frac{1}{\sqrt{2}} |x^2-z^2\rangle_j\equiv |x\rangle_j 
                     \,, 
\end{align}
where the occupied (empty) orbitals for this type of AO order are 
denoted as $|x\rangle$ ($|z\rangle$) on both sublattices.

The reason why these particular pairs of basis orbitals 
(\ref{eq:basis}) are interesting here is that this is the only choice 
of occupied $e_g$ flavors which forms a two-sublattice AO order with
the interorbital hopping between occupied orbitals vanishing by 
symmetry, and where the interactions described by pseudospin 
operators do not allow for any quantum fluctuations. This resembles the 
$t_{2g}$ case discussed in this paper. There is, however, one subtle 
difference: two occupied $\{|x\rangle_i,|x\rangle_j\}$ orbitals on 
sublattices $A$ and $B$ are not orthogonal and do not form the global 
basis in the $e_g$ orbital space. The choice made in Eq. 
(\ref{eq:basis}) means that we 
consider two different pairs of orbitals for both sublattices, and the
interorbital hopping between the \emph{unoccupied} orbitals is also 
rather small but remains finite.\cite{noteint} Hence, the respective 
strong-coupling Hamiltonian is richer than the one for the $t_{2g}$ 
case and we need to check under which conditions it can be reduced to 
a similar polaron Hamiltonian as Eq. (\ref{t2geff}).

The $e_g$ orbital $t$-$J$ Hamiltonian for the FM planes without 
the three-site terms but including the crystal field was given e.g. 
in Ref. \onlinecite{Bal02}. Here we rewrite the kinetic term in a 
slightly different form (there it was written already using the 
polaron representation) and substitute $\phi =\pi / 12$ to obtain:
\begin{equation}
{\cal H}_{e_g} = H_t +H_J+H_z\,,
\end{equation}
where
\begin{eqnarray}
\label{Ht_fl}
H_{t} &=&- \frac{1}{2}t\sum_{i}
( \tilde{z}^\dag_i \tilde{z}_{i+\hat{a}}+
\tilde{z}^\dag_i \tilde{z}_{i+\hat{b}} +\mbox{h.c.}) \nonumber \\
&-&\frac{\sqrt{3}}{2}t \sum_{ i\ \in A}
( \tilde{z}^\dag_i \tilde{x}_{i+\hat{a}}+
\tilde{z}^\dag_i \tilde{x}_{i+\hat{b}} +\mbox{h.c.})\nonumber \\
&-&\frac{\sqrt{3}}{2}t\sum_{ i\ \in B}
( \tilde{x}^\dag_i \tilde{z}_{i+\hat{a}}+
\tilde{x}^\dag_i \tilde{z}_{i+\hat{b}} +\mbox{h.c.})         \;, \\
\label{HJ_fl}
H_{J} &=& \frac12 J \sum_{\langle ij\rangle || \hat{a} }
\left(T^z_i T^z_j + \sqrt{3} T^z_i T^x_j\right) \nonumber \\
&+&\frac12 J \sum_{\langle ij\rangle || \hat{b} }
\left(T^z_i T^z_j - \sqrt{3} T^x_i T^z_j\right) \;, \\
H_{z} &= &-\frac{1}{4} J \sum_{ i \in A} (T^z_i+\sqrt{3}T^x_i)
\nonumber \\
&+&\frac{1}{4} J \sum_{ i \in B} (T^z_i-\sqrt{3}T^x_i) \;.
\end{eqnarray}
Here $T^z_i = \frac{1}{2} (\tilde{n}_{iz}-\tilde{n}_{ix})$ for $i \in A$,
$T^z_j = \frac{1}{2} (\tilde{n}_{jx}-\tilde{n}_{jz})$ for $j \in B$,
and $T^x_i=\frac{1}{2}(\tilde{x}^\dag_i\tilde{z}_i+\tilde{z}^\dag_i
\tilde{x}_i)$ for every site $i$, see Ref. \onlinecite{vdB99}. 
As before, a tilde above a fermion operator indicates that the Hilbert
space is restricted to unoccupied and singly occupied sites,
e.g. $\tilde{x}^\dag_i=x_i^\dag(1-n_{iz})$.
The last term $H_z$ represents the above mentioned crystal field 
with the strength of the interaction written according to 
Eq. (\ref{eq:phi}) with $\phi =\pi/12$.

However, we are not aware of any work where the three-site terms 
complementing such a $t$-$J$ model have been derived. We use second 
again order perturbation theory\cite{Cha77,Esk94} applied to the Hubbard 
model for spinless $e_g$ electrons in a FM plane,\cite{Fei05} with the 
basis rotated by $\phi=\pi/12$, following Eqs. (12) and (13) of
Ref. \onlinecite{vdB99}. This leads to the following three-site terms 
for the $e_g$ strong-coupling model (with $\phi=\pi/12$):
\begin{equation}
H_{\tau}=H^a_{\tau}+H^b_{\tau}+H^{ab}_{\tau},
\end{equation}
where
\begin{widetext}
\begin{eqnarray}
\label{Heg_3sa} H^a_{\tau}= &-&\!\! \frac{1}{4}\tau
\sum_{i \in A} \Big[ \tilde{z}^\dag_{i-\hat{a}} \tilde{n}_{ix}
\tilde{z}_{i+\hat{a}} +\underline{3 \tilde{x}^\dag_{i-\hat{a}}
\tilde{n}_{ix} \tilde{x}_{i+\hat{a}}} +
\underline{\underline{\sqrt{3} \tilde{x}^\dag_{i-\hat{a}}
\tilde{n}_{ix} \tilde{z}_{i+\hat{a}}}} +
\underline{\underline{\sqrt{3} \tilde{z}^\dag_{i-\hat{a}}
\tilde{n}_{ix} \tilde{x}_{i+\hat{a}}}}
+ \mbox{h.c.} \Big] \nonumber \\
&-&\frac{1}{4}\tau \sum_{i\in B}
\Big[\tilde{z}^\dag_{i-\hat{a}} \tilde{n}_{ix}
\tilde{z}_{i+\hat{a}} +3 \tilde{z}^\dag_{i-\hat{a}} \tilde{n}_{iz}
\tilde{z}_{i+\hat{a}} -\sqrt{3}
\tilde{z}^\dag_{i-\hat{a}}\tilde{z}^\dag_{i}\tilde{x}_{i}
\tilde{z}_{i+\hat{a}} -\sqrt{3} \tilde{z}^\dag_{i-\hat{a}}
\tilde{x}^\dag_{i}\tilde{z}_{i} \tilde{z}_{i+\hat{a}} + \mbox{h.c.}
\Big]\,,
\end{eqnarray}
\begin{eqnarray}
\label{Heg_3sb}
H^b_{\tau}&= &- \frac{1}{4}\tau \sum_{i \in A}
\Big[
 \tilde{z}^\dag_{i-\hat{b}} \tilde{n}_{ix} \tilde{z}_{i+\hat{b}}
+3 \tilde{z}^\dag_{i-\hat{b}} \tilde{n}_{iz} \tilde{z}_{i+\hat{b}}
-\sqrt{3}
\tilde{z}^\dag_{i-\hat{b}} \tilde{z}^\dag_{i}\tilde{x}_{i}
\tilde{z}_{i+\hat{b}}
-\sqrt{3} \tilde{z}^\dag_{i-\hat{b}}
\tilde{x}^\dag_{i}\tilde{z}_{i} \tilde{z}_{i+\hat{b}} 
+ \mbox{h.c.} \Big] \nonumber \\
&-& \frac{1}{4}\tau \sum_{i \in B} \Big[
\tilde{z}^\dag_{i-\hat{b}} \tilde{n}_{ix} \tilde{z}_{i+\hat{b}}
+\underline{3\tilde{x}^\dag_{i-\hat{b}} \tilde{n}_{ix}
\tilde{x}_{i+\hat{b}}} 
+ \underline{\underline{\sqrt{3}\tilde{x}^\dag_{i-\hat{b}}
\tilde{n}_{ix} \tilde{z}_{i+\hat{b}}}}
+ \underline{\underline{\sqrt{3}
\tilde{z}^\dag_{i-\hat{b}} \tilde{n}_{ix} \tilde{x}_{i+\hat{b}}}}
 + \mbox{h.c.} \Big]\,,
\end{eqnarray}
\begin{eqnarray}
\label{Heg_3sab} 
H^{ab}_{\tau}&= &- \frac{1}{4}\tau \sum_{i
\in A} \Big[ \tilde{z}^\dag_{i\pm \hat{a}} \tilde{n}_{ix}
\tilde{z}_{i\pm \hat{b}} -3\tilde{x}^\dag_{i\pm\hat{a}}
\tilde{x}^\dag_{i}\tilde{z}_{i} \tilde{z}_{i\pm \hat{b}} +
\underline{\underline{\sqrt{3}\tilde{x}^\dag_{i \pm\hat{a}}
\tilde{n}_{ix} \tilde{z}_{i \pm\hat{b}}}}
-\sqrt{3}\tilde{z}^\dag_{i\pm \hat{a}}
\tilde{x}^\dag_{i}\tilde{z}_{i}
\tilde{z}_{i \pm \hat{b}} \nonumber \\
&+& \tilde{z}^\dag_{i\pm \hat{a}} \tilde{n}_{ix} \tilde{z}_{i\mp
\hat{b}} -3 \tilde{x}^\dag_{i\pm\hat{a}}
\tilde{x}^\dag_{i}\tilde{z}_{i} \tilde{z}_{i\mp \hat{b}} +
\underline{\underline{\sqrt{3} \tilde{x}^\dag_{i \pm\hat{a}}
\tilde{n}_{ix} \tilde{z}_{i \mp \hat{b}}}} -\sqrt{3}
\tilde{z}^\dag_{i\pm \hat{a}} \tilde{x}^\dag_{i}\tilde{z}_{i}
\tilde{z}_{i \mp \hat{b}}
+ \mbox{h.c.} \Big] \nonumber \\
&-& \frac{1}{4}\tau \sum_{i \in B}
\Big[\tilde{z}^\dag_{i\pm\hat{a}} \tilde{n}_{ix}
\tilde{z}_{i\pm\hat{b}}-3 \tilde{z}^\dag_{i\pm \hat{a}}
\tilde{z}^\dag_{i}\tilde{x}_{i} \tilde{x}_{i\pm \hat{b}} +
\underline{\underline{\sqrt{3} \tilde{z}^\dag_{i\pm \hat{a}}
\tilde{n}_{ix} \tilde{x}_{i\pm \hat{b}}}} -\sqrt{3}
\tilde{z}^\dag_{i\pm\hat{a}} \tilde{z}^\dag_{i}\tilde{x}_{i}
\tilde{z}_{i\pm \hat{b}} \nonumber \\
&+& \tilde{z}^\dag_{i\pm\hat{a}} \tilde{n}_{ix}
\tilde{z}_{i\mp\hat{b}} -3 \tilde{z}^\dag_{i\pm \hat{a}}
\tilde{z}^\dag_{i}\tilde{x}_{i} \tilde{x}_{i\mp \hat{b}} +
\underline{\underline{\sqrt{3} \tilde{z}^\dag_{i\pm \hat{a}}
\tilde{n}_{ix} \tilde{x}_{i\mp \hat{b}}}} -\sqrt{3}
\tilde{z}^\dag_{i\pm\hat{a}} \tilde{z}^\dag_{i}\tilde{x}_{i}
\tilde{z}_{i\mp \hat{b}} + \mbox{h.c.} \Big]\,.
\end{eqnarray}
%\end{subequations}
\end{widetext}
Here we underlined (doubly underlined) terms which do not require
orbital excitations (require orbital excitations), respectively, i.e.
\begin{equation}
H^{(0)}_{\tau}=\underline{H_{\tau}}, \hskip 1cm
H^{(1)}_{\tau}=\underline{\underline{H_{\tau}}}.
\end{equation}

Next, we perform the same standard transformation to obtain the polaron
Hamiltonian from the strong-coupling model \cite{Mar91} for the lightly
doped ordered states as done in Sec. \ref{sec_t_2g_vc_scba}, i.e. we introduce
boson operators $\alpha_i$ (orbitons) and fermion operators $h_i$ 
(holons) which are related to the $x_i$ and $z_i$
operators in the following way:
\begin{equation}
\tilde{x}^\dag_i \equiv h_i(1-\alpha^\dag_i \alpha_i),
\hskip 1cm
\tilde{z}^\dag_i \equiv h_i \alpha_i^\dag.
\end{equation}
Please note, however, that here we did not have to perform rotation of
the pseudospins since we defined distinct electron operators for the 
occupied and empty orbitals, cf. Eq. (\ref{eq:basis}).

Again we implement a linear orbital-wave approximation\cite{vdB99} 
(we keep only linear terms in orbiton operators) and we skip 
$(1-h^\dag_i h_i)$ operators (which in fact is not an approximation if 
there is only one hole in the entire plane). This means that e.g. the 
three-site terms are reduced only to the terms which were either 
underlined or doubly underlined in Eqs. (\ref{Heg_3sa})--(\ref{Heg_3sab}), 
i.e. to either $H^{(0)}_{\tau}$ or $H^{(1)}_{\tau}$. Here, however, we 
have to use yet another approximation which was unnecessary for the
$t_{2g}$ model: we skip terms $H^{(1)}_{\tau}$ which were absent in
Sec. \ref{sec_homa}. This approximation is allowed since these terms 
contribute to the vertex as $\propto\tau$ and not as $\propto t$,
resulting typically in much reduced energy scale for the new vertex 
contributions. Furthermore, we showed in Sec. \ref{sec:NNN} that such 
terms [cf. Eq. (\ref{Htp}) and Fig. \ref{fig:spec_t_ab}] do not change 
the energy of the QP and merely modify the incoherent spectrum. 
Eventually, we arrive at the polaron Hamiltonian for the holes doped 
into the $e_g$ orbitals of the fluorides, with the hopping terms:
\begin{eqnarray}
\label{Heg_pol1}
H_t&=& \sqrt{3}t\frac{1}{\sqrt{N}}\sum_{{\bf k},{\bf q}}
\left\{\cos(k_x-q_x) h^\dag_{{\bf k}A}h_{{\bf k}-{\bf q},B}^{}
\alpha_{{\bf q}A}^{}\right. \nonumber \\
&+&\left.\cos(k_y\!-\!q_y) h^\dag_{{\bf k}B}h_{{\bf k}-{\bf q},A}^{}
\alpha_{{\bf q}B}^{}+{\rm h.c.}\right\}\,,
\\
\label{Heg_pol2}
H^{(0)}_{\tau}&=& \frac{3}{2}\tau
\sum_{k}\left\{\cos(2k_y)h_{{\bf k} A}^\dag h_{{\bf k} A}^{} 
+ \cos(2 k_x) h_{{\bf k} B}^\dag h_{{\bf k} B}^{}\right\}\,,\nonumber\\
\label{app:1}
\end{eqnarray}
and the remaining terms resulting in the energy renormalization
\begin{equation}
\label{Heg_pol3}
H_J+H_z=\frac{3}{4}J\sum_{k}
\left(\alpha_{{\bf k}A}^\dag \alpha_{{\bf k}A}^{} 
+ \alpha_{{\bf k}B}^\dag\alpha_{{\bf k}B}^{}\right)\,.
\end{equation}
Therefore, the Hamiltonian given by Eqs. 
(\ref{Heg_pol1})--(\ref{Heg_pol3}) reduces to the polaron Hamiltonian 
(\ref{t2geff}) after substituting $\sqrt{3}t/2 \rightarrow t$, and 
consequently $3J/4 \rightarrow J $ and $3\tau/4 \rightarrow \tau$.
This substitution stems from the different definitions of the
hopping $t$ in the $e_g$ and in the $t_{2g}$ systems --- in the former
case it is the $(dd\sigma)$ hopping between the $3z^2-r^2$ orbitals
along the $\hat{c}$ direction, whereas in the latter case it is the 
hopping element between a pair of active $t_{2g}$ orbitals, e.g. $yz$ 
orbitals in the ($a,b)$ plane.

In summary, we have shown that the Hamiltonian given by Eqs. 
(\ref{Heg_pol1})--(\ref{Heg_pol3}) provides the framework to analyze 
the behavior of certain lightly doped $e_g$ systems, with FM planes and
AO order which suppresses the interorbital hopping between occupied
orbitals. Its equivalence to the polaron model (\ref{t2geff}) 
demonstrates that the results obtained and discussed in Sec. 
\ref{sec:t2g} should also apply to the case of a hole doped into the 
fluoride plane with the AO ordered $e_g$ orbitals.

%%%%%%%%%%%%%%%%%%%%%%%%%%%%%%%%%%%%%%%%%%%%%%%%%%%%%%%%%%%%%%%%%%%
%%                             REFERENCES
%%%%%%%%%%%%%%%%%%%%%%%%%%%%%%%%%%%%%%%%%%%%%%%%%%%%%%%%%%%%%%%%%%%


\begin{thebibliography}{99}
\bibitem{notebet} For a minority of sites on a square lattice with AO 
                  order this number is equal to $7$ or $11$.
\bibitem{notenob} The Green's function $G_b(k,\omega)$ does not show
                  new qualitative features as compared with the 
                  solution obtained for the 1D orbital chain of Sec. 
                  \ref{sec:1Dmodel}.
\bibitem{notesub} There is however some subtlety related to this statement.
		In the 1D model there are actually two interrelated reasons why 
                the hole cannot move coherently when doped into the $b$ orbital: 
                (i) the creation of the defect after the first $t$ step which has 
                to be erased by the hole before the hole itself is annihilated, and 
                (ii) the fact that this defect blocks the hole motion by three-site 
                terms in one direction. However, in the SCBA treatment the 
                latter constraint is neglected [as we skipped projection operators 
                $(1-\alpha_i^\dag \alpha_i)$ in the three-site terms when 
                deriving the effective Hamiltonian (\ref{t2geff})], so it is 
                the point (i) which suffices alone to confine the hole.
\bibitem{noteint} The physical reason for this is that finite crystal field
                  cannot fully eliminate the interorbital hopping.
\end{thebibliography}
\end{document}